\documentclass[prx,twocolumn,floatfix,superscriptaddress,longbibliography,notitlepage, nofootinbib]{revtex4-1}

\usepackage{graphicx, color, graphpap}% Include figure files
\usepackage{enumitem}
\usepackage{amssymb}
\usepackage{amsthm}
\usepackage{dsfont}
\usepackage{mathtools}
\usepackage[dvipsnames]{xcolor}
\usepackage{multirow}
\usepackage[colorlinks=true,citecolor=blue,linkcolor=magenta]{hyperref}
\usepackage[T1]{fontenc}
\usepackage{bbm}
\usepackage{thmtools,thm-restate}
\usepackage{verbatim}
\usepackage{mathtools}
\usepackage{titlesec}
\usepackage{amsmath}
\usepackage{subfigure}
\usepackage[normalem]{ulem}

\usepackage{listings}% http://ctan.org/pkg/listings
\usepackage{csquotes}

\makeatletter 
\renewcommand\onecolumngrid{
\do@columngrid{one}{\@ne}
\def\set@footnotewidth{\onecolumngrid}
\def\footnoterule{\kern-6pt\hrule width 1.5in\kern6pt}
}
\renewcommand\twocolumngrid{
        \def\footnoterule{
        \dimen@\skip\footins\divide\dimen@\thr@@
        \kern-\dimen@\hrule width.5in\kern\dimen@}
        \do@columngrid{mlt}{\tw@}
}
\makeatother
\long\def\ca#1\cb{} %Use for commenting out: \ca...\cb

\newcommand{\braket}[2]{\langle #1 \hspace{1pt} | \hspace{1pt} #2 \rangle}

\newcommand{\ketbra}[2]{|  #1 \rangle \! \langle #2  |}

\newcommand{\ket}[1]{|#1\rangle}               %ket
\newcommand{\bra}[1]{\langle #1|}              %bra
\newcommand{\dya}[1]{\ket{#1}\!\bra{#1}}
\newcommand{\poly}{\operatorname{poly}}

\newcommand{\rank}{\text{rank}}

\newcommand{\Tr}{{\rm Tr}}

\renewcommand{\geq}{\geqslant}
\renewcommand{\leq}{\leqslant}

\DeclareMathOperator*{\argmin}{arg\,min}
\renewcommand{\vec}[1]{\boldsymbol{#1}}  % Bold vectors instead of arrow vectors

\newcommand{\ad}{^\dagger}
\newcommand*{\id}{\openone}

\newcommand{\jakarta}{ibmq\_jakarta}
\newcommand{\casablanca}{ibmq\_casablanca}
   
   \theoremstyle{plain}
\newtheorem{thm}{Theorem}
\newtheorem{corol}[thm]{Corollary}

\newtheorem{propos}[thm]{Proposition}
\theoremstyle{definition}
\newtheorem{defn}{Definition}
\theoremstyle{remark}
\newtheorem{rmk}[thm]{Remark}

{}%         Body font
{}%         Indent amount (empty = no indent, \parindent 
\newcommand{\zo}[1]{{\color[RGB]{66, 135, 245}{ZH: #1}}} 
\newcommand{\eb}[1]{{\color[RGB]{250, 75, 43}{EB: #1}}}
\begin{document}

\title{Quantum Mixed State Compiling}

\author{Nic Ezzell}
\email[]{naezzell@proton.me}
\affiliation{Department of Physics \& Astronomy, University of Southern California, Los Angeles, California 90089, USA}
\affiliation{Information Sciences, Los Alamos National Laboratory, Los Alamos, New Mexico 87545, USA}

\author{Elliott M. Ball}
\affiliation{Physics Department, Lancaster University, Lancaster, United Kingdom, LA1 4YB}

\author{Aliza U. Siddiqui}
\affiliation{Hearne Institute for Theoretical Physics, Department of Physics and Astronomy,
and Center for Computation and Technology, Louisiana State University, Baton
Rouge, Louisiana 70803, USA}
\affiliation{Division of Computer Science and Engineering, Louisiana State University, Baton Rouge, LA 70803, United States of America}

\author{Mark M. Wilde} 
\affiliation{Hearne Institute for Theoretical Physics, Department of Physics and Astronomy,
and Center for Computation and Technology, Louisiana State University, Baton
Rouge, Louisiana 70803, USA}
\affiliation{School of Electrical and Computer Engineering, Cornell University, Ithaca, New York 14850, USA}

\author{Andrew T. Sornborger} 
\affiliation{Information Sciences, Los Alamos National Laboratory, Los Alamos, New Mexico 87545, USA}
\affiliation{Quantum Science Center, Oak Ridge, TN 37931, USA}

\author{Patrick~J.~Coles} 
\affiliation{Theoretical Division, Los Alamos National Laboratory, Los Alamos, New Mexico 87545, USA}
\affiliation{Quantum Science Center, Oak Ridge, TN 37931, USA}

\author{Zo\"{e} Holmes}
\affiliation{Information Sciences, Los Alamos National Laboratory, Los Alamos, New Mexico 87545, USA}
\affiliation{Institute of Physics, Ecole Polytechnique F\'{e}d\'{e}rale de Lausanne (EPFL), CH-1015 Lausanne, Switzerland}

\begin{abstract}
The task of learning a quantum circuit to prepare a given mixed state is a fundamental quantum subroutine. We present a variational quantum algorithm (VQA) to learn mixed states which is suitable for near-term hardware. Our algorithm represents a generalization of previous VQAs that aimed at learning preparation circuits for pure states. We consider two different ans\"{a}tze for compiling the target state; the first is based on learning a purification of the state and the second on representing it as a convex combination of pure states. In both cases, the resources required to store and manipulate the compiled state grow with the rank of the approximation. Thus, by learning a lower rank approximation of the target state, our algorithm provides a means of compressing a state for more efficient processing. As a byproduct of our algorithm, one effectively learns the principal components of the target state, and hence our algorithm further provides a new method for principal component analysis. We investigate the efficacy of our algorithm through extensive numerical implementations, showing that typical random states and thermal states of many body systems may be learnt this way. Additionally, we demonstrate on quantum hardware how our algorithm can be used to study hardware noise-induced states. 

%In the case of the convex combination ansatz, one in effect learns the principal components of the target state, and hence our algorithm further provides a new method for principal component analysis. 

% We learn mixed states using two strategies which are amenable to NISQ devices and one of which scales. In both cases, it's possible to learn a lower rank approximation of the full state, and given QLRAP paper, we compare empirical performance to optimal lower rank approximation. In both methods, the lower rank approximation gives the principal components of target state. 
\end{abstract}
 
\maketitle

\section{Introduction}

% \textbf{State tomography} 
The task of learning an unknown $d \times d$ quantum state $\rho$ is a fundamental primitive for quantum computing. The well known method of full quantum state tomography learns the matrix elements of $\rho$ directly. The original  scheme employs $\mathcal{O}(d^2 / \epsilon^2)$ Pauli measurements to learn the state up to additive error $\epsilon$ in trace distance~\cite{nielsen2000quantum}. Subsequent refinements require only $\mathcal{O}((\rank(\rho) d / \epsilon^2) \cdot \ln(d / \epsilon))$ measurements~\cite{haah2017sample}, or $\mathcal{O}(d / \epsilon^2)$ allowing for a small failure probability~\cite{o2016efficient}. A recent improvement has found it is necessary to use at least $\Omega(\rank(\rho) d / \epsilon)$ measurements, and it was also conjectured to be sufficient~\cite{yuen2022improved}. In all cases, tomography aims to obtain a classical description of a quantum state, and as such, the number of measurements it requires scales exponentially with the number of qubits of the target state.

% , and if we accept an error larger than $\epsilon$ \emph{with low probability}, then $\mathcal{O}(d / \epsilon^2)$ measurements are sufficient~\cite{odonnell_efficient_2016}.
% Methods inspired by compressed sensing can even achieve the $\mathcal{O}(\rank(\rho) d / \epsilon^2)\ln(d / \epsilon)$ scaling where the data itself reveals whether $\rank(\rho)$ was too large for the number of distinct measurements operators used to succeed~\cite{gross_quantum_2010}.
% Methods inspired by compressed sensing can even achieve the $\mathcal{O}(\rank(\rho) d / \epsilon^2)\ln(d / \epsilon)$ scaling where the data itself reveals whether $\rank(\rho)$ was too large for the number of distinct measurements operators used to succeed~\cite{gross_quantum_2010}.

% \textbf{Classical Shadows}
An alternative approach is to give a more operational meaning to learning. In practice, we are often not interested in the exact form of $\rho$ but instead in its properties; i.e., we wish to estimate $o_i \coloneqq \Tr[\rho O_i]$ for some observable $O_i$. Ref.~\cite{aaronson2019shadow} proves that if each observable $O_i$, for $1 \leq i \leq M$, is restricted to two-outcome measurements, then only $\widetilde{\mathcal{O}}(\epsilon^{-4} \log^4 M \log d)$ copies of $\rho$ are sufficient to estimate each $o_i$ up to additive error $\epsilon$ by using a method called \textit{shadow tomography}~\cite{huang2020predicting, elben2022randomized, huang2021provably}.
% ~\footnote{The $\Tilde{\mathcal{O}}$ indicates that some $\poly(\log\log)$ factors are excluded for compactness.}
These results were expanded to simpler, experimentally tractable Clifford measurements and random Pauli measurements in Refs.~\cite{huang2020predicting,paini2021estimating}. In particular, it was shown that a polynomial number of Clifford (Pauli) measurements is sufficient to estimate any low rank (local) observable. The polynomial scaling achieved by shadow tomography is gained at the cost of providing only a partial description of the quantum state.

% An alternative approach is to give a more operational meaning to learning. In practice, we are often not interested in the exact form of $\rho$ but instead in properties of it, i.e., we wish to estimate $o_i = \Tr[\rho O_i]$ for some observable $O_i$. In Aaronson's influential \enquote{Shadow Tomography of Quantum States} paper~\cite{aaronson_shadow_2018}, he proves that if the $O_i, 1 \leq i \leq M$ are restricted to 2-outcome measurements, then only $\Tilde{\mathcal{O}}(\epsilon^{-4} \log^4 M \log d)$ copies of $\rho$ are sufficient to estimate the $o_i$ up to error $\epsilon$.huang2020predicting
% % ~\footnote{The $\Tilde{\mathcal{O}}$ indicates that some $\poly(\log\log)$ factors are excluded for compactness.}
% Hence, a polynomial number of measurements is sufficient to estimate an exponential number of observables. These results were further expanded for qubits by Huang, Kueng, and Preskill~ \cite{huang2020predicting}, and Paini \textit{et al}.~\cite{paini2021estimating} who considered simpler, experimentally tractable single qubit Clifford or Haar random measurements respectively. In short, the same polynomial requirement remains depending on the choice of measured $O_i$ and those we wish to estimate. 

% \textbf{Compiling}
Yet another approach to quantum state learning, and the one we take here, is instead to solve the problem of state \emph{compilation}. This involves learning a quantum circuit with which we may prepare an approximation of the target state. In this paper, we propose and demonstrate a near-term algorithm for approximately compiling an unknown quantum state by variationally training a parameterized quantum circuit. Similar to the shadow tomography case, one can use the final cost function value in our algorithm to place bounds on the deviation of observables estimated from the learnt state from those of the target state.

While the variational compilation of pure states has been explored in Refs.~\cite{khatri2019quantum, Jones2022robustquantum, sharma2019noise}, here we focus on mixed state compilation. Specifically, we propose two different ans\"{a}tze that can compile a mixed state either into a purification using ancilla or as a convex combination of pure states. The latter is similar to the Hamiltonian-based model approach in Ref.~\cite{verdon2019quantum}. In addition, our approach shares similarities with  Refs.~\cite{larose2019variational,cerezo2020variational}, which present algorithms to learn the diagonalization of a target state, but in contrast to our approach, these methods only apply to low-rank quantum states. %The proposal in Ref.~\cite{verdon2019quantum} is also similar here.

% Due to the variational nature of the algorithm, the exact scaling of the number of copies of $\rho$ is hard analyse, we find numerically that the scaling is substantially better than $\mathcal{O}(d^2 / \epsilon^2)$, and is even easier for small $R$ and highly structured problems. 

% Quantum state tomography is one way of approaching this task. However, as tomography aims to obtain a classical description of a quantum state, the number of measurements it requires in general scales exponentially with the size of the target. In many instances, it is thus more appropriate to learn a quantum circuit that can prepare the target state. The quantum description obtained from compiling a state in this way need not scale exponentially.

\begin{figure}[ht]
    \centering
    \includegraphics[width=0.49\textwidth]{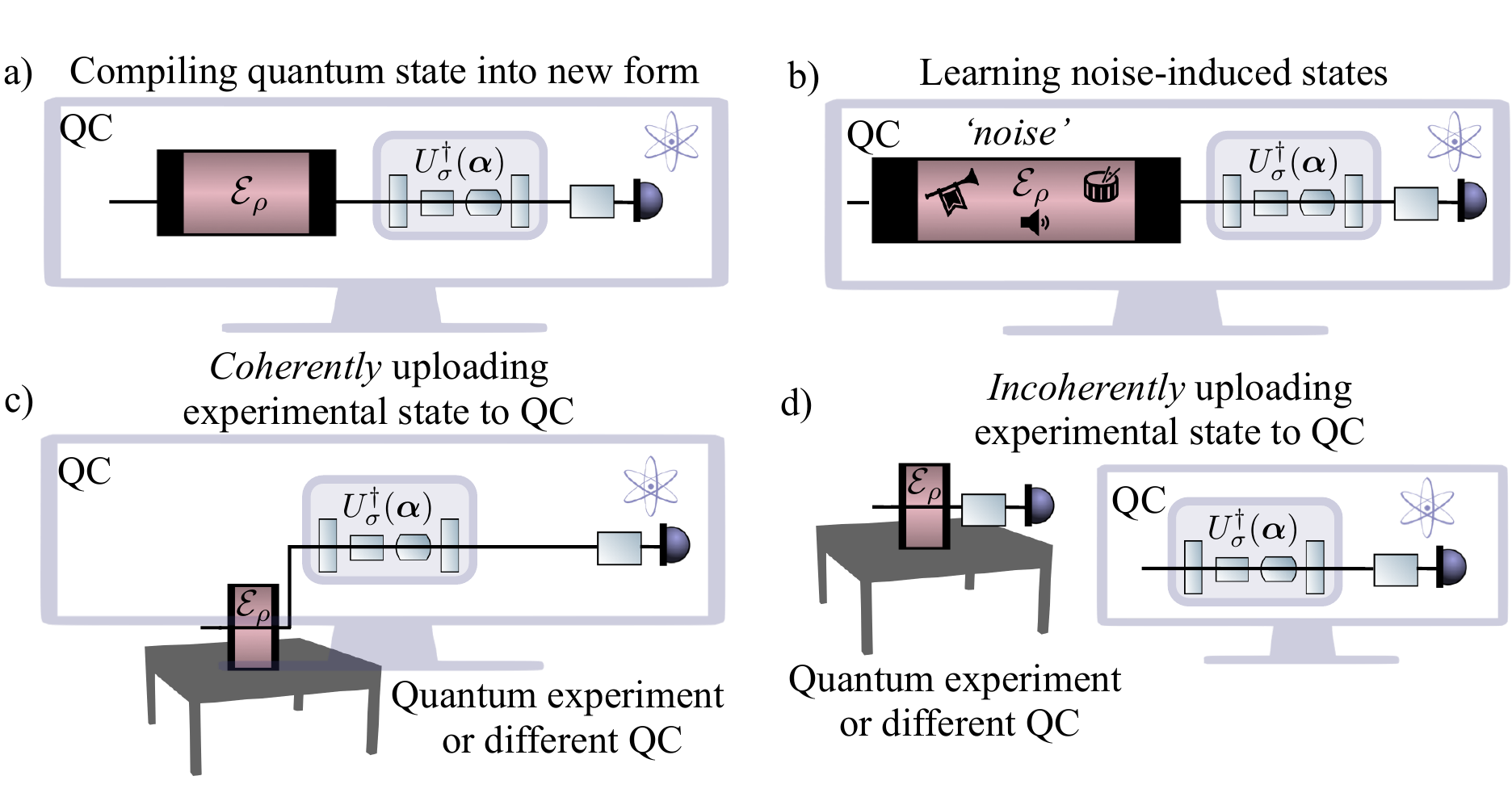}
    \caption{\textbf{Applications of mixed state compilation.} a) A circuit to prepare the target state is already known and quantum compilation is used to learn a more efficient circuit to prepare that state. b) The target state is the (unknown) output of (unknown) noise processes on quantum hardware. c), d) Quantum compiling is used to upload a quantum state from an experimental system (or different quantum computer) to a quantum computer. This could be performed either coherently using a Loschmidt echo/SWAP test to compute the cost (c) or incoherently by combining our algorithm with classical shadows techniques (d). In all sub-figures $\mathcal{E}_\rho$ denotes the channel that prepares $\rho$ and $U^\dagger_\sigma(\vec{\alpha})$ denotes a single parameterized unitary to learn $\rho$ via the CCPS ansatz.}
    \label{fig:applications}
\end{figure}

% \textbf{Applications of mixed state compilation.}
Quantum state compilation, in providing a quantum circuit description of a state, serves a different purpose from tomography or classical shadows, as summarized in Fig.~\ref{fig:applications}. In the first case, it may be used to `upload' an unknown state of an experimental quantum system to a quantum computer. In other words, given a quantum system in an unknown target mixed state, our algorithms can learn a circuit representation of it for implementation on any digital quantum device. In this case, we can use the compilation to more efficiently store the state for later processing, compute properties of the state that are hard to measure directly, or use the state as the input to another algorithm, say for quantum simulation. Alternatively, a circuit to prepare the target state might already be known, but the aim would be to learn a more efficient (i.e., shorter depth and/or more noise resilient) circuit to prepare that state. These applications are particularly interesting when the unknown mixed state is generated by unknown noise as in (b). In this case, the compilation serves as a snapshot of how a noisy quantum computer corrupted a desired input state. This snapshot can then be used later or perhaps even on a more coherent machine for reliable processing. 
%Quantum compilation may also be used as an alternative method to benchmark sources of noise.
% As we will discuss, by learning a state that has been subject to noise one can, at least partially, characterise that noise.

% \textbf{Learning lower rank approximations.} 
In some cases, it may be of interest not to  learn the target state perfectly but rather attempt to learn a low(er) rank approximation of it. In general, the resources required to manipulate a quantum state on quantum hardware grow with its rank. Thus, learning a lower rank approximation provides a means of compressing a state to store it more efficiently. The compression of quantum data has a long history tracing back to the early days of quantum information theory~\cite{schumacher1995quantum,cleve1996schumacher,nielsen2000quantum}. More recently, the question of how well any given state may be approximated using a lower rank state was addressed analytically in Ref.~\cite{ezzell2022quantum} for the Hilbert--Schmidt and trace distances. Interestingly, the optimal lower rank approximation essentially corresponds to performing principal component analysis (PCA) (jump to Eq.~\eqref{eqn_optimalstate_dhs} to look ahead) for the Hilbert--Schmidt distance (but not for trace distance). Hence, as we will show, one can also use our learned compilation to perform PCA with a desired cut-off rank. At the same time, the analytical expression for the optimal state from this work provides a natural benchmark for our learning task.

\begin{figure*}[t!]
    \includegraphics[width=\textwidth]{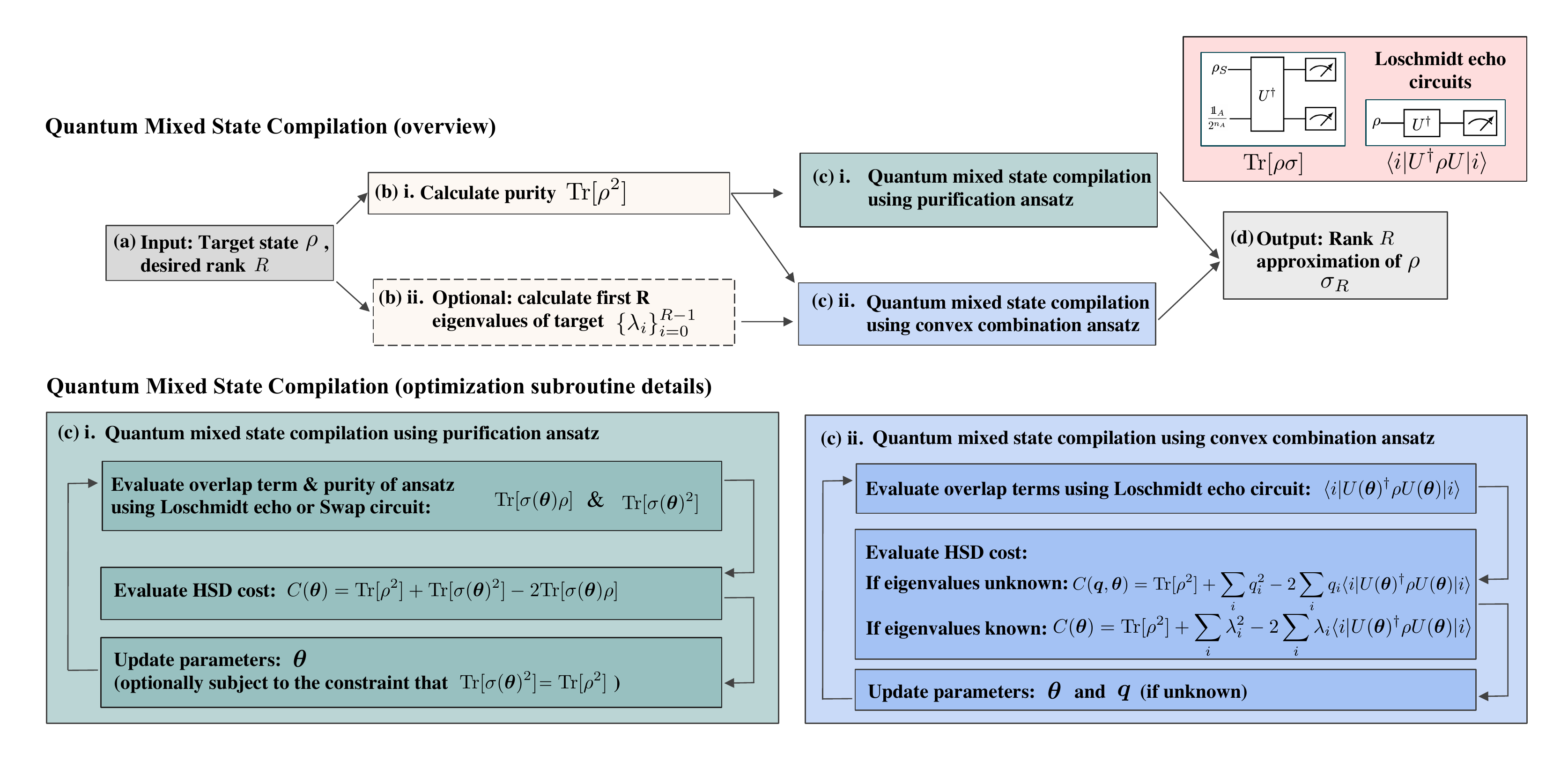}
    \caption{\textbf{The Quantum Mixed State Compiling Algorithm.} (a) The algorithm takes as inputs a target state $\rho$ and a desired approximation rank $R$. The next step is to (b) i.~estimate the purity of the target state or (optionally) ii.~estimate the first $R$ eigenvalues of $\rho$. The core of the algorithm consists of (c) variationally compiling the mixed state using either i.~the state purification (SP) ansatz or ii.~the convex combination of pure states (CCPS) ansatz (detailed in the green and dark blue boxes respectively). (d) The algorithm outputs a rank-$R$ approximation of $\rho$. The pink box details the Loschmidt echo circuits used to evaluate the overlap terms when using the SP ansatz (left) or the CCPS ansatz (right). A detailed description of all the circuits necessary to evaluate the cost for both ans{\"a}tze is given in Appendix~\ref{app:ansatze}.}
    \label{fig:flowchart}
\end{figure*}

% \textbf{What we do.}
Our algorithm, which is summarized in Fig.~\ref{fig:flowchart}, involves variationally minimizing a cost function that is formulated in terms of the Hilbert--Schmidt distance between the target state and ansatz state. This cost can be efficiently measured using either a SWAP test or a Loschmidt echo circuit. In contrast to Ref.~\cite{lloyd2018quantumgenerative}, which sketches a quantum generative adversarial neural network that might be used to learn a mixed state, we further present local variants of our cost functions to mitigate the trainability barrier posed by barren plateaus~\cite{mcclean2018barren, cerezo2020cost, holmes2020barren,holmes2021connecting,arrasmith2021equivalence,larocca2021diagnosing,sharma2020trainability, patti2020entanglement, thanasilp2021subtleties, uvarov2020barren,marrero2020entanglement}.  

%We consider two different possible ans\"{a}tze to represent the mixed state: one based on learning a purification of it and the second on decomposing it into a convex combination of pure states. 

In the numerical simulations of our proposed algorithm, we demonstrate the applicability of both the purification and convex combination ans\"{a}tze for learning full and lower rank approximations of a target mixed state. In particular, we numerically simulate the learning of typical random states, as well as of random thermal states of the Heisenberg XY model. We additionally implement our algorithm on quantum hardware in order to compile an unknown state generated by hardware noise. 

\section{Quantum Mixed State Compiling Algorithm}\label{sec:algs}

% The overall structure of the Variational Mixed State Learning algorithm is shown in Fig.~\ref{fig:flowchart}. The algorithm is a hybrid quantum classical algorithm (VQA).  quantum hardware used to efficiently compute a chosen cost function, whilst classical hardware is used to perform optimization over the cost landscape. 

%\subsection{Overview of algorithm}\label{sec:overview}

The Quantum Mixed State Compiling (QMSC) algorithm takes as input a mixed state $\rho$ and a desired approximation rank $R$. Though not necessary, we assume $R \leq \rank(\rho) \eqqcolon r$ in defining our algorithm since this is both sensible and simplifies our discussion. In our algorithm, we take this a step further and constrain $R \leq r_\epsilon$, where $r_\epsilon$ is the \emph{$\epsilon$-rank}\footnote{We note that the notion of $\epsilon$-rank was also used in Ref.~\cite{cerezo2020variationalfidelity}, but its formal definition is slightly different than the one we choose to use here. A smooth version of the max-entropy (or $\alpha =0$ Renyi entropy) provides yet another definition for $\epsilon$-rank that is distinct from ours~\cite{renner2004smooth,renner2005security}.} of $\rho$, which counts the number of eigenvalues greater than $\epsilon > 0$. This notion of approximate rank is designed to capture the intuition that the contribution of very small but non-zero eigenvalues can often in practice be ignored. The goal then is to optimize the classical parameters $\vec{\alpha}$ of a parameterized trial state $\sigma(\vec{\alpha}, R)$ satisfying $\rank (\sigma(\vec{\alpha^*},R)) = R$, such that, for the optimized parameters $\vec{\alpha^*}$, the output state $\sigma(\vec{\alpha^*},R)$ well approximates the target $\rho$. 
% Since the learnt state $\sigma_R(\vec{\alpha^*})$ is generated from some optimized quantum circuit, we gain a classical description of $\sigma_R(\vec{\alpha^*})$ that provides us with a means to prepare it.

In order to assess the closeness between our training state $\sigma$ and our target state $\rho$, we employ the following cost function in terms of the Hilbert--Schmidt distance between the two states:
\begin{equation}
\begin{aligned}
    \label{eq:Hilbert--Schmidt-distance}
    C(\vec{\alpha}, R) &\equiv  \left\|\rho - \sigma(\vec{\alpha}, R)\right\|_2^2 \\
    & \coloneqq \Tr[\rho^2] + \Tr[\sigma(\vec{\alpha}, R)^2] - 2 \Tr[\rho\sigma(\vec{\alpha}, R)] \, .
\end{aligned}
\end{equation}
As discussed in more detail in Sections~\ref{sec:ansatzpurification} and \ref{sec:ansatzconvex}, the Hilbert--Schmidt distance can be efficiently computed using a combination of SWAP tests and/or a Loschmidt-echo-like circuit. In Appendix~\ref{app:localcost} we detail how this cost can be reformulated such that it requires only local measurements~\cite{cerezo2020cost} to mitigate the barrier to trainability posed by barren plateaus~\cite{mcclean2018barren, cerezo2020cost, holmes2020barren,holmes2021connecting,arrasmith2021equivalence,larocca2021diagnosing,sharma2020trainability, patti2020entanglement, thanasilp2021subtleties, uvarov2020barren,marrero2020entanglement}.

If trained well, the learned state $\sigma(\vec{\alpha^*}, R)$ should be close to the solution to the Quantum Low-Rank Approximation Problem~\cite{ezzell2022quantum}. As discussed in Ref.~\cite{ezzell2022quantum}, the unique optimal state that minimizes the Hilbert--Schmidt distance, subject to a rank constraint, i.e., the state $\sigma(\vec{\alpha_{\rm opt}}, R)$ where 
\begin{equation}
    \vec{\alpha_{\rm opt}} \coloneqq \argmin C(\vec{\alpha}, R) \, ,
\end{equation}
takes the form
\begin{align}
    \label{eqn_optimalstate_dhs}
    \sigma(\vec{\alpha_{\rm opt}}, R) & = \tau_R + \left(\frac{1-\Tr[\tau_R]}{R}\right)\Pi_R, 
    \\
    \tau_R & \coloneqq \Pi_R \rho \Pi_R \, .
\end{align}
% \NE{Note that the discussion above implicitly assumes that our parameterization of $\sigma(\alpha, R)$ is sufficiently expressible to contain $\alpha_{\text{opt}}$, even though this may not be true in general.  We have chosen this notation for simplicity, rather than defining an $\alpha$-independent optimum.}
Here $\Pi_R$ is a projector onto the eigenstates corresponding to the $R$ largest  eigenvalues of $\rho$. That is, $\Pi_R$ is a projector onto the first $R$ principal components of $\rho$. We note here that the projection $\Pi_R$ is the same conceptually as the typical subspace projection \cite[page 31, Theorem 1.18]{ohya2004quantum} used in quantum data compression \cite{schumacher1995quantum} (given that it projects onto the high probability subspace of the state $\rho$), and the quantum channel in \eqref{eqn_optimalstate_dhs} (i.e., $(\cdot) \to \Pi_R (\cdot) \Pi_R + \operatorname{Tr}[(I-\Pi_R)(\cdot)] \frac{\Pi_R}{R}$) is essentially the same as the encoding channel used in quantum data compression \cite[equation (12.50)]{nielsen2000quantum}.

With this analytical solution, we can define a natural performance metric for the Quantum Mixed State Compilation algorithm as the difference between the found cost and optimal possible cost,
\begin{equation}
    \label{eq:fig-of-merit}
    \Delta_R \coloneqq C(\vec{\alpha}^*, R) - C(\vec{\alpha_{\text{opt}}}, R).
\end{equation}
More explicitly, this expression takes the form 
\begin{equation}
         \Delta_R 
         = C(\vec{\alpha}^*, R) - \sum_{i=R }^{r-1} \lambda_i^2 + \sum_{i=0}^{R-1} (\lambda_{\sigma i} - \lambda_i)^2 ,
\end{equation}
where $\{ \lambda_i \}_{i=0}^{r-1}$ denotes the set of eigenvalues of the target state. Of course when $\rho$ is truly unknown, we can only report the optimal cost $C(\vec{\alpha}^*, R)$, but in this work, we actually compute $\Delta_R$ to verify that  our algorithm is working as intended.

In this paper, we develop a practical, noisy intermediate-scale quantum (NISQ)-friendly algorithm to find $\sigma(\vec{\alpha^*}, R)$, an approximation of $\sigma(\vec{\alpha_{\text{opt}}}, R)$. In doing so, we also find the first $R$ principal values and components of $\rho$, and so we are essentially performing quantum principal component analysis (PCA), analogous to some previously proposed algorithms~\cite{lloyd2014quantum,larose2019variational,cerezo2020variational}. The precise form in which $\sigma(\vec{\alpha^*}, R)$ (or the first $R$ principal components) is obtained depends on our choice of ansatz. We consider two different ans\"{a}tze, one based on learning an approximate purification of $\rho$ and the second on decomposing an approximation of it into a convex combination of pure states. We also discuss the complexity of computing our cost function, as well as its operational meaning. These are detailed in the following sections. A summary of our algorithm, which also describes the two ans\"{a}tze choices, is given in Fig.~\ref{fig:flowchart}.

\subsection{Complexity of Cost Function}\label{sec:cost_function_complexity}

For all VQAs, the purpose of using the quantum computer in the optimization loop is to estimate a cost function efficiently, which otherwise would be difficult to estimate classically. Hence, it is helpful to establish that the cost function is indeed classically hard to estimate. Previously proposed VQAs for quantum compiling~\cite{khatri2019quantum} and linear system solving~\cite{bravo2020variational} have established classical hardness via the DQC1 hardness of estimating the relevant cost function. (In the previous sentence, DQC1 stands for deterministic quantum computation with one clean qubit~\cite{knill1998power}).  We can make this same argument for the cost function in Eq.~\eqref{eq:Hilbert--Schmidt-distance}, as follows. 

A special case of computing Eq.~\eqref{eq:Hilbert--Schmidt-distance} is when the two states happen to be Choi states associated with unitary processes acting on a $d$ dimensional Hilbert space. In this case, we write $\rho = \dya{\phi_U}$ and $\sigma = \dya{\phi_V}$, where $\ket{\phi_U} = (\id \otimes U) \ket{\phi}$, $\ket{\phi_V} = (\id \otimes V) \ket{\phi}$, and $\ket{\phi}=(1/\sqrt{d}) \sum_j \ket{j}\ket{j}$ is the standard Bell state. Then the cost function in \eqref{eq:Hilbert--Schmidt-distance} becomes $C = 2 - 2 \Tr(\rho \sigma) $ where $\Tr(\rho\sigma)= |\Tr(U\ad V)|^2 / d^2$. Hence, in this special case we have that $C = 2C_{\text{HST}}$ where $C_{\text{HST}}$ is the Hilbert--Schmidt test cost of Ref.~\cite{khatri2019quantum}. Estimating the latter was shown to be DQC1-hard in Ref.~\cite{khatri2019quantum}, and hence our cost function $C$ is also DQC1-hard. Since efficient classical simulation of DQC1 would imply a collapse of the polynomial hierarchy~\cite{fujii2018impossibility,morimae2017hardness}, standard complexity assumptions imply the classical hardness of estimating $C$.

\subsection{Operational Meaning of Cost Function}\label{sec:cost_function_OM}

\subsubsection{Observable estimation}

There is a close connection between the Hilbert--Schmidt distance and trace distance for low-rank states~\cite{coles2019strong}, which gives operational meaning to our cost function in this case. Namely when at least one of the states is low rank, then the two distance measures are essentially equivalent~\cite{coles2019strong}. This can be seen from the following inequality relating the 2-norm to the 1-norm:
\begin{equation}
\|\rho - \sigma\|_2^2 \leq \|\rho - \sigma\|_1^2   \leq 4\mathcal{R}\cdot \|\rho - \sigma\|_2^2
\end{equation}
for any two density matrices $\rho$ and $\sigma$. Here, $\mathcal{R} = \rank(\rho)\rank(\sigma)/[\rank(\rho)+\rank(\sigma)]$ is a quantity called the reduced rank, which is analogous to the reduced mass employed in physics.

In this sense, one can use the Hilbert--Schmidt distance, and hence our cost function, as a strong upper bound on the trace distance. In turn, the trace distance has operational meaning in terms of the difference of observable expectation values on the two states~\cite{nielsen2000quantum}, $\|\rho - \sigma\|_1  = 2 \max_P \Tr(P(\rho - \sigma))$ where the maximization is over all POVM (Positive Operator Valued Measure) elements $P$, i.e., operators satisfying $0\leq P\leq \id$. Hence, our cost function inherits this operational meaning:
\begin{equation}
    C(\vec{\alpha}, R) \geq \frac{1}{\mathcal{R}}[\Tr(P\rho) -\Tr(P \sigma(\vec{\alpha}, R))]^2
\end{equation}
for any POVM element $P$, where $\mathcal{R}$ is the reduced rank for $\rho$ and $\sigma(\vec{\alpha}, R)$.

%In this sense, one can use the Hilbert--Schmidt distance, and hence our cost function, as a strong upper bound on the trace distance. In turn, the trace distance has operational meaning in terms of the difference of observable expectation values on the two states~\cite{nielsen2000quantum}, $\|\rho - \sigma\|_1  = 2 \max_P \Tr(P(\rho - \sigma))$ where the maximization is over all projectors $P$. Hence, our cost function inherits this operational meaning:
%\begin{equation}
%    C(\vec{\alpha}, R) \geq %\frac{1}{\mathcal{R}}[\Tr(P\rho) %-\Tr(P \sigma(\vec{\alpha}, R))]^2
%\end{equation}
%for any projector $P$, where $\mathcal{R}$ is the reduced rank for $\rho$ and $\sigma(\vec{\alpha}, R)$.

%In this sense, one can use the Hilbert--Schmidt distance, and hence our cost function, as a strong upper bound on the trace distance. Then, since the trace distance has operational meaning in terms of the difference of observable expectation values on the two states~\cite{nielsen2000quantum}, our cost function in turn inherits this operational meaning.

\subsubsection{Eigenvalue and Eigenvector Estimation}
\medskip

We also establish an operational meaning for our cost function in the context of minimizing the errors in the eigenvectors and eigenvalues of the compiled state. This is particularly relevant in the context of using our algorithm for PCA, since PCA precisely aims to extract the eigenvectors with the largest eigenvalues.
% Hence meaningful measures of algorithmic performance would quantify eigenvalue error and eigenvector error, i.e., the amount of error in the estimated eigenvalues and eigenvectors. 

First, we note that our proposed cost is an upper bound on the squared difference between the true eigenvalues of $\rho$ and the eigenvalues of the learnt state $\sigma$, as follows. Let us denote the eigenvalues of $\rho$ in increasing order as $\{ \lambda_i \}_{i = 0}^{d-1}$ and the eigenvalues of $\sigma$ as $\{ \mu_i \}_{i = 0}^{d-1}$, with $d$ the Hilbert space dimension. Then a measure of eigenvalue error is 
\begin{equation}\label{eqn_deltalambda}
  \Delta_{\lambda} \coloneqq  \sum_{i=0}^{d-1} (  \lambda_i  - \mu_i )^2 \, .
\end{equation}
It then follows from the Hoffman--Wielandt theorem that 
\begin{equation}
  \Delta_{\lambda} \leq \left\| \rho - \sigma \right\|_2^2  = C\, .
\end{equation}
Thus if one can achieve a small cost value, the average error in the learnt eigenvalues is guaranteed to be small. 

We remark that a small cost value might not be achievable since the optimal cost $C(\vec{\alpha_{\text{opt}}}, R)$ is often non-zero. In this case, one could consider an alternative measure of eigenvalue error that, unlike \eqref{eqn_deltalambda}, would vanish for the optimal state in \eqref{eqn_optimalstate_dhs}. Namely, one can consider the set $S_R$ of states that have the correct $R$-largest eigenvalues up to an additive constant, and then define a measure of eigenvalue error as $\Delta_{S_R}(\sigma) = \min_{\tau \in S_R } \left\| \sigma - \tau \right\|_2^2$. Then it is clear that our cost function upper bounds this error as well: $C \geq \Delta_{S_R}$.

%Thus on successfully minimizing the cost, the average error in the learnt eigenvalues, i.e., the learnt principal values of $\rho$, is guaranteed to be small. 

Second, we note that our proposed cost function is an upper bound on the eigenvector error measure introduced in Ref.~\cite{larose2019variational}. Specifically, a natural measure of the difference between the true eigenvectors of the target state, $\{ |v_i \rangle \}_{i=0}^{r-1}$, and the learnt eigenvectors, $ \{ | u_i \rangle \}_{i=0}^{R-1}$, is given by 
\begin{equation}
    \Delta_v \coloneqq  \sum_{i=0}^{R-1} \langle \delta_i | \delta_i \rangle 
\end{equation}
where
\begin{equation}
    | \delta_i \rangle \coloneqq  \rho |u_i \rangle - \mu_i |u_i \rangle \, . 
\end{equation}
Expanding this out we have that
\begin{equation}
    \Delta_v = \sum_{i=0}^{R-1} \left( \langle u_i | \rho^2 | u_i \rangle + \mu_i^2 - 2 \mu_i \langle u_i | \rho |u_i \rangle \right) \leq \left\| \rho - \sigma \right\|_2^2 \, 
\end{equation}
where for the inequality we use the fact that
\begin{equation}
\sum_{i=0}^{R-1} \langle u_i | \rho^2 | u_i \rangle \leq \sum_{i=0}^{d-1} \langle u_i | \rho^2 | u_i \rangle = \Tr[ \rho^2 ],    
\end{equation}
(i.e., the trace can be taken with respect to an arbitrary orthonormal basis).
Thus a small cost function guarantees a small eigenvector error by this measure.

\subsection{State Purification Ansatz}\label{sec:ansatzpurification}

% The variational state purification algorithm (VSPA) constructs trial states of the form
% \begin{equation}
%     \sigma(\vec{\theta};n_A) = \Tr_A[ U_{\vec{\theta}}(\ketbra{0}{0})^{\otimes(n +n_A)} U_{\vec{\theta}}^\dagger ],
% \end{equation}
% where $U$ acts on the $n$ system qubits (upon which $\sigma$ resides after tracing out) plus $n_A$ ancilla qubits. In using $n_A$ ancilla, we guarantee that $\rank(\sigma) \leq 2^{n_A}$, or in other words, VSPA is only compatible with choosing $R$ as powers of 2, $R = 2^{n_A}$, limiting the algorithm's fine-tuneability when compared to VC3A. That the use of ancilla is necessary to control output rank, $R$, VSPA requires a greater number of registers than VC3A.

% \NE{Elliot description just copied over but not formatted consistnetly with above}
% \eb{Shall we chat about format/notation soon? If there's something agreed upon I can change my notation and everything for consistency.} 

Let us now move onto the different types of ansatz constructions. The state purification (SP) ansatz constructs trial states of the form
\begin{equation}
    \sigma_{\mbox{\tiny SP}}(\vec{\theta},n_A) \coloneqq \Tr_A[ U_{\vec{\theta}} (\ketbra{0}{0})^{\otimes(n +n_A)} U_{\vec{\theta}}^\dagger ],
\end{equation}
where $U_{\vec{\theta}}$ acts on the $n$ system qubits (the same qubits where the state $\rho$ resides) plus an ancilla register $A$ composed of $n_A$ qubits. In using $n_A$ ancillas, we guarantee that $\rank(\sigma) \leq 2^{n_A}$, but for a typical parameterized circuit $U_{\vec{\theta}}$ with randomly generated parameters, one has precisely $\rank(\sigma) = 2^{n_A}$. Thus, in practise, the SP ansatz is only compatible with controlling $R$ in powers of two, so that $R = 2^{n_A}$.
While there are many different possible choices in the ansatz for $U_{\vec{\theta}}$, we suggest using either a problem-inspired approach obeying the symmetries of the target state~\cite{gard2020efficient, gibbs2021long, larocca2022group} and/or an adaptive approach~\cite{bilkis2021semi} to mitigate the problem posed by barren plateaus. A discussion of specific choices in our work for different ensembles is given in Appendix~\ref{app:ansatze}. 
%limiting the ansatz's fine-tuneability when compared to CCPS. As the use of ancilla is necessary to control output rank, $R$, such ans\"{a}tze require a greater number of registers than CCPS-based methods.

Upon substitution of the SP ansatz into our proposed Hilbert--Schmidt distance-based cost function, we obtain
\begin{equation}
    C_{\mbox{\tiny SP}}(\vec{\theta},n_A) = \Tr[\rho^2] + \Tr[\sigma(\vec{\theta},n_A)^2] 
    - 2\Tr[ \rho \sigma(\vec{\theta},n_A)],
\end{equation}
where we use $S$ to denote the system of the states $\rho$ and $\sigma(\vec{\theta},n_A)$.
Below we describe efficient ways of computing each of these terms. 

\medskip

\paragraph*{Purity of target}
%As in the convex combination ansatz case, NOTE: this comes after
The purity $\Tr[\rho^2]$ of the target $\rho$ may be measured using a SWAP test \cite{barenco1997stabilization} or its destructive variant~\cite{garcia2013swap}. For the estimate to be within $\varepsilon$ additive error of the true value with probability not smaller than $1-\delta$,
the Hoeffding bound implies that it suffices to take $O(\varepsilon^{-2}\log\delta^{-1})$ samples. Alternatively, as this term remains constant throughout the procedure, we may opt to neglect it and only focus on optimizing the remaining two parameter-dependent terms.

\medskip 

\paragraph*{Purity of ansatz}
The purity $\Tr[\sigma(\vec{\theta},n_A)^2]$ of the ansatz $\sigma(\vec{\theta},n_A)$  may again be measured via a SWAP test. It is also possible to make use of the fact that we are representing it via its purification and measure the purity of the ansatz via the Loschmidt-echo circuit shown in Fig.~\ref{fig:flowchart}. Again, from the Hoeffding bound, it suffices to take $O(\varepsilon^{-2}\log\delta^{-1})$ samples.

\medskip

\paragraph*{Overlap term}
The overlap term $\Tr[ \rho \sigma(\vec{\theta},n_A)]$ can be evaluated using either a SWAP test or the Loschmidt-echo type circuit pictured in Fig.~\ref{fig:flowchart}. The latter requires fewer qubits and controlled unitaries and hence is more NISQ friendly. To see why this circuit works, note that since we prepare the trial state via a higher-dimensional purification found by evolving an initial all-zero state to the purification of~$\rho$, we can rewrite the overlap term as follows
\begin{equation}
\begin{aligned}\label{eq:GlobalCostTerm}
    \Tr[\rho \sigma] &= \Tr_S[ \rho \Tr_A[U_{\vec{\theta}} (\ket{0}\!\bra{0})^{\otimes (n +n_A)} U_{\vec{\theta}}^\dagger]] \\ 
     &= \Tr_{SA}[(\rho \otimes \id_A) (U_{\vec{\theta}} (\ket{0}\!\bra{0})^{\otimes (n +n_A)} U_{\vec{\theta}}^\dagger)] \\ 
    % &= \Tr_{AB}[U(\vec{\theta})^\dagger \left(\rho \otimes \mathbb{I}_B \right) U_{\vec{\theta}}(\ket{0}\!\bra{0})^{\otimes (n+n_A)} ] \\
    &= 2^{n_A} \Tr_{SA}\!\left[U_{\vec{\theta}}^\dagger \left(\rho \otimes \frac{\id_A}{2^{n_A}} \right) U_{\vec{\theta}}(\ket{0}\!\bra{0})^{\otimes (n +n_A)} \right] \, .
\end{aligned}
\end{equation}
This corresponds to preparing a maximally mixed state on the $n_A$ qubit ancillary system alongside $\rho$ which is prepared on the system register $S$, evolving the system and ancilla registers under $U_{\vec{\theta}}$, and then performing an all-zero measurement. Due to the factor of $2^{n_A}$, the number of shots required to measure this cost within additive error $\epsilon$ will scale exponentially in the number of ancilla qubits $n_A$. Hence the Loschmidt echo method for computing the overlap is only appropriate for learning low rank approximations to $\rho$ where $n_A \in \text{poly}(\text{log}(n))$.
% We note that this is a global measurement, resulting in a cost-function that is susceptible to barren plateaus as the number of qubits scales up. To avoid barren plateaus, we can replace this with a local operator. We explore such potential avenues in the appendices. \eb{Or will we include a bit of discussion here?} \zo{Commented out because I mention this when we first introduce the cost}

As discussed earlier, the above method may also be used to calculate the purity of our ansatz simply by replacing the initial state $\rho$ with the trial state $\sigma$. However, the purity of our target may not be measured this way, as it requires knowledge of 
a purification of $\rho$, which is precisely what we are trying to find.

\subsection{Convex Combination of Pure States Ansatz}

\label{sec:ansatzconvex}

The convex combination of pure states (CCPS) ansatz constructs trial states of the form
\begin{equation}
    \label{eq:ccps-ansatz}
     \sigma_{\mbox{\tiny CCPS}}( \vec{\alpha} , R) \coloneqq \sum_{i=0}^{R-1} p_{\vec{\phi}}(i) U_{\vec{\theta}} \ketbra{i}{i} U_{\vec{\theta}}^\dagger \, ,
\end{equation}
Here $\{\ket{i}\}_{i=0}^{R-1}$ denotes a subset of the computational basis of $n$ qubits, $\vec{\alpha} = ( \vec{\theta}, \vec{\phi} )$ is a vector of parameters, $U_{\vec{\theta}}$ is a parameterized quantum circuit, and $p_{\vec{\phi}}$ is a parameterized probability distribution. An appealing feature of this ansatz is that learning a rank-$R$ approximation gives any rank $R' < R$ approximation for free since one can always `drop' the eigenstates corresponding to the smallest $R - R'$ eigenvalues and then re-normalize (see Eq.~\eqref{eq:truncated-CCPS-result}). 

Since the state that minimizes our cost, $\sigma( \vec{\alpha_{\rm opt}}, R)$ (i.e., the solution to the Quantum Low-Rank Approximation Problem given in Eq.~\eqref{eqn_optimalstate_dhs}), is proportional to the first $R$ principal components of the target state $\rho$, it follows that this ansatz can be used to learn the principal components of $\rho$.
% defining the optimal parameters via 
% \begin{align}\label{eq:optimization}
%     \{U_{\vec{\theta}_{\rm opt}},U_{\vec{\phi}_{\rm opt}}\} \coloneqq\underset{\vec{\theta},\vec{\phi}}{\text{ arg min }} C(\rho, \vec{\theta}, \vec{\phi}) \, ,
% \end{align}
More precisely, we have that the first $R$ principal values of $\rho$ are given by $\{p_{\vec{\phi}}(i) \}_{i=0}^{R-1}$ and its principal components are $\{ U_{\vec{\theta}_{\rm opt}} | i \rangle \}_{i = 0}^{R-1}$. 

When the rank of the trial state is low, i.e., $R \in \Omega(\text{poly}(n))$, such that we are learning a low rank approximation of the target state $\rho$, or for small scale problems, the vector of probabilities $\vec{p}_{\vec{\phi}} \coloneqq (p_{\vec{\phi}}(0),  \ldots  ,p_{\vec{\phi}}(R-1))$ may be stored simply as a classical vector. For learning high rank approximations to larger full rank states, the probability vector is exponentially large and thus cannot be explicitly stored efficiently. Rather the process will need to be sampled from. Such samples can be generated via classical neural networks, such as generative neural networks~\cite{yang2022learning} or Boltzmann machines~\cite{montufar2016restricted}. In this case, there are similarities between the CCPS ansatz and the Hamiltonian models considered in Ref.~\cite{verdon2019quantum}. Similarly to the SPA, for $U_{\vec{\theta}}$ we suggest using either a problem inspired ansatz obeying the symmetries of the target state~\cite{gard2020efficient, gibbs2021long, larocca2022group} and/or an adaptive approach~\cite{bilkis2021semi}. A discussion of specific choices in our work for different ensembles is given in Appendix~\ref{app:ansatze}.

Upon substituting the expression for the convex combination of pure states ansatz into our proposed Hilbert--Schmidt distance cost function, we obtain 
\begin{multline}
    C_{\mbox{\tiny CCPS}}( \vec{\alpha}, R)  = \Tr[\rho^2] + \sum_i p_{\vec{\phi}}(i)^2  \\
    - 2 \sum_i p_{\vec{\phi}}(i) \langle i | U^\dagger_{\vec{\theta}} \rho U_{\vec{\theta}} | i \rangle \, ,
\end{multline}
with $\vec{\alpha} = ( \vec{\theta}, \vec{\phi} )$. 
Here we describe how each of the terms in the above cost may be efficiently computed. 

\medskip

\paragraph*{Purity of target.}
As for the state purification ansatz, the first term, i.e., the purity of the target state, may be computed using a SWAP test \cite{barenco1997stabilization} or its destructive variant~\cite{garcia2013swap} (see also Ref.~\cite{subacsi2019entanglement}). 

\medskip 

\paragraph*{Purity of ansatz.}
The second term, the purity of the guessed state, is equal to the sum of the square of the probabilities of the parameterized distribution, $p_{\vec{\phi}}(i)$.
For low rank approximations, i.e., $R \in \text{poly}(n)$, one may store the probability vector classically and thus this term can be computed by basic arithmetic.

When the probability vector is too large to be stored explicitly, but rather is handled via sampling, the purity of the ansatz can be estimated with a classical version of the SWAP test.
Indeed, the approach is to take two independent samples from the distribution
$p_{\vec{\phi}}(i)$ (call the random samples $I$ and $J$). We then set an indicator random variable $\chi_P$ as
$\chi_P = 0$ if the samples are not equal, and $\chi_P = 1$ if the samples are equal. The
expectation of this random variable is then given by
\begin{align}
\mathbb{E}[\chi_P]  & =0\cdot\Pr[I\neq J]+1\cdot\Pr[I=J]\\
& =\Pr[I=J]\\
& =\sum_{i,j}p_{\vec{\phi}}(i)p_{\vec{\phi}}(j)\delta_{i,j}\\
& =\sum_{i}\left(  p_{\vec{\phi}}(i)\right)  ^{2}.
\end{align}
Thus, the random variable $\chi_P$ is an unbiased estimator of the collision
probability, $\sum_{i}\left(  p_{\vec{\phi}}(i)\right)  ^{2}$, and takes
values between zero and one. The Hoeffding bound then applies, and we can take
$O(\varepsilon^{-2}\log\delta^{-1})$ independent samples of $\chi_P$ in order to
estimate the collision probability~$\sum_{i}\left(  p_{\vec{\phi}}(i)\right)
^{2}$ to within additive error $\varepsilon$ with probability not smaller than $1-\delta$. 
% In more detail, let $X_{k}$ denote the $k$th sample, where
% $k\in\left\{  1,\ldots,n\right\}$. Then the estimator of $\sum_{i}\left(
% p_{\vec{\phi}}(i)\right)  ^{2}$ is $n^{-1}\sum_{k=1}^{n}X_{k}$.

% That is, one can sample twice from the distribution $\mathcal{P}$, assigning the random variable $X = 1$ to the case where the two outputs agree and $X = 0$ otherwise. The average of $X$ over $\mathcal{P}$ is precisely the norm of $\vec{p}$ as required $\langle X \rangle_{\mathcal{P}} = \sum_i p_{\vec{\phi}}(i)^2$. 

\medskip

\paragraph*{Overlap term.}
A naive approach to computing the third term, the overlap between the guess $\sigma$ and target~$\rho$, would be to estimate each of the $\langle i | U^\dagger_{\vec{\theta}} \rho U_{\vec{\theta}} | i \rangle$ terms using the Loschmidt echo circuit shown in Fig.~\ref{fig:flowchart}, followed by classical post processing. That is, we compute $\langle i | U^\dagger_{\vec{\theta}} \rho U_{\vec{\theta}} | i \rangle$ by preparing the state $\rho$, performing the unitary $U^\dagger_{\vec{\theta}}$, and then measuring in the computational basis.
The total overlap term could simply be computed by weighting each of the probabilities $\langle i | U^\dagger_{\vec{\theta}} \rho U_{\vec{\theta}} | i \rangle$ by the corresponding classical probability $p_{\vec{\phi}}(i)$ and taking their sum. This naive approach will work for small problems; however, even in the case where $p_{\vec{\phi}}(i)$ can be explicitly stored, this method will not be efficient for larger problems. The problem is that one needs enough shots to estimate all $2^n$ probabilities $\langle i | U^\dagger_{\vec{\theta}} \rho U_{\vec{\theta}} | i \rangle$. Combining this observation with Hoeffding's equality, it is apparent that this method requires an exponential number of shots, $O(2^n \varepsilon^{-2}\log\delta^{-1})$. 

Instead, we propose computing the overlap term using a generalization of the classical SWAP test.
Let us define the distribution%
\begin{equation}
    \label{eq:classical-swap-q-dist}
q_{\vec{\theta}}(i)\coloneqq\langle i|U^\dagger_{\vec{\theta}}\rho U_{\vec{\theta}}|i\rangle,
\end{equation}
which can be sampled from via the Loschmidt echo circuit. 
Then we see that%
\begin{equation}
\operatorname{Tr}[\rho\sigma]=\sum_{i}p_{\vec{\phi}}(i)q_{\vec{\theta}}(i).
\end{equation}
This quantity can be estimated by taking a sample from $p_{\vec{\phi}}(i)$ and an independent sample from
$q_{\vec{\theta}}(i)$ (call the samples $I$ and $J$) and setting an indicator random
variable $\chi_O$ if the samples are not equal and $\chi_O=1$ if the samples are
equal. The expectation of this random variable is given by%
\begin{align}
\mathbb{E}[\chi_O]  & =0\cdot\Pr[I\neq J]+1\cdot\Pr[I=J]\\
& =\Pr[I=J]\\
& =\sum_{i,j}p_{\vec{\phi}}(i)q_{\vec{\theta}}(j)\delta_{i,j}\\
& =\sum_{i}p_{\vec{\phi}}(i)q_{\vec{\theta}}(i).
\end{align}
Thus, the random variable $\chi_O$ is an unbiased estimator of the collision
probability $\operatorname{Tr}[\rho\sigma]$, and it takes values between zero
and one. The Hoeffding bound then applies, and we can take $O(\varepsilon
^{-2}\log\delta^{-1})$ independent samples of $\chi_O$ in order to estimate this
collision probability to within additive error $\varepsilon$ with probability not smaller than $1-\delta$.

\subsection{Performing PCA with our Ans\"{a}tze}

Both ans\"{a}tze also allow for principal component analysis (PCA) of $\rho$. To make this precise, we first write the target state as
\begin{equation}
    \rho = \sum_{i=1}^r \lambda_i \ketbra{v_i}{v_i}
\end{equation}
 where $\lambda_1 > \lambda_2 > \cdots > \lambda_r$ are the ordered principal values of $\rho$ with associated principal components $\ket{v_i}$. 
 
By inspection, the CCPS ansatz is directly an ansatz for the principal components of $\rho$. In other words, learning a CCPS representation of $\sigma(\vec{\alpha}^*, R)$ provides an estimate of the principal components, $\ket{u_i} \coloneqq U(\vec{\theta}^*) \ket{i}$, explicitly. It also provides an estimate of the principal values, $p_{\vec{\phi}}(i)$, but due to normalization, these are expected to be different from $\lambda_i$ by an additive constant when $\Delta_R$ is small. As $R$ approaches $r_\epsilon$, this additive constant goes to zero, and here $p_{\vec{\phi}}(i) \approx \lambda_i$. Nevertheless, even when $R < r_\epsilon$, the values of $p_{\vec{\phi}}(i)$ are such that $\sigma(\vec{\alpha}^*, R)$ acts as the closest rank $R$ proxy, so in this operational sense, it is still appropriate to call $p_{\vec{\phi}}(i)$ the principal value estimates.
 
 One very useful property of these explicit principal component/value estimates is that they allow us to construct any numerically optimal $R' < R$ approximation by truncation,
\begin{subequations}
\begin{align}
    \label{eq:truncated-CCPS-result}
    \sigma(\vec{\alpha}^*, R') &= \sum_{i=0}^{R' - 1} \Tilde{p}_{\vec{\phi}}(i) U(\vec{\theta}^*) \ketbra{i}{i} U^{\dagger}(\vec{\theta}^*) , \\
    %------------------------------------
    \Tilde{p}_{\vec{\phi}}(i) &= p_{\vec{\phi}}(i) + \left(1 - \sum_{i=0}^{R'-1} p_{\vec{\phi}}(i)\right) / R'.
\end{align}
\end{subequations}
Indeed, this is perhaps one convincing way to view each $p_{\vec{\phi}}(i)$ as the appropriate operational sense of \enquote{principal value} when we truncate the rank. 
 
 Though less obvious, the SP ansatz can also be used for PCA by using a carefully designed circuit for the purification ansatz as described in Appendix~\ref{app:ansatze}. The result is that computational basis measurements on the ancilla system prepare the principal vectors on the target system with probabilities given by the principal values. Thus, the knowledge of principal values/components here is implicit, and hence cannot be directly used to obtain lower rank approximations by truncation.
 
Finally, we remark that while PCA is a general procedure, it has an intuitive meaning for physically relevant classes of states. As an example, consider an XY thermal state. In general, the eigenvalues follow a Boltzmann distribution, and the eigenvectors are those of the XY Hamiltonian itself. At low temperature, we often say a state is \emph{approximately in its ground-state.} More precisely, we mean,
\begin{equation}
   \rho \approx \rho_\epsilon \equiv \sum_{i=1}^{q} \lambda_i \ketbra{v_i}{v_i}, \ \ \ \sum_{i=1}^{q} \lambda_i = 1 - \epsilon
\end{equation}
for $\epsilon$ small. In our language, we would say the state is approximately rank $q$, and when $q \ll r$ as is typical for low temperature thermal states, it's approximately low rank~\footnote{We comment that this exact notion of low rank actually precisely agrees with the definition of epsilon-rank used in Ref.~\cite{cerezo2020variationalfidelity}. We employ a slightly different definition already given that is more amenable to NISQ experiments.}. By performing PCA with a target rank $R = q$, we find approximations of the eigenvectors with low energies $\{\ket{v_i}\}_{i=1}^q$ and corresponding Boltzmann weights $\{\lambda_i\}_{i=1}^q$. Hence, our PCA algorithm can be thought of as a way to learn the Boltzmann weights and a means to prepare low-lying energy eigenvectors of a quantum thermal state which has also been explored in other NISQ friendly works~\cite{guo2023variational}.

\subsection{Comparison of Ans\"{a}tze}

% \textit{Paragraph comparing the strengths and weaknesses of the two algorithms}
% Convex combo:
% \begin{enumerate}
%     \item List pros (i.e., no ancilla needed, rank finely controlled, purity finely controlled if willing to make constraints)
%     \item List cons (i.e., must impose probability constraint)
% \end{enumerate}
% Purification:
% \begin{enumerate}
%     \item List pros (finds a purification)
%     \item List cons (needs ancilla and less fine control over rank)
% \end{enumerate}

The reliance on an ancillary system to compute the cost function for the SP ansatz naturally increases the resources required for computation. Furthermore, in general, for a typical choice of $U_{\vec{\theta}}$, the guess state will have rank $R = 2^{n_A}$. That is, one is limited to ranks of powers of two. In contrast, CCPS both requires no ancilla and allows for fine control over both output ranks. However, the need to learn $\vec{p_{\phi}}$ under the constraint of convexity, in addition to $\vec{\theta}$, potentially increases the complexity of our optimization subroutine. Thus the choice as to whether to use CCPS or SP will depend in large part on whether classical resources (optimization power) or quantum resources (qubits available) are more constrained.

Of course, the choice as to whether to use the CCPS or SP ansatz may depend not only on the required resources but also the end goal of the subroutine. For example, if the end goal is principal component analysis, this is more readily performed using CCPS since it automatically finds the first $R$ principal components of the target. Alternatively, one can imagine situations where it is desirable to learn the purification of the target. For example, the purification of a state opens up methods for computing entanglement measures between subsystems of that state~\cite{Vedral1998Entanglement, Schumacher1996Sending, wilde2013quantum}, the fidelity between two states (given Ulhmann's theorem)~\cite{chen2021variational,agarwal2021estimating}, as well as symmetry measures~\cite{laborde2021testing}.

% \eb{also face different challenges as system size increases (need to either store exponentially growing prob.s or perform increasing number of samples for CCPS}

% \subsection{Incoherent access model}

\subsection{Summary of Algorithm}

The Quantum Mixed State Compiling algorithm is summarized in Fig.~\ref{fig:flowchart}. At its core it is composed of the following steps.

\begin{enumerate}
    \item Start with a target state $\rho$, and a desired rank $R$ for the compiled state $\sigma_R$ used to approximate~$\rho$.
    
    \item Choose whether to learn the target using the purification or convex combination ansatz. In general, this decision will depend on the purpose for which the state is being learned and the resources available. A discussion of the circuits one must run in both cases is given in Appendix~\ref{app:ansatze}.
    
    \item Minimize the Hilbert--Schmidt distance cost $C$, Eq.~\eqref{eq:Hilbert--Schmidt-distance}, using a hybrid quantum-classical optimization loop to find the trained parameters $\vec{\theta^*}$ (in the case of the SP ansatz) and $( \vec{\theta^*}, \vec{\phi^*} )$ (in the case of the CCPS ansatz) that approximately minimize $C$. If using a gradient based optimizer one can analytically compute the gradient of the cost using the parameter shift rule \cite{mitarai2018quantum,schuld2019evaluating}. For the case of the purity of the ansatz term for the state purification ansatz, $\Tr[\sigma(\vec{\theta},n_A)^2]$, this rule needs modifying to account for correlations. For more details on computing the gradients, see Appendix~\ref{app:grads}.
\end{enumerate} 

Here we briefly describe two ways in which prior information about the structure of the target state could be used to make the mixed state learning algorithm more efficient. 

In the first instance, the purity of the target state may be computed and then we need only consider `guess' states $\sigma_R$ with the correct purity. That is, one may use the target purity as a constraint during the minimization of the Hilbert--Schmidt distance cost, Eq.~\eqref{eq:Hilbert--Schmidt-distance}. This amounts to maximizing the overlap term $\Tr[\rho \sigma]$ subject to the constraint $\Tr[\sigma^2] = \Tr[\rho^2]$. We note that this constrained optimization may not be compatible with the rank constraint if the desired rank is much lower than the true rank of the target, i.e., $R \ll r$. 

Going a step further, one could also simplify the task of learning a rank-$R$ approximation to $\rho$ by first learning the $R$ largest eigenvalues of $\rho$ and then using the mixed state learning algorithm to learn their corresponding eigenvectors. The largest eigenvalues of any state $\sigma$ can be learnt non-variationally using the quantum algorithm proposed in Ref.~\cite{subacsi2019entanglement}. Denoting the measured eigenvalues as $\{ \lambda_i \}_{i=0}^{R-1}$, the eigenvectors may be learnt by maximizing the overlap term $\sum_i \lambda_i \langle i | U^\dagger_{\vec{\theta}} \rho U_{\vec{\theta}} | i \rangle$. 

As discussed further in Appendix~\ref{app:localcost}, one theoretical advantage of these two modifications is that they require optimizing only a single overlap term rather than the difference between two purity terms and an overlap term. This overlap term takes the form of a standard Variational Quantum Eigensolver (VQE) cost and therefore can be readily transformed into a local cost for which we have trainability guarantees~\cite{cerezo2020cost}.

\section{Numerical Simulations}
We begin our numerical simulations discussion with a brief summary of the target states and chosen optimizer. Additional details can be found in our open source code~\cite{ezzell2022qmsc} which includes our raw data and the scripts we used to generate it. We then discuss the results of compiling each of the listed states in separate sections.  

\subsection{Description of States and Optimizer}

We discuss the performance of our QMSC algorithm for three main types of states: \begin{enumerate}
    \item Random states drawn from the Bures measure.
    \item Thermal states of an XY chain with random coefficients at both low and high temperature.
    \item Noise-induced states generated by simple circuits on NISQ hardware. (We shall call them NISQ states for short.)
\end{enumerate}
%(i) random states drawn from the Bures measure, (ii) thermal states of an XY chain with random coefficients at both low and high temperature, and (iii) noise-induced states generated by simple circuits on NISQ hardware (we shall call them NISQ states for short). 
In the first two cases, we perform the entire optimization using idealized classical simulations. That is, we evaluate the cost functions using matrix operations on a classical machine with no error model and no shots (i.e., infinite precision). Henceforth, the use of \enquote{idealized classical simulations} will continue to have this precise meaning. This serves both as a proof-of-principle as well as a means to estimate an empirical idealized scaling. For the NISQ states, discussed in Section~\ref{subsec:results-qh}, we evaluate all cost functions on quantum hardware but use classical computation for the parameter updates, which is the standard variational quantum algorithm approach.

In each case, we consider a low rank approximation by setting $n_A = 1$ (or, equivalently, $R = 2$) or a full (epsilon) rank approximation with $n_A = \lceil \log_2 r_\epsilon \rceil$  (or $R = 2^{\lceil \log_2 r_\epsilon \rceil}$). It is natural to wonder why we choose to use $R = 2^{\lceil \log_2 r_\epsilon \rceil}$ instead of $R = r_\epsilon$. The reason is straightforward: we simply want both the SP and the CCPS ansatz to attempt to learn the same state to the same rank approximation to make a fair comparison, and we can only control the rank of the SP ansatz in powers of two. 

We use the gradient-free Powell optimizer~\cite{powell1964efficient} provided in the \verb^scipy^~\cite{virtanen2020scipy} optimization library. We find that Powell is more robust and generally outperforms the common \verb^scipy^ black box alternatives such as BFGS~\cite{nocedal2006numerical}, Nelder-Mead~\cite{nelder1965simplex}, SLSQP~\cite{kraft1988software}, and COBYLA~\cite{powell1994direct} for our problems. Of course, the performance could be improved by using advanced VQA optimizers such as SPSA~\cite{spall1998overview} or  ICANs~\cite{kubler2020adaptive, arrasmith2020operator}, but we do not pursue this refinement since we found Powell to give reasonable results. 

\subsection{Bures Random States Results}

We first study the Bures random state distribution because it is a reasonable sampling distribution when nothing about the quantum state is known~\cite{hall1998random, zyczkowski2011generating} (see Appendix~\ref{app:bures} for more details). For some intuition, note that one way to generate $n$ qubit Bures random states is by preparing a Haar random state on $n + n$ qubits, applying an $n$-qubit \enquote{local Haar random} unitary on the system qubits, and then tracing out the ancilla. In this sense, learning the purification is similar to learning a Haar random unitary on $2n$ qubits which we know to be intractable for VQAs due to an intrinsic, ansatz-independent vanishing gradient problem~\cite{holmes2020barren}.

Given their lack of structure we see Bures random states as a good test case to compare the state purification and convex combination of pure states ans\"{a}tze.
However, due to the unavoidable vanishing gradient problem---along with the large number of parameters needed for an unstructured state (see Appendix~\ref{app:ansatze})---we only test modest sizes. Specifically, we test from $n = 1$ to $n = 4, 5$ for low rank approximations and from $n = 1$ to $n = 3$ for full $\epsilon$-rank approximations (i.e., we learn up to a six-qubit random purification). 

\begin{figure*}
    \subfigure[Summary of Bures results learned using SP ansatz]{
    \includegraphics[width=0.48\textwidth]{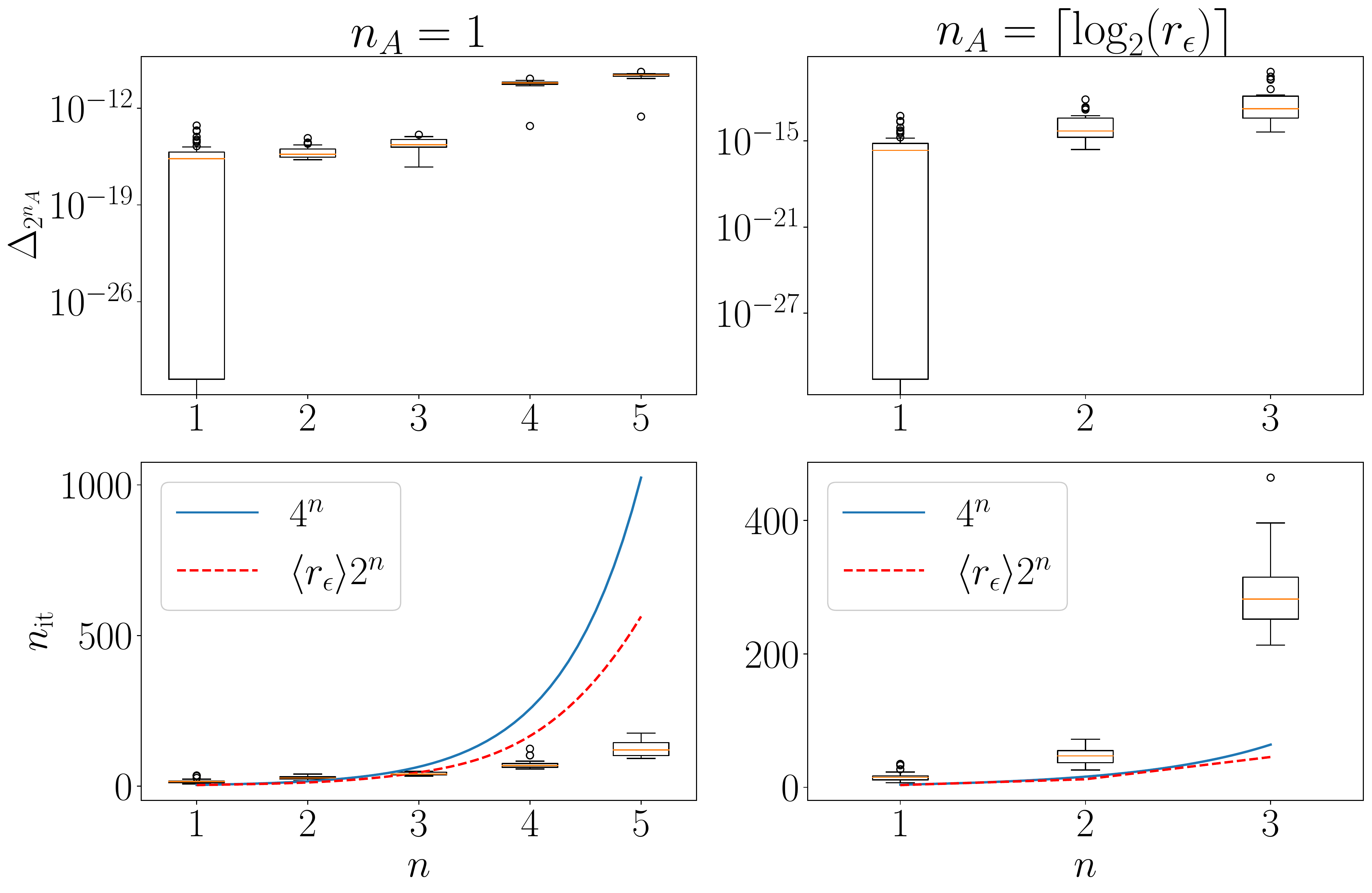}}
    %--------------
    \subfigure[Summary of Bures results learned using CCPS ansatz]{
    \includegraphics[width=0.48\textwidth]{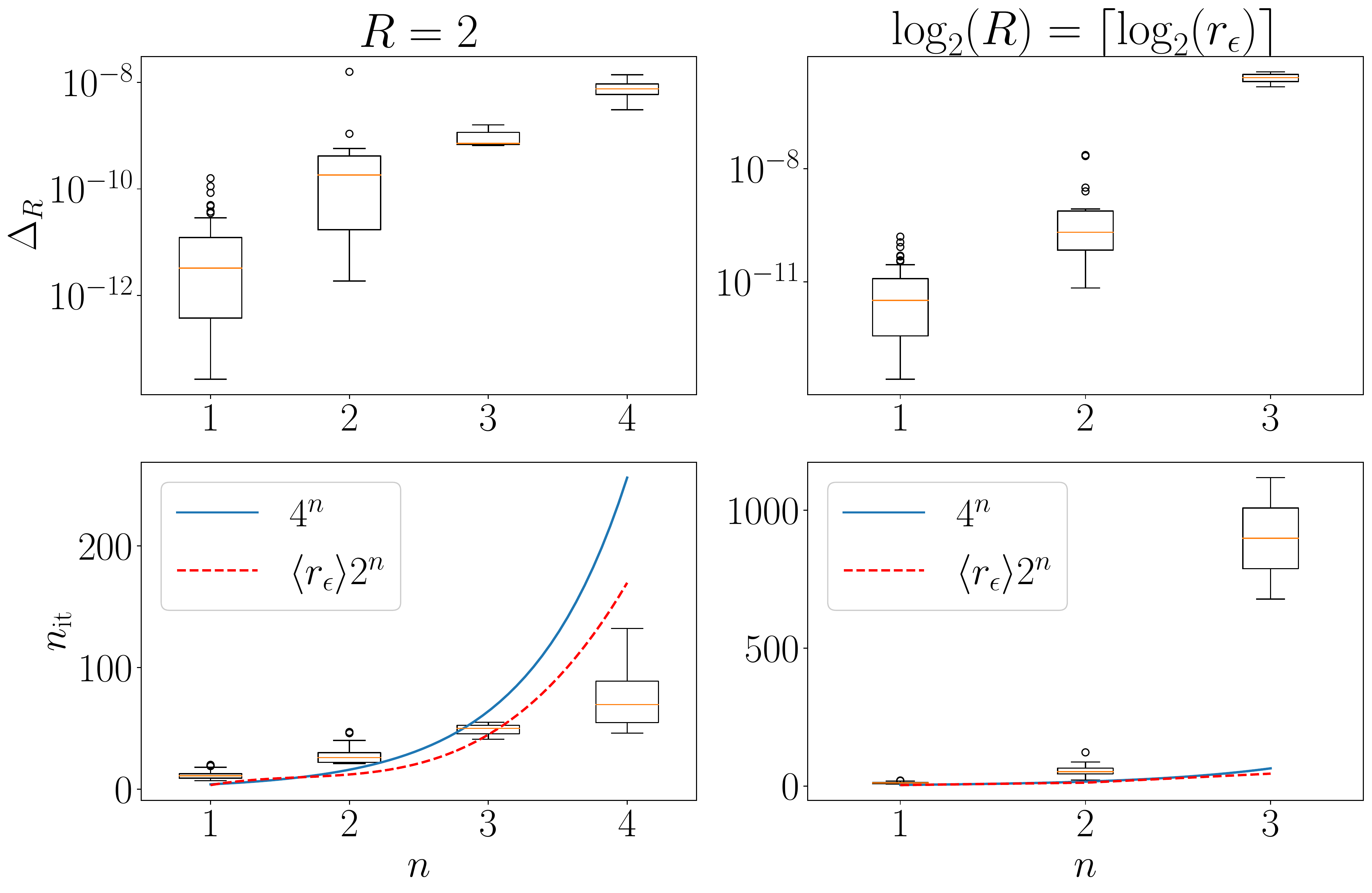}}
    %======================
    \caption{\textbf{Compiling Bures random states with idealized classical numerics.} Each box plot bins the quartiles from 25 Bures random states. Specifically, the orange line is the median, the box contains 75\% of the runs, the top and bottom lines show the max and min, and circles represent outliers. The left column of each four-panel figure corresponds to a low-rank approximation whereas the right corresponds to a full $\epsilon$-rank approximation. The first row shows how close the optimized state is to the best possible state, $\Delta_R$, as a function of the number of qubits, $n$. The bottom row shows the number of iterations, $n_{\text{it}}$, it takes to perform the optimization alongside the naive scaling of full quantum state tomography $4^n$ and enhanced quantum sensing tomography $\langle r_\epsilon \rangle 2^n$ where the average is across the 25 random instances.}
    \label{fig:bures-results}
\end{figure*}

We tested our algorithm on $25$ Bures random states for each $n$. The results are shown in Fig.~\ref{fig:bures-results}. Here, SP results are shown in (a) and the CCPS results in (b). We plot the difference between the optimized cost and the lowest possible cost $\Delta_R$ (see Eq.~\eqref{eq:fig-of-merit} or Appendix~\ref{app:computing-fig-of-merit} for more details) as a function of system size, $n$, for both low rank ($n_A = 1, R = 2)$ and full $\epsilon$-rank $(n_A = \lceil \log_2 r_\epsilon \rceil, R = 2^{\lceil \log_2 r_\epsilon \rceil})$ approximations along with the number of iterations $n_{\text{it}}$ necessary to reach $\Delta_R$. Here the compilation is performed classically using the Powell optimizer. Our ansatz, which uses alternating layers of arbitrary two-qubit gates, is explained in Appendix~\ref{app:ansatze}.

We focus our discussion first on the top row, where we plot the performance metric $\Delta_R$. Here, we see that our algorithm is capable of learning completely (Bures) random states provided $n$ is small enough. Indeed, for the SP ansatz, the value $\Delta_R$ stays below $10^{-10}$ for all values of $n$ tested when compiling both a low rank and a  full $\epsilon$-rank approximation. Interestingly, the values of $\Delta_R$ reached for the CCPS ansatz are substantially higher, reaching values of up to $10^{-8}$ for the low-rank and up to $10^{-6}$ for the full $\epsilon$-rank approximations. This suggests that the CCPS optimization is more difficult than the SP one. This is plausibly due to the fact that the optimization was performed over both angles and probabilities, which needed to satisfy a normalization constraint.

In the second row of Fig.~\ref{fig:bures-results}, we plot the number of iterations, $n_\mathrm{it}$, needed to reach the $\Delta_R$ values above. For reference, we also plot the curve showing $4^n$ scaling for naive full tomography as well as a curve $\langle r_\epsilon \rangle 2^n$ for improved quantum sensing tomography. Here, $\langle r_\epsilon \rangle$ denotes the average $\epsilon$-rank across the 25 random instances.
%which corresponds to a numerical measure of $\rank(\rho)$. 
While it is difficult to draw definitive claims for the small values of $n$ accessible, we see an interesting split in the results. For low-rank $(n_A = 1, R=2$) approximations, the number of iterations required seems to scale more favorably for our method than for both forms of tomography for both the SP and CCPS ans\"{a}tze. The opposite appears to be true for the full-$\epsilon$-rank approximation where it appears that even full tomography is a better strategy at $n = 3$ for both ans\"{a}tze. This could plausibly be explained by the barren plateau phenomenon for learning random states that was proven in Ref.~\cite{holmes2020barren}; hence reconfirming that variational methods are not well suited to fully learning typical random states. On the other hand, with no shot noise and such small $n$, this could also be due to the presence of many local minima~\cite{anschuetz2022beyond, bittel2021training}. Regardless of the cause, our simulations suggest that while our algorithm can learn small unstructured random states, it cannot scale beyond modest $n$.

\subsection{XY Model Results}

This discussion naturally raises the question of what happens when structure is present. This leads us to the study of thermal states in the Heisenberg XY model, given by
\begin{subequations}
    \begin{align}
        \label{eq:xy-thermal-state}
        \rho^{(\text{XY})}_n &\coloneqq \frac{e^{- \beta H_{\text{XY}}}}{\Tr[e^{-\beta H_{\text{XY}}}]} , \\
        %------------------
        H_{\text{XY}} &\coloneqq \sum_{i=1}^{n-1} J_i X_i X_{i+1} + K_i Y_i Y_{i+1} ,
    \end{align}
\end{subequations}
where $J_i, K_i \sim \mathcal{N}(0, 1)$ are i.i.d.~standard Gaussian random variables and $\beta = 1 / k_B T$ is the inverse temperature. By controlling the temperature, we can control the $\epsilon$-rank of the generated mixed states (see Appendix~\ref{app:xy-thermal} for more details). Structurally, this model is clearly invariant under any global rotation of all spins, and the number of spins is constant. Thus, even with random coefficients, it exhibits important symmetries which allow us to greatly simplify our ansatz (see Appendix~\ref{app:ansatze}). For this reason, we are able to test from $n = 2$ to $n = 8$ qubits relatively easily, which is sufficient for an initial study of empirical resource scaling.

\begin{figure*}[ht]
    \subfigure[Summary of XY thermal states learned using SP ansatz]{
    \includegraphics[width=0.48\textwidth]{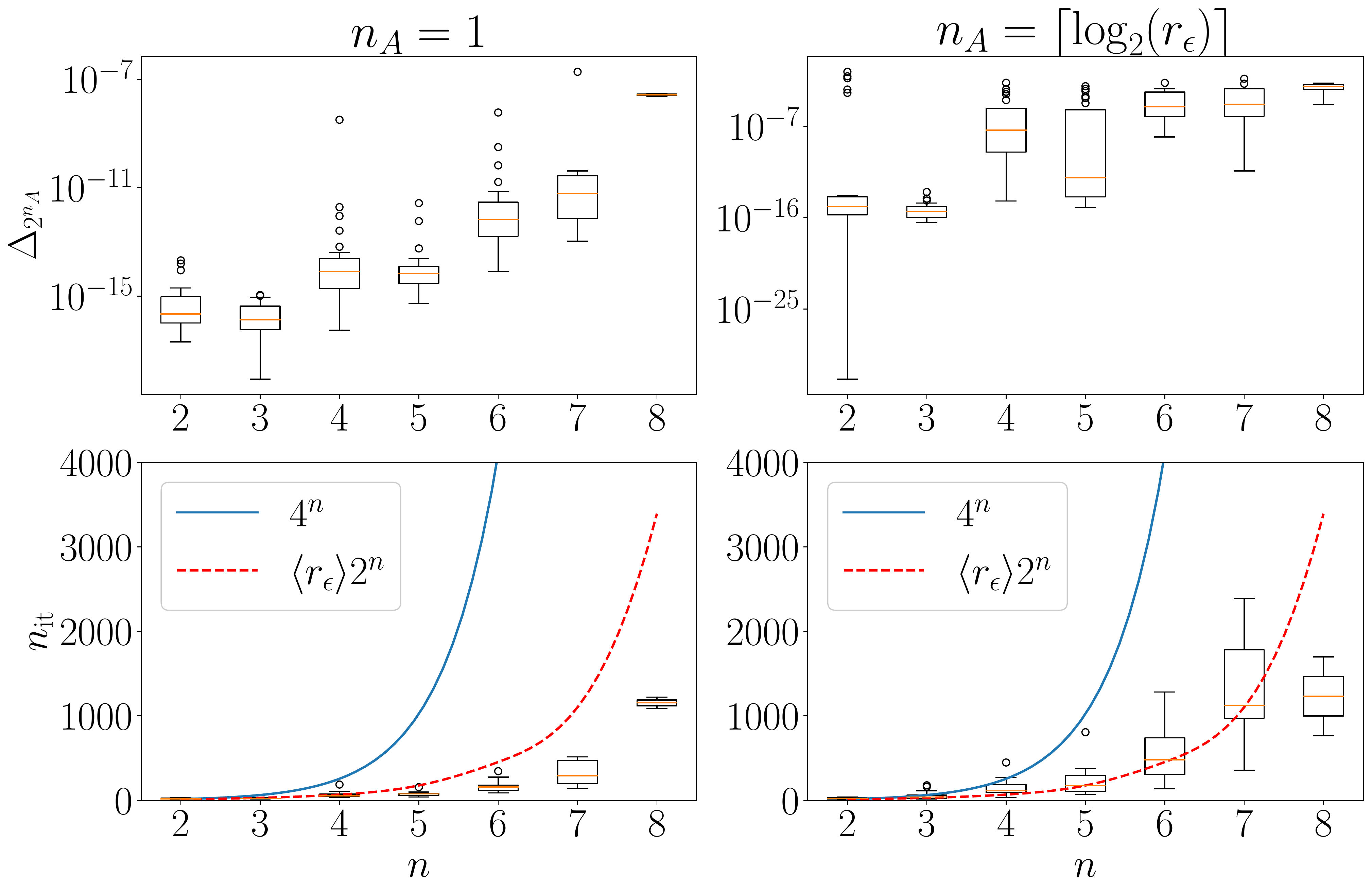}}
    %--------------
    \subfigure[Summary of XY thermal states learned using CCPS ansatz]{
    \includegraphics[width=0.48\textwidth]{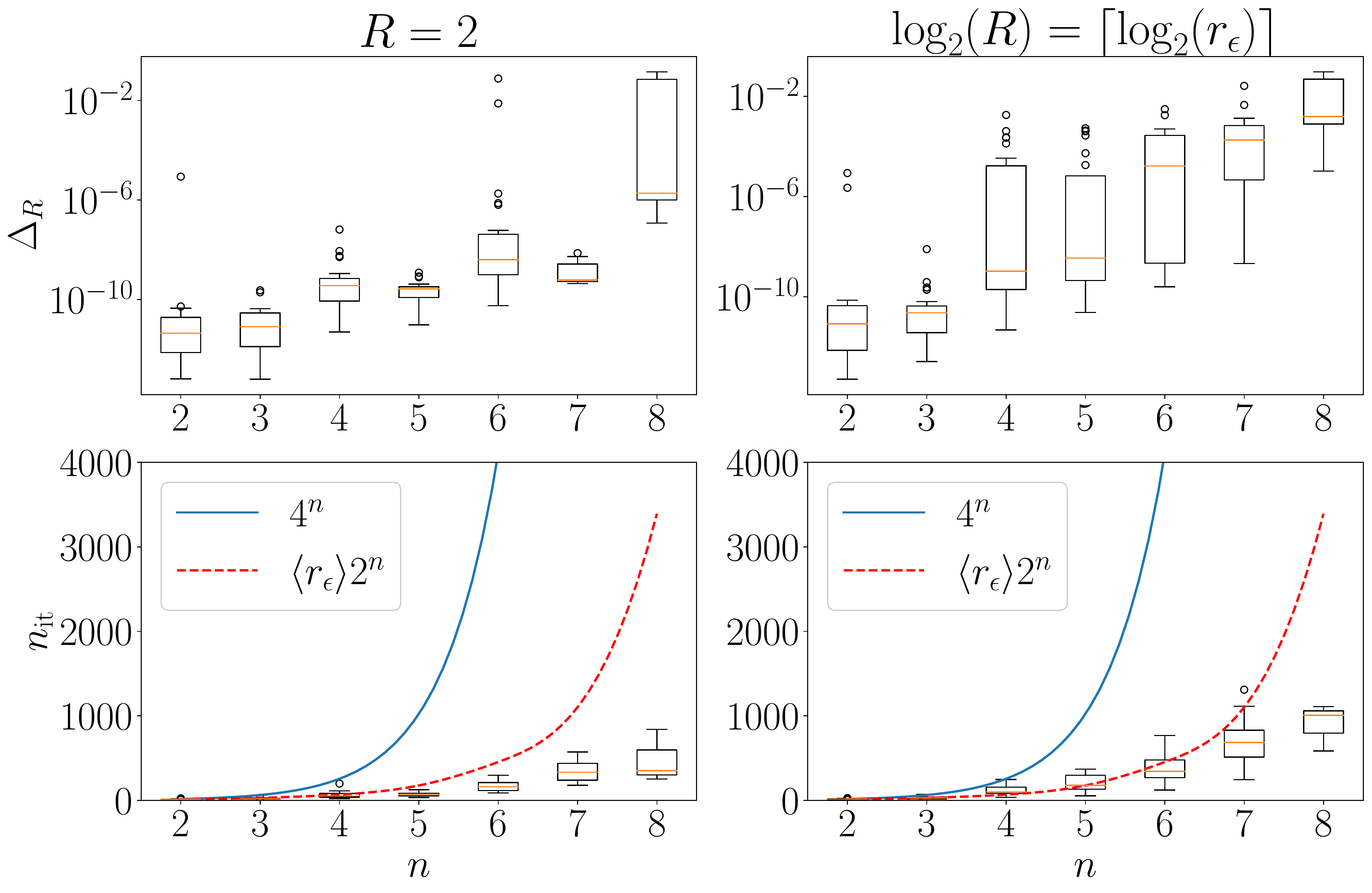}}
    %======================
    \caption{\textbf{Compiling XY thermal states with idealized classical numerics.} We summarize the XY thermal state results for the SP ansatz (left) and the CCPS ansatz (right). Each box plot bins the quartiles from 50 thermal states of the XY model where $J_i, K_i \sim \mathcal{N}(0, 1)$ are i.i.d.~standard Gaussian random variables: 25 at a low temperature $\beta = 20$ and 25 at a higher temperature $\beta = 2$. Specifically, the orange line is the median, the box contains 75\% of the runs, the top and bottom lines show the max and min, and circles represent outliers. The left column of each four-panel figure corresponds to a low-rank approximation whereas the right corresponds to a full $\epsilon$-rank approximation. The first row shows how close the optimized state is to the best possible state, $\Delta_R$, as a function of the number of qubits, $n$. The bottom row shows the number of iterations, $n_{\text{it}}$, it takes to perform the optimization alongside the naive scaling of full quantum state tomography $4^n$ and enhanced quantum sensing tomography $\langle r_\epsilon \rangle 2^n$ where the average is across the 25 random instances.}
    \label{fig:xy-results}
\end{figure*}

We tested our algorithm on 50 random XY chain thermal states (see Eq.~\eqref{eq:xy-thermal-state}) for each $n$: 25 at a large inverse temperature of $\beta = 20$ (i.e., low temperature regime) and 25 at a relatively smaller inverse temperature of $\beta = 2$ (i.e., moderate to high temperature regime). The results are shown in Fig.~\ref{fig:xy-results}, whose format is the same as that in Fig.~\ref{fig:bures-results}. Similar to the Bures results, we generated the results using noiseless classical optimization of an alternating layer ansatz with the Powell optimizer. Unlike Bures, however, the ansatz consisted of so-called Givens gates (see Appendix~\ref{app:ansatze} for exact definition) which respect the symmetries of the XY model.

Our first observation is that the performance (i.e., both $\Delta_R$ and $n_{\text{it}}$) does not have a noticeable dependence on $\beta$. For this reason, we chose to simplify the presentation of the results and combine all 50 states (for each $n$) into a single box plot. Note that this was not expected \emph{a priori}. As discussed in Appendix~\ref{app:xy-thermal}, $\beta = 20$ corresponds to the limit when $\rho_n^{(\text{XY})}$ is approximately in the ground state of $H_{\text{XY}}$ with a low $\epsilon$-rank, whereas $\beta = 2$ samples intermediate to large $\epsilon$-ranks. Hence, what we have found is that our optimization is insensitive to the underlying $\epsilon$-rank of the target state in this case.

For the SP ansatz, the performance (as measured by $\Delta_R$) is quite good, with a worst case of $n = 8,n_A = \lceil \log_2 r_\epsilon \rceil$, $\Delta_R\approx 10^{-6}$. The same point for the CCPS ansatz only reached $\Delta_R\approx 10^{-4}$ in median performance, and indeed, the CCPS performance is noticeably worse across the entire data set. That is, when solving the same problem, the final $\Delta_R$ for the CCPS ansatz is often noticeably larger than the SP ansatz. However, even this worst median $\Delta_R \approx 10^{-4}$ is an acceptable \enquote{4 nines} result (i.e., $C(\vec{\alpha}^*, R)$ differs from $C(\vec{\alpha}_{\text{opt}}, R)$ only in the fourth decimal place). 

Across the entire data set, whether low or full rank and SP or CCPS, the actual number of iterations scales slightly better than the compressed tomography scaling of $\langle r_\epsilon \rangle 2^n$ and is substantially better than na\"ive tomography. For example, at $n=8$, both low and high rank results use $\approx 56$ times fewer iterations (relative to $4^n = 4^8$ for full naive tomography) with SP. However, this positive result must be considered along with the result that the quality of the solution, $\Delta_R$, deteriorates with $n$. A fair summary of the result can be understood by setting an acceptable cut-off, $\delta_c$. Supposing that $\delta_c = 10^{-3}$ (a \enquote{three nines criterion}), what our data shows is that we can reach $\Delta_R < \delta_c$ in fewer iterations than both na\"ive tomography and compressed sensing tomography.

\subsection{\label{subsec:pca-example} PCA and state compression example}

\begin{figure}
    \centering
    \includegraphics[width=0.45\textwidth]{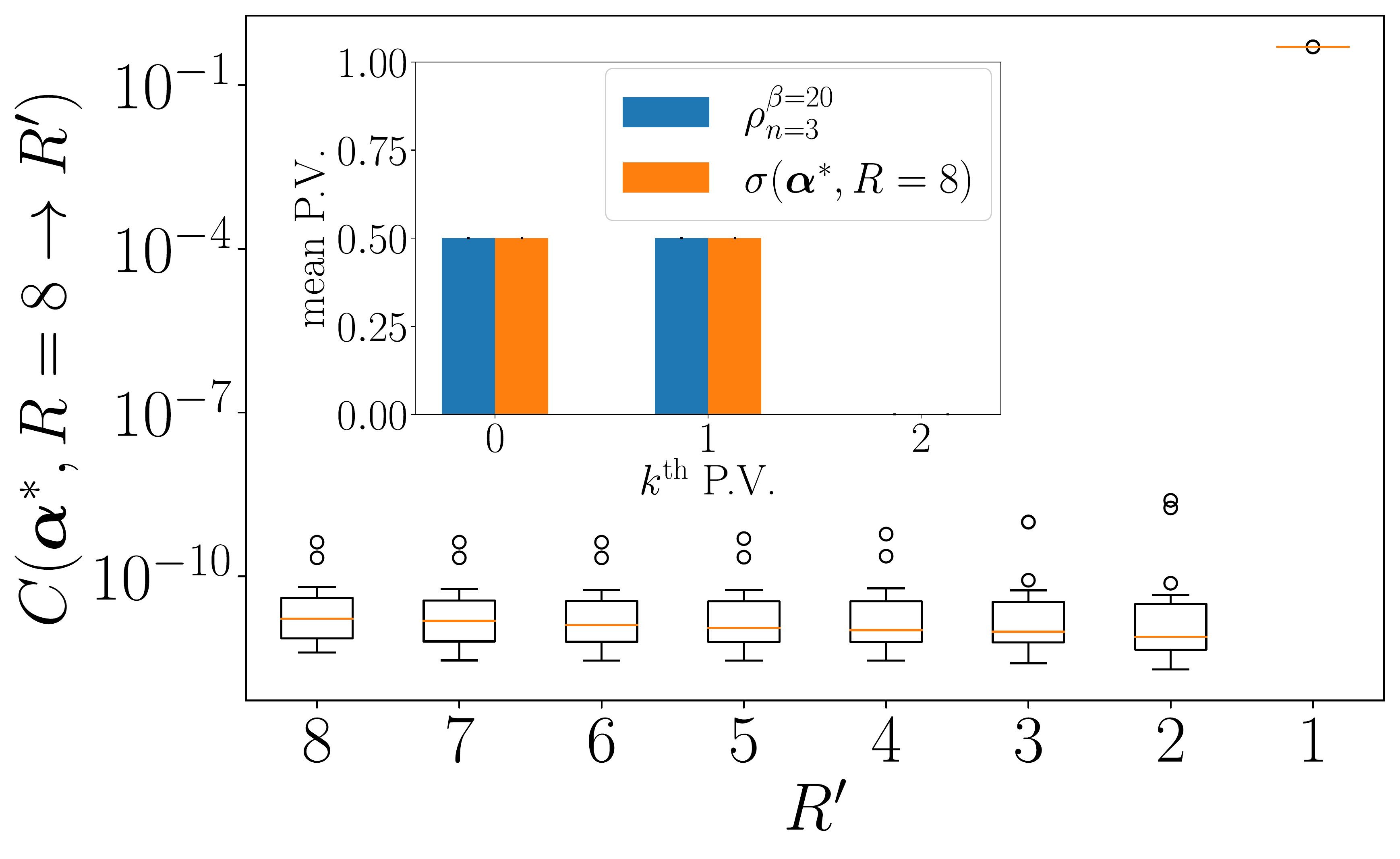}
    %---------------------------------
    \includegraphics[width=0.45\textwidth]{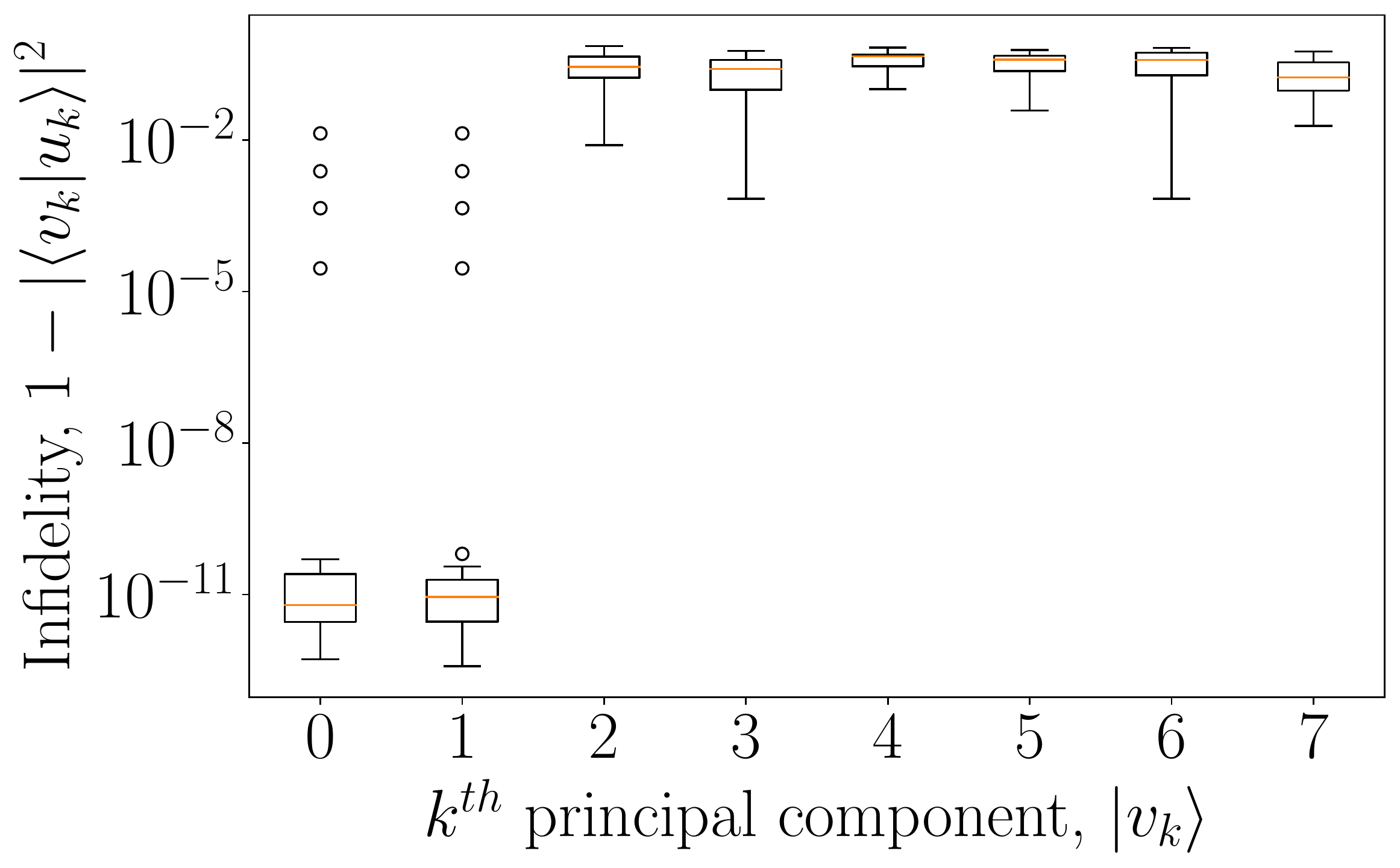}
    %---------------------------------
    \caption{\textbf{PCA and truncation with the CCPS ansatz.} We learn a full rank ($R = 8)$ approximation of 25 random three-qubit XY thermal states, achieving a median $\Delta_{R=8} \approx 10^{-10}$. \textbf{Top:} In the \emph{main plot}, we show the Hilbert--Schmidt cost between the target and learned CCPS ansatz when truncating to keep the first $R'$ states, i.e., $\ket{\psi_k}$ from $k \in\{ 1, \ldots, R'\}$. In the \emph{inset}, we show the distribution of principal values (i.e., ordered eigenvalues of $\rho$) for these XY thermal states alongside the found principal values of $\sigma$. \textbf{Bottom:} We show the infidelity between the $k^{\text{th}}$ principal component of $\rho$, $\ket{v_k}$, and the CCPS learned estimate $\ket{u_k}$. }
    \label{fig:pca-example}
\end{figure}

As discussed, the CCPS ansatz (see Eq.~\eqref{eq:ccps-ansatz}) takes the form of a convex combination of the $R$ estimated principal components of $\rho$ weighted by the associated principal values. Hence, learning $\sigma_{\text{CCPS}}(\vec{\alpha}^*, R)$ is tantamount to performing principal component analysis (PCA). As a consequence, we get the form of any $R' < R$ approximation from the rank $R$ solution for free by truncation (see Eq.~\ref{eq:truncated-CCPS-result}). We explore the practical meaning of these two statements by example in Fig.~\ref{fig:pca-example}. In this case, we learn a full rank CCPS ansatz approximation of a random three qubit XY thermal state at low temperature, $\rho_{n=3}^{\beta=20}$. As before, we consider 25 such optimizations over randomly drawn coefficients (see Eq.~\eqref{eq:xy-thermal-state}).

In the top plot of Fig.~\ref{fig:pca-example}, we explore the effect of truncation from $R = 8$ to $R' < R$ on the cost as a function of $R'$. At $R = 8$, the median cost is $C(\vec{\alpha}^*, R) = \Delta_{R=8} \approx 10^{-10}$ which means we have successfully learned $\rho_{n=3}^{\beta=20}$. But even as we truncate down to $R' = 2$, the cost hardly changes--only jumping to a large value of $0.5$ at $R' = 1$. This suggests that only two principal components are necessary to approximate $\rho_{n=3}^{\beta=20}$, and we corroborate this intuition by showing the principal values (or spectrum) of both $\rho_{n=3}^{\beta=20}$ and $\sigma(\vec{\alpha}^*, R = 8)$ in the inset. Here, the bar-plot shows the median $k^{\text{th}}$ principal value as a function of $k$. For $\rho_{n=3}^{\beta=20}$, the values are $\lambda_0 \approx 1/2$, $\lambda_1 \approx 1/2$, and $\lambda_3 \approx 10^{-13}$, so $r_\epsilon = 2$ for $\epsilon > 10^{-13}$ for at least half of the instances (and all for $\epsilon > 10^{-4}$). 

In the bottom plot of Fig.~\ref{fig:pca-example}, we show the pure state infidelity between the $k^{\text{th}}$ principal component of $\rho_{n=3}^{\beta=20}$, $\ket{v_k}$, and the associated estimate contained in the CCPS ansatz, $\ket{u_k} = U_{\vec{\theta}^*} \ket{k}$ as a function of $k$. Note that the $k$ labels are ordered by decreasing principal value, i.e., $k = 0$ corresponds to the principal component with the largest principal value and so on. Clearly, the infidelity for $k = 0$ and $k = 1$ is very small $\sim 10^{-11}$ whereas the infidelity for $k \geq 2$ can be rather large. This again is due to $\rho_{n=3}^{\beta=20}$ having an effective $\epsilon$-rank of two. But by plotting the pure state infidelity we have also made the notion of \enquote{learning the principal components of $\rho$} more explicit. Namely, a good approximation of $\rho$ with the CCPS ansatz relies on having a high quality and explicit estimate of its important principal components as weighted by the relative importance of the principal values. 

Finally, this discussion suggests an alternative way to use our algorithm as a means to find the approximate rank of an unknown state. By training for different values of $R$ until $C \ll 1$ or $C$ converges, we can estimate that $\rank(\rho) \approx R$. This procedure also clarifies what we mean by claiming that our algorithm allows us to \enquote{learn a lower rank approximation that allows for more efficient processing.} In this example, we mean that learning a rank $R = 2$ approximation is sufficient, and therefore $\sigma_{\text{CCPS}}(\vec{\alpha}^*, R=2)$ is a low-rank approximation/compression of $\rho_{n=3}^{\beta=20}$.

\section{\label{subsec:results-qh}Quantum Hardware Implementation}
\begin{figure*}[ht]
    %-----------------
    \subfigure[Random state results on IBM hardware]{
    \centering
    \label{fig:hardware-result-a}
    \includegraphics[width=0.48\textwidth]{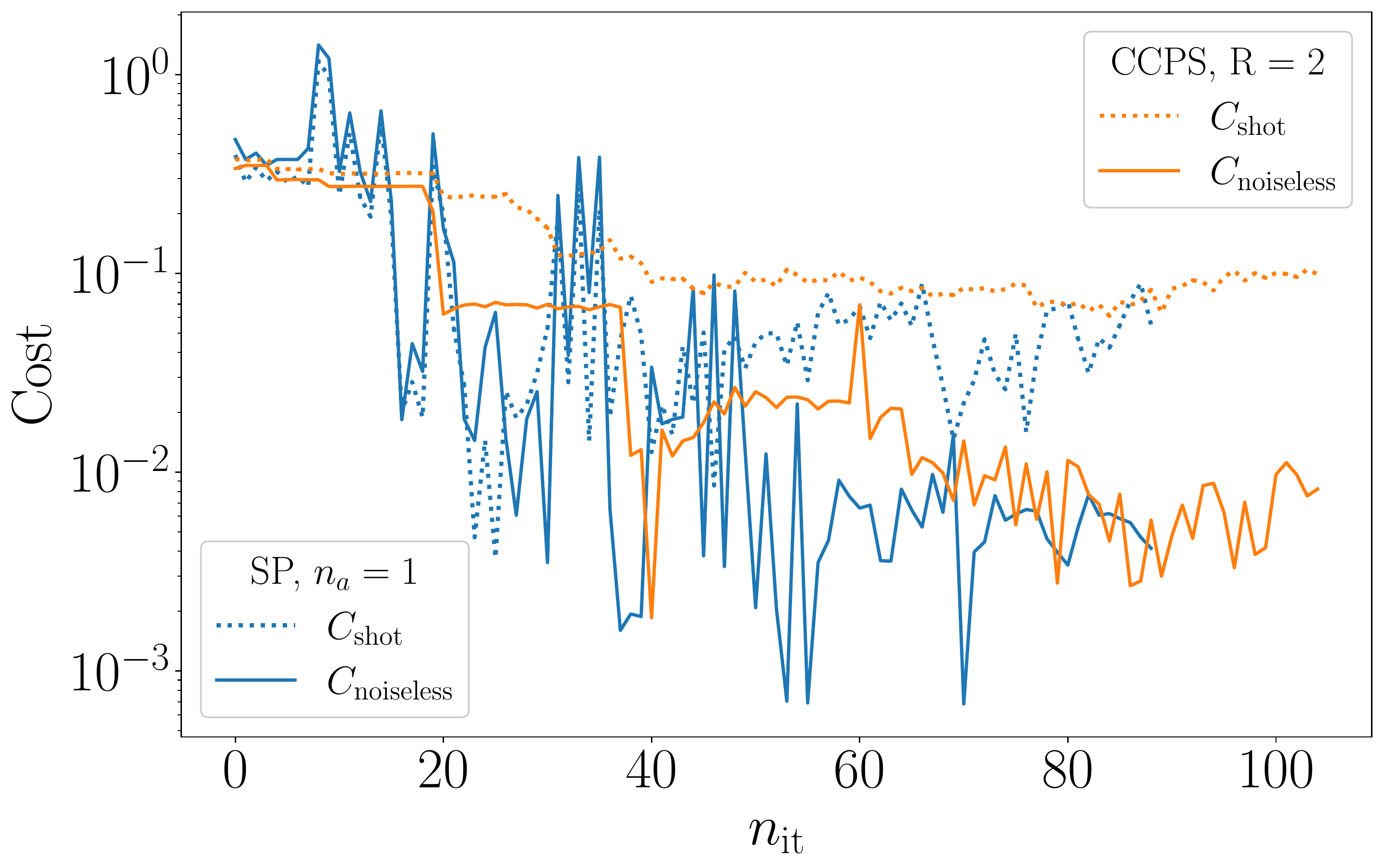}
     }
    %---------------------------------
    \subfigure[Hardware noise-induced state results on IBM hardware]{
    \centering
    \label{fig:hardware-result-b}
    \includegraphics[width=0.48\textwidth]{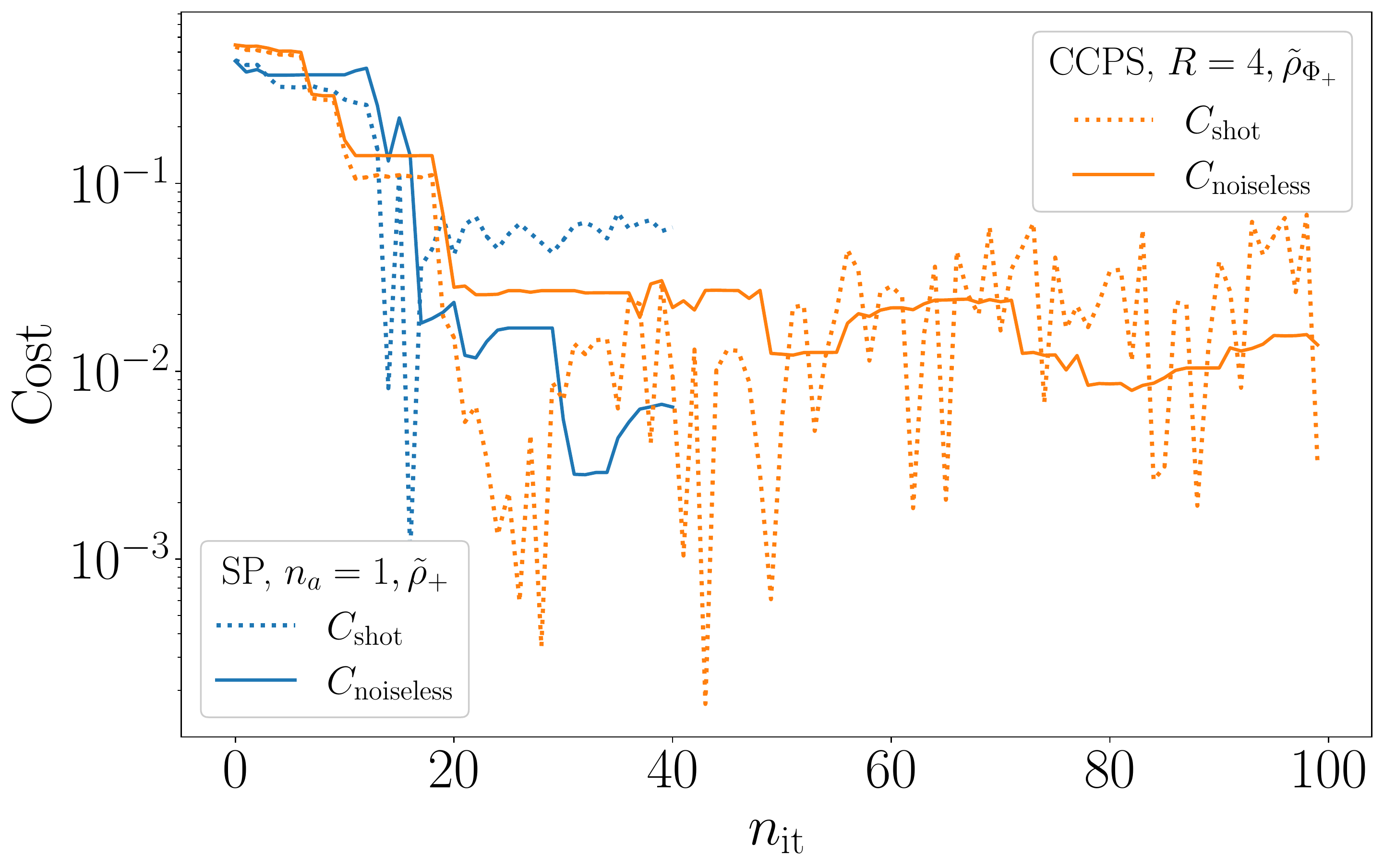}
    }
    %---------------------------------
    \caption{\textbf{Hardware Implementation of QMSC} We demonstrate the ability to compile quantum states on IBM superconducting devices by evaluating the cost function on them. In (a), we compile random single qubit states generated by Eq.~\eqref{eq:hardware-random-state} with both the SP and CCPS ansatz successfully.  In (b), we learn two hardware-noise induced states which are described in Eqs.~\eqref{eq:noisy-plus-state} and \eqref{eq:noisy-bell-state} using the SP and CCPS ansatz, respectively. Optimizations were performed on the seven-qubit devices \casablanca\ (random SP result) and \jakarta\ (all others) using $N = 10^5$ shots to evaluate $C_{\text{shot}}$ (dotted line). A noiseless cost $C_{\text{noiseless}}$ was also computed classically for verification. In each case, $10^{-3} \leq C_{\text{noiseless}} \leq 10^{-2}$ is reached which denotes a successful compiling given $10^5$ shots.}
    \label{fig:hardware-results}
\end{figure*}

Finally, we consider the most important task for our algorithm: compiling a quantum state \emph{on a quantum device.} In Fig.~\ref{fig:hardware-results}, we demonstrate the ability to learn two classes of quantum states on the IBM superconducting qubit devices with both ansatz choices. In Fig.~\ref{fig:hardware-result-a}, we successfully compile a random single qubit mixed state with both the SP and CCPS ansatz. Here, the random state is generated by tracing over one qubit in a two-qubit Haar random state
\begin{equation}
    \label{eq:hardware-random-state}
    \rho_{\operatorname{HS}} = \Tr_A[U_{\text{Haar}}\ketbra{00}{00}_{AB} U^{\dagger}_{\text{Haar}}].
\end{equation}
The Hilbert--Schmidt (HS) subscript signifies that such a state is uniformly sampled from the Hilbert--Schmidt metric~\cite{zyczkowski2011generating}. Finding a compilation for a HS random state has a similar proof-of-principle goal to the Bures random states, but this class is easier to prepare on NISQ devices. We remark that the state $\rho_{\operatorname{HS}}$ is known \emph{a priori} in this example in the sense that we supply the quantum device with $U_{\text{Haar}}$.

In Fig.~\ref{fig:hardware-result-b}, we learn \emph{unknown} states generated by noisy state preparation which we call \enquote{NISQ} states. In particular, we learn a noisy $\ket{+}$ state and a noisy $\ket{\Phi^+}$ state, 
\begin{subequations}
    \begin{align}
        \label{eq:noisy-plus-state}
        \Tilde{\rho}_+ &= \mathcal{E}_1 ({\ketbra{+}{+}}) \\
        %---------------------
        \label{eq:noisy-bell-state}
        \Tilde{\rho}_{\Phi^+} &= \mathcal{E}_2 ({\ketbra{\Phi^+}{\Phi^+}}),
    \end{align}
\end{subequations}
a single- and two-qubit state, respectively. For $\Tilde{\rho}_+$, we apply a single Hadamard and then idle for the duration of 20 Hadamards. For $\Tilde{\rho}_{\Phi^+}$, we follow the same procedure but then apply a final CNOT at the end. Under perfect conditions, these would generate the states $\ket{+}$ and $\ket{\Phi^+}$. Due to intrinsic $ZZ$ cross-talk on superconducting devices~\cite{gambetta2012characterization, mundada2019suppression, tripathi2021suppression, sheldon2016procedure} along with other secondary sources of noise, the qubits undergo a complicated dephasing process which we summarize as some unknown quantum channels $\mathcal{E}_1$ and $\mathcal{E}_2$, respectively. For some sense of the strength of noise on these devices, we remark that $\Tr[\Tilde{\rho}_+^2] = 0.90(1)$ and $\Tr[\Tilde{\rho}_{\Phi^+}^2] = 0.78(4)$ with $1\sigma$ confidence intervals generated from $10$ bootstrapped tomography experiments. Since we do not know the quantum channels in advance, compiling the NISQ states is also tantamount to learning the states. For example, we can learn a low-depth circuit to prepare $\Tilde{\rho}_+$, which is short enough to be unaffected by $ZZ$ cross-talk. This serves as a permanent snapshot to probe $\Tilde{\rho}_+$ even when $\mathcal{E}_1$ inevitably drifts due to two-level system and calibration effects.

With the states now defined, we discuss the results presented in Fig.~\ref{fig:hardware-results} more closely. For each optimization we plot two costs, $C_{\text{shot}}$ (dotted line) and $C_{\text{noiseless}}$ (solid line). The shot cost is computed on quantum hardware using $10^5$ shots, and it is the cost we optimize over using the Powell optimizer. The noiseless cost is computed classically in post-processing for verification. For the random states this amounts to classically storing the circuits to prepare $\rho$ and $\sigma$ throughout the optimization and computing the cost with matrix operations. For the NISQ states, we rely on full quantum state tomography to compute $\rho$ as a \enquote{trusted third party} method since the states are generated by unknown noise. We terminate the optimization when either $C_{\text{shot}}$ flattens for at least 10 iterations or 100 iterations are reached. In all cases, the final noiseless cost reaches $10^{-3} \leq C_{\text{noiseless}} \leq 10^{-2}$. We remark that this is consistent with the use of $N = 10^5$ shots since we expect a reported precision to scale as $\sim 1 / \sqrt{N}$. In all cases, we show the result of learning a full rank approximation of the target state, so $\Delta_R = C_{\text{noiseless}}$ and hence $10^{-3} \leq \Delta_R \leq 10^{-2}$. For all but the $\Tilde{\rho}_{\phi^+}$ optimization we note that the noiseless cost is an order of magnitude lower than the cost evaluated on the hardware which is indicative of optimal parameter noise resilience~\cite{sharma2019noise}. That is, it suggests that the position of the global cost minimum of the cost landscape is (approximately) invariant under the action of noise.

Overall, our results show that we can successfully learn full-rank approximations of hardware relevant states. As we might expect from the Bures and XY state results, we can also learn lower rank approximations which we show explicitly in App.~\ref{app:additional-hardware-results}. The net result is very similar: we can learn lower rank approximations to within a precision of $10^{-3}$ to $10^{-2}$, and as in the idealized classical experiments, it takes fewer iterations to learn lower rank approximations. Alternatively, we may choose to learn a high rank approximation of $\rho$ with the CCPS ansatz and obtain lower rank approximations via truncation (as in Sec.~\ref{subsec:pca-example}) which we also explore for the hardware data in App.~\ref{app:additional-hardware-results}.

\section{Discussion and Outlook}

%\section{Discussion and Future Work}

We have presented an algorithm to learn an unknown mixed state $\rho$. In particular, we have developed a procedure to learn a rank-$R$ approximation to $\rho$, where $\rank(\rho) \geq R$ is assumed but not essential. Put precisely, our algorithm is a practical variational way to solve the quantum low-rank approximation problem~\cite{ezzell2022quantum}. Applications of this algorithm are numerous and include PCA, state compression, learning noise-induced states, and uploading of states onto quantum computers, as described in Fig.~\ref{fig:applications}.

We considered two ansatz constructions. If the purification ansatz is chosen, the end result is a unitary $V(\vec{\theta}^*)$ such that $V(\vec{\theta}^*) \ket{0}^{\otimes n} \otimes \ket{0}^{\otimes n_A}$ generates a purification of the rank-$R$ approximation of $\rho$. By tracing out the $n_A$ ancilla qubits, we get the desired rank $R$ mixed state. For the convex combination of pure states (CCPS) ansatz, the final output is a classical vector $\vec{p}$ containing the $R$ principal values of $\rho$ and a unitary $U(\vec{\theta}^*)$ such that $U(\vec{\theta}^*) \ket{i}$ for $i \in\{ 1, \ldots, R\}$ gives the $R$ principal vectors of $\rho$.

Our numerical simulations and hardware implementations indicate that our algorithm works well for a variety of state ensembles, including random XY spin chain thermal states at arbitrary temperatures and unknown states generated by hardware noise on superconducting qubit devices. Unsurprisingly, we found that learning XY thermal states was easier than random states because the additional structure of the problem opened up the possibility of using simpler ans\"{a}tze. Additionally, while the SP ansatz performs better in numerical simulations (because its optimization problem is simpler), the CCPS ansatz, as expected, allows for larger hardware implementations because it requires fewer qubits.  %, and (iii) totally random states generated uniformly from the Bures measure.

For both ans\"{a}tze, our algorithm provides a means of compressing the target state when $\rank(\rho) > R$. For the purification ansatz, the reduction is in terms of the number of qubits; for the convex combination ansatz, the reduction is in terms of the number of pure states required to simulate the effect of the state. While the compression of states \cite{schumacher1995quantum,jozsa1998universal,schumacher2001indeterminate,horodecki2007quantum,plesch2010efficient,abdelhadi2020second}, and indeed data sets encoded in states~\cite{pepper2019experimental}, has been explored previously, much of the compression-based literature focuses on finding the compressed state via maximizing the degree to which the original state can be reconstructed via a successful decompression process~\cite{romero2017quantum}.

A particularly timely application of our algorithm is for quantum PCA. While quantum PCA was orginally proposed to have an exponential speedup over classical methods for low rank states~\cite{lloyd2014quantum}, it was later dequantized for the case of classical data analysis~\cite{tang2021quantum}, reducing speedups for this case to being modest ones~\cite{arrazola2019quantum}. However, recently it was shown that these dequantization arguments break down for quantum data analysis~\cite{cotler2021revisiting} and that quantum PCA for quantum data can indeed achieve an exponential quantum speedup~\cite{huang2021quantum}. Moreover a simple method to encode the covariance matrix into a density matrix was recently proposed~\cite{gordon2022covariance}, making quantum PCA an easily accessible application for near-term quantum computers. Hence, our approach for extracting the principal components of a density matrix is especially timely, in the quest for near-term quantum advantage.

%Relatedly, our algorithm further provides a new means of extracting the principal components of the target state---an important primitive for quantum machine learning~\cite{dunjko2018machine, li2021resonant}, as well as other applications~\cite{lloyd2014quantum}, and an area which has recently attracted much attention in both the fault tolerant~\cite{lloyd2014quantum, YuPCA_2019, cotler2021revisiting} and NISQ~\cite{huang2021quantum, gordon2022covariance} settings. 

In this article we have focused on coherent access models for quantum compilation. That is, computing the cost using the Loschmidt echo or SWAP test requires coherent interaction between the state we wish to compile and the device on which we wish to compile it. For this to be possible, the target quantum state either needs to be already prepared (potentially by some unknown process) on the quantum computer or we require a quantum sensor to mediate the interaction between the target quantum system and the quantum computer. The former is practically viable and reasonable to assume for applications such as learning noise-induced states or state compression. However, further developments in quantum sensors will be required to upload the unknown quantum state of an experimental system to a quantum computer in this coherent access model.

An alternative approach would be to use Clifford shadow tomography~\cite{huang2020predicting, elben2022randomized, huang2021provably} to efficiently compute the overlap between the target state and a large set of guess compilations via independent measurements and then classical post-processing. This incoherent version of quantum compilation could be applied in situations where the unknown state is prepared on a platform that is very different from the hardware on which we wish to compile it. In this manner, this approach opens up new techniques for uploading the quantum state of experimental systems to quantum hardware. In Fig.~\ref{fig:applications} we sketch the difference between the coherent and incoherent access models. Additionally, Ref.~\cite{huang2020predicting} provides a means of upper bounding the copy complexity of such a compilation task. In the case of efficiently preparable target states and ansatz states, i.e., those that can be prepared via a circuit with $T\in\mathcal{O}(\poly (n))$ local gates, then  $\mathcal{O} (T\log(T/\epsilon)/\epsilon^2)\leq\widetilde{\mathcal{O}}(\poly (n)/\epsilon^2)$ copies of the target state $\rho$ suffices to approximately evaluate the cost and its partial derivatives to precision $\epsilon$ arbitrarily often~\cite{caro2022outofdistribution}. Thus $\widetilde{\mathcal{O}}(\poly (n)/\epsilon^2)$ copies of the target $\rho$ in theory suffice to compile it to precision $\widetilde{\mathcal{O}}(\epsilon)$.

% Classical shadow tomography~\cite{huang2020predicting, elben2022randomized, huang2021provably} provides a way towards a copy complexity bound that is independent of the number of optimization steps.
% Namely, exploiting covering number bounds for the space of pure output states of polynomial-size quantum circuits (compare~\cite{caro2021generalization, huang2021quantum}), polynomial-size classical shadows can be used to perform tomography among such states. 
% In the case of an efficiently implementable target unitary $U$ and QNN $V(\alv)$ that both admit a circuit representation with $T\in\mathcal{O}(\poly (n))$ local gates, $\mathcal{O} (T\log(T/\epsilon)/\epsilon^2)\leq\Tilde{\mathcal{O}}(\poly (n)/\epsilon^2)$ copies of each of the input states $\ket{\Psi^{(j)}}$ and output states $\ket{\Phi^{(j)}}$ suffice to approximately evaluate the cost (both the global and local variants) and its partial derivatives arbitrarily often.

The framework investigated in this article for mixed state compilation may more generally be applied to the compiling of quantum channels. 
This follows from the fact that a channel may be represented via its Choi representation as a mixed state. That is, a channel may be fully characterized via the mixed state generated by applying a quantum channel to one half of a Bell state. Therefore one means of compiling a quantum channel would be to minimize the Hilbert--Schmidt distance between the Choi state corresponding to a target channel and an ansatz mixed state, formed by applying a parameterized channel to half a Bell state. In this sense, our algorithm further open up new avenues for learning unknown quantum processes.

% Yu, Gao, Lin, and Wang recently proposed an algorithm for compression of approximately low-rank datasets via PCA~\cite{YuPCA_2019}, in a non-VQA framework. Our paper tackles a slightly more general task from a different angle, seeking to achieve PCA via low-rank quantum compression, within a VQA framework, for states of any initial rank.
% A great deal of literature exists on the tasks of compilation, compression; however, much of the compression-based literature has an emphasized focus on decompression algorithms, i.e., examining how well one can reconstruct the original state from the compressed output.

% \section{Discussion}

\section{Data and Code Availability}
The data and source code that support the findings of this study are openly available in a Zenodo repository~\cite{ezzell2022qmsc}, a static and citable version of an available Github repository.

\begin{acknowledgments}

NE was supported by the U.S. Department of Energy (DOE) Computational Science Graduate Fellowship under Award Number DE-SC0020347. EMB acknowledges support from The Engineering and Physical Sciences Research Council (EPSRC) in the UK. AUS acknowledges support from the Center for Computation and Technology (CCT) at Louisiana State University. MMW acknowledges support from the National Science Foundation under grant no.~1907615. ATS acknowledges support from the QSC (see below).  PJC was initially supported by the Los Alamos National Laboratory (LANL) ASC Beyond Moore's Law project. PJC also acknowledges later funding from the Laboratory Directed Research and Development (LDRD) program of
LANL under project number 20210116DR. ZH acknowledges initial support from the LANL Mark Kac Fellowship and subsequent support from the Sandoz Family Foundation-Monique de Meuron
program for Academic Promotion. This work was also supported by the Quantum Science Center (QSC), a National Quantum Information Science Research Center of the U.S. Department of Energy (DOE).

%This work was also supported by the U.S. Department of Energy (DOE), Office of Science, Office of Advanced Scientific Computing Research, under the Accelerated Research in Quantum Computing (ARQC) program.

\end{acknowledgments}

\bibliography{quantum.bib}

\onecolumngrid

%\pagebreak 

\appendix
% \addcontentsline{toc}{chapter}{Appendices}
% \include{Appendix}
\vspace{0.2in}

\begin{center}
	{\Large \bf Appendices} 
\end{center}

\begin{comment}
%\section{Algorithm pseudo-code}
\begin{lstlisting}
    Initialize $\rho_{\text{tar}}, G^{(0)}(\theta_0), N_{\text{max}}, \delta, T_0$
    Set $j = 0$
    While $j < N_{\text{max}}$
        Minimize $D_{\operatorname{HS}}(G^{(j)}(\theta_j), \rho_{\text{tar}})$ with respect to $\theta_j$
        If $D_{\operatorname{HS}}(G^{(j)}(\theta_j^*), \rho_{\text{tar}}) < \delta$
            break
        end
        Generate $G^{(j + 1)}(\theta_{j+1})$ by adding identity perturbation to $G^{(j)}(\theta_j)$
        Minimize $G^{(j + 1)}(\theta_{j+1})$ with respect to $\theta_{j+1}$
        $\Delta_RD_{\operatorname{HS}} = D_{\operatorname{HS}}(G^{(j + 1)}(\theta^*_{j+1}), \rho_{\text{tar}}) - D_{\operatorname{HS}}(G^{(j)}(\theta^*_j), \rho_{\text{tar}})$
        If $\Delta_RD_{\operatorname{HS}} < 0$
            $G^{(j + 1)}(\theta^*_{j+1}) \longleftarrow G^{(j)}(\theta^*_j)$
        Else
            $p = e^{- | \Delta_RD_{\operatorname{HS}}| / T_j}$
            accept = Bernoulli(p)
            If accept == 0
                $G^{(j + 1)}(\theta^*_{j+1}) \longleftarrow G^{(j)}(\theta^*_j)$
            Else
                Set $G^{(j + 1)}(\theta_{j+1}) = G^{(j)}(\theta^*_j)$
            end
        end
        update $T_{j+1} \longleftarrow T_j$ by annealing schedule
        j += 1
\end{lstlisting}
\end{comment}

\section{Local Cost Function Analysis (HS)}\label{app:localcost}

As discussed briefly in Section~\ref{sec:algs}, barren plateaus form a barrier for trainability of VQAs for large-scale problems. Here, we formulate and analyse different proposals of localised versions of the Hilbert--Schmidt distance.

\subsection{Marginal Local Cost Function}

A simple, na\"ive choice in 1-local cost could be formulated in terms of the Hilbert--Schmidt distance between the local reduced states. We call this cost function the marginal 1-local cost function, and it takes the same form for both the state purification and convex combination of pure states ansatz. 

\begin{defn}
Given two quantum states $\rho$ and $\sigma$, let us define a \emph{Marginal $1$-local cost} as follows:
    \begin{align}
        C_{M}^{(1)}\left(\rho,\sigma\right) &\coloneqq  \frac{1}{n} \sum_{j=1}^n \left\|\rho_j -\sigma_j \right\|^2_2  \, ,
    \end{align}
where $\rho_j = \Tr_{\overline{j}}[\rho]$ and  $\sigma_j = \Tr_{\overline{j}}[\sigma]$, with $\overline{j}$ denoting all system qubits except for qubit $j$.
\end{defn}

This can be measured using the methods outlined in Section~\ref{sec:algs}. (Note, in contrast to the main text, in this appendix we stress the dependence of the cost on the target state $\rho$ and guess state $\sigma$.) 

\medskip

A successful candidate for a local cost function needs to remain faithful to the global cost, i.e., the minimal value of the cost corresponds to the case when $\rho = \sigma$. This cost function is trivially faithful for learning product states using a product state ansatz, i.e., if $\sigma = \bigotimes_{j=1}^n \sigma_j$ and $\rho = \bigotimes_{j=1}^n \rho_j$. However, it does not take much thought to see that this local cost function is not faithful more generally. In simple terms, different entangled states may have the same reduced states and so the local cost may vanish even when the global states are non-identical. For example, consider the case in which $\rho$ and $\sigma$ are orthogonal Bell states, i.e., $\rho = \ket{\psi_+}\bra{\psi_+}$ and $\sigma(\vec{\theta},n_A) = \ket{\psi_-}\bra{\psi_-}$. Now, in this case, the local cost vanishes because the reduced states are all maximally mixed 
    \begin{align}
        C_{M}^{(1)}(\rho,\sigma) &= \frac{1}{n} \sum_{j=1}^n \left\|\rho_j -\sigma_j \right\|^2_2  \\&  = \frac{1}{n} \sum_{j=1}^n \left\|\id/2 - \id/2 \right\|^2_2 = 0 \, .
    \end{align}
However, $ \ket{\psi_+}\bra{\psi_+} \neq  \ket{\psi_-}\bra{\psi_-}$ and so the local cost is not faithful. In contrast, the equivalent global cost, that is the HS distance between $\ket{\psi_+}\bra{\psi_+}$ and $\ket{\psi_-}\bra{\psi_-}$, is non zero, 
    \begin{align}
        D_{\rm HS}(\rho,\sigma) &\coloneqq \sqrt{ \Tr[\rho^2] + \Tr[\sigma^2] -2\Tr[\rho \sigma]} \\ 
        &= \sqrt{1 + 1 - 0} = \sqrt{2} \, ,
    \end{align}
as it should be.

% \begin{thm}
% $C_{HS}^{(1)}(\rho,\sigma(\vec{\theta},n_A)) = \frac{1}{n} \sum_j \|\rho_j -\sigma_j \|_2$ is faithful for tensor-product states
% \end{thm}

% \zo{This proof isn't quite right in its current form. Or at least, it needs to made clear that $U_{\vec{\theta}i}$ act on the $i_{\rm th}$ qubit of $S$ and $A$. Also, this would also work for product states and the proof would be even more trivial. So I'm starting to think maybe we don't even need to state this proof.}

% \textbf{Proof}
% Consider target and training states of the form $\rho = \bigotimes_{i=1}^{n} \rho_i$ and $\sigma(\vec{\theta},n_A) = \Tr_A[U_{\vec{\theta}} (\ket{0}\!\bra{0})^{\otimes (n+n_A)} U_{\vec{\theta}}^{\dagger}]$ with $U_{\vec{\theta}} = \bigotimes_{i=1}^{n+n_A} U_{\vec{\theta}i}$

% Then  
% \begin{equation}
% C_{HS}^{(1)}(\rho,\sigma(\vec{\theta},n_A)) = \frac{1}{n} \sum_j \|\rho_j - U_{S_jA_j}^{\vec{\theta}}\ket{0}\!\bra{0} U_{S_jA_j}^{\vec{\theta}}^{\dagger} \|_2,
% \end{equation}
% which vanishes iff $\rho_j = U_{S_jA_j}^{\vec{\theta}} \ket{0}\!\bra{0} U_{S_jA_j}^{\vec{\theta}}^{\dagger}$ for all $j$
% \qedsymbol

The lack of faithfulness does not necessarily preclude us using it for training. We could start training on the $1$-local cost and as we approach the solution, add in $k$-local cost terms where we look at the distance between the reduced states on $k$ qubits. That is, we propose to use a cost of the following form.

%  Whilst we must ensure each qubit is measured (and an equal amount of times), the number of measurements per qubit is also left to be chosen freely, i.e., we could choose a number of partitions such that, in total, we perform measurements on all ${}^n C_k$ choices of $k$ qubits.

\begin{defn}
Given two quantum states $\rho$ and $\sigma$, and some (possibly multiple) partitioning(s)\footnote{Choosing partitions is relatively arbitrary. One simple choice would be to partition the $n$ qubits into $\lfloor n/k \rfloor$ subsets of $k$ qubits and one subset of $n$ \% $k$ qubits. It would be natural to assign the qubits to the subsets to minimize the distance between qubits in a subset as determined by the target state's/hardware's geometry. However, one can imagine sets of more complex partitions including those where each qubit belongs to multiple subsets. For example, you could consider the set of all possible subsets of $k$ qubits; however, this would become resource intensive to measure in practise for $1 \ll k \ll n$.} $\mathcal{P}_k$ of the set of $n$ qubits into $N_k$ subsets of at most $k < n$ qubits, where at least one set has cardinality $k$, let us define the \emph{Generalized Marginal Cost} as follows:
\begin{equation}
    C_{\rm M}^{\rm gen}(\rho, \sigma) \coloneqq \sum_{k=1}^{n} \alpha_k(t)  C^{(k, \mathcal{P}_k)}_{\rm M}(\rho, \sigma),
\end{equation}
where 
\begin{equation}
    C^{(k, \mathcal{P}_k)}_{\rm M}(\rho, \sigma) \coloneqq  \frac{1}{N_k} \sum_{i=1}^{N_k} \left\Vert \rho_{i}^{(k)} - \sigma_{i}^{(k)} \right\Vert_2^2 ,
\end{equation}
with $\{ \rho_{i}^{(k)} , \sigma_{i}^{(k)} \}_{i=1}^{N_k}$ the set of marginals of $\rho$ and $\sigma$ as determined by $\mathcal{P}_k$, and the iteration dependent weightings $\alpha_k(t)\geq 0$ satisfying $\sum_k \alpha_k(t) = 1$.
\end{defn}

We note that an analysis of the general behaviour of $C_{\rm M}^{(k)}(\rho, \sigma)$ is non-trivial as the behaviour depends on the partitions $\mathcal{P}_k$. One (simple) potential choice of partition is to only allow $k$ such that $k \mid n$, and choose $N_k$ sets of $k$ qubits. In this case, for states that are $k$-local product states on the chosen subsets, $C^{(k)}_{\rm M}(\rho,\sigma)$ would be trivially faithful. By starting with the one local cost, i.e., with $\alpha_1 = 1$, and by slowly ramping to the global cost $C_{\rm M}^{(n)} \equiv  D^{(n)}_{\rm M}(\rho, \sigma) $ by increasing the weightings of the various $\alpha_k$ terms until $\alpha_n =1$, it should be possible to train to the minimum of the global cost. We expect this proposal to prove most useful for learning mixed states with relatively localized entanglement.

%We note that the $k$-local cost $D^{(k)}_{\rm M}(\rho, \sigma)$, is trivially faithful for $k$-local product states. 
%Therefore, by starting with the one local cost, i.e., with $\alpha_1 = 1$, and by slowly ramping to the global cost $C_{\rm M}^{(G)} \equiv  D^{(n)}_{\rm M}(\rho, \sigma) $ by increasing the weightings of the various $\alpha_k$ terms until $\alpha_n =1$, it should be possible to train to the minimum of the global cost. We expect this proposal to prove most useful for learning mixed states with relatively localized entanglement. 

% \paragraph*{Summary.}
% \medskip
% Advantages of cost:
% \begin{enumerate}
%     \item Faithful for tensor $k$-product states.
%     \item The cost is iteratively refinable.
%     \item The cost is an operationally meaningful distance measure.
% \end{enumerate}

% Disadvantages of cost:
% \begin{enumerate}
%     \item The cost will not be faithful for highly entangled states.
% \end{enumerate}
% This na\"{i}ve proposal of a local cost function remains operationally meaningful, simply being the Hilbert--Schmidt distance between marginal states. It is provably faithful for tensor-product states, however is limited in general use due to its incompatibility with highly entangled states, unless paired with the global cost, allowing iterative refinements to avoid barren plateaus, at the cost of requiring further measurements to compute all required global and ($k$-)local costs.

%\subsection{Low Impurity Local Cost Function}
\subsection{Local Measurements on Global States}
An alternative approach to formulating a local cost function is to continue to use the full $n$-qubit input states, but replace the global measurements with \emph{local measurements}  (similarly to Ref.~\cite{khatri2019quantum}). As each ansatz includes specifically prescribed measurement operators, we treat each of our ans\"atze separately, and derive provable guarantees of faithfulness for special cases.

\subsubsection{Local costs for CCPS ansatz}

Consider the global Hilbert--Schmidt cost function in the CCPS ansatz framework:
% \begin{equation}
%     C_{\mbox{\tiny CCPS}}( \vec{\alpha}, R)  = \Tr[\rho^2] + \sum_i p_{\vec{\phi}}(i)^2 - 2 \sum_i p_{\vec{\phi}}(i)\Tr\left[U^\dagger_{\vec{\theta}} \rho U_{\vec{\theta}} \ket{i}\bra{i}\right] \, ,
% \end{equation}
% which can be rewritten as
\begin{equation}
    C_{\mbox{\tiny CCPS}}( \vec{\alpha}, R)  = \Tr[\rho^2] + \sum_i p_{\vec{\phi}}(i)^2 - 2 \sum_i p_{\vec{\phi}}(i)\Tr\left[U^\dagger_{\vec{\theta}} \rho U_{\vec{\theta}} H_{\textrm{G}}^{(i)}\right]\, ,
\end{equation}
where we write $H_{\textrm{G}}^{(i)} := \ket{i}\bra{i}_S$ to emphasise the globality of the measurement.
Since the purity of the target state is not optimized and the purity of the guess state is optimized fully classically, these terms can be left global without impeding trainability. However, for the overlap term, we seek to replace the global $H_{\textrm{G}}^{(i)}$ term with some set of local measurements. There is freedom in how we may choose our local measurements, but one simple approach is to choose the average of all $1$-local measurements, as follows
\begin{defn} 
For the CCPS ansatz, we can define a \emph{Local cost function} as
\begin{equation}
    C_{\textrm{L}}^{\rm  CCPS} = \Tr[\rho^2] + \sum_i p_{\vec{\phi}}(i)^2  - 2 \sum_i p_{\vec{\phi}}(i) \Tr_{S}\left[U_{\vec{\theta}}^\dagger  \rho  U_{\vec{\theta}}  H_{\rm L}^{\rm CCPS \, (i)} \right] \,\, ,
\end{equation}
with 
\begin{equation}
    H_{\rm L}^{\rm CCPS \, (i)} = \frac{1}{n} \sum^n_{j=1} \ket{i_j}\bra{i_j}_{S_j} \otimes \id_{S_{\overline{j}}}   \, ,
\end{equation}
where, as in \eqref{eq:ccps-ansatz}, $\ket{i}=\bigotimes_{j=1}^n \ket{i_j}$ denotes an element of the computational basis of $n$ qubits, and $\ket{i_j}$ denotes the $j$-th bit of the bit string $i$.
\end{defn}
In order to prove faithfulness in the case of pure states, we first state and re-derive a result introduced in~\cite{khatri2019quantum}.
\begin{propos}\label{QAQCThm}
Let $\rho$ be an arbitrary quantum state, and let $U_{\vec{\theta}}$ be a parameterized unitary matrix. Denoting $\rho_{U} = U_{\vec{\theta}}^{\dagger} \rho U_{\vec{\theta}}$, and $H_{\rm L} ^{(i)} = H_{\rm L} ^{\rm{CCPS} (i)}$, we have that
\begin{equation}\label{QAQCrel}
    1 - \Tr\left[\rho_{U} H_{\rm L}^{(i)}\right] \leq 1 - \Tr\left[\rho_{U} H_{\rm G}^{(i)}\right] \leq n \left(1 - \Tr\left[\rho_{U} H_{\rm L}^{(i)}\right] \right)
\end{equation}
\end{propos}
\begin{proof}
We can write $H_{\textrm{L}}^{(i)} = \frac{1}{n} \sum_j^n H_{\textrm{L},j}^{(i)}$, where
\begin{equation}
    H_{\textrm{L},j}^{(i)} = \ket{i_j}\bra{i_j}_{S_j} \otimes \id_{S_{\overline{j}}} 
\end{equation}
are projectors that mutually commute. Note that $\prod_{j=1}^{n} H_{\textrm{L},j}^{(i)} = H_{\rm{G}}^{(i)}$. We can associate events $E_j$ with the projectors $H_{\textrm{L},j}$ such that $\Pr[E_j] = \Tr\left[\rho_{U} H_{\textrm{L},j}^{(i)}\right].$ Then, $\Tr\left[\rho \prod_{j=1}^n H_{\textrm{L}, j}^{(i)}\right] = \Pr \cap_{j=1}^n E_j$.
Recall, from basic probability theory, that for any set of events $\mathcal{A} \coloneqq \left\{ A_1,A_2, \ldots , A_n \right\}$, it holds that
\begin{equation}
    \Pr\!\left[\cup_{i=1}^n A_i\right] \geq \frac{1}{n} \sum_{i=1}^{n} \Pr\!\left[A_i\right].
\end{equation}

Choosing $A_i = \overline{E_j}$, we see
\begin{equation}
\begin{aligned}
    \Pr\!\left[\cup_{j=1}^n \overline{E_j}\, \right] &\geq \frac{1}{n} \sum_{i=j}^n \Pr\!\left[\, \overline{E_j}\, \right] \\
    \implies 1 - \Pr\!\left[\cap_{j=1}^n E_j\right] &\geq \frac{1}{n}\sum_{j=1}^n \left( 1 - \Pr\!\left[E_j\right] \right) \\
    \implies 1 - \Pr\!\left[\cap_{j=1}^n E_j\right] &\geq 1 - \frac{1}{n} \sum_{j=1}^n \Tr\!\left[\rho_U H^{(i)}_{\textrm{L},j}\right] \\
    \implies 1 - \Tr\!\left[\rho_U H_G^{(i)}\right] &\geq 1 - \Tr\!\left[\rho_U H^{(i)}_{\textrm{L}}\right].
\end{aligned}
\end{equation}
This is precisely the first desired inequality $1 - \Tr\left[\rho_U H^{(i)_{\rm L}}\right] \leq 1 - \Tr\left[\rho_U H_G^{(i)}\right]$. To prove the remaining inequality, observe that, via the union bound, we have
\begin{equation}
\begin{aligned}
    \Pr\!\left[\cup_{j=1}^n \overline{E_j}\,\right] &\leq \sum_{j=1}^n \Pr\!\left[\overline{E_j}\right] \\
    \implies 1 - \Pr\!\left[\cap_{j=1}^n E_j\right] &\leq \sum_{j=1}^n \left(1 - \Pr\!\left[E_j\right]\right) \\
    \implies 1 - \Tr\!\left[\rho_U H_G^{(i)}\right] &\leq n\left(1 - \Tr\!\left[\rho_U H^{(i)}_L\right]\right)
\end{aligned}
\end{equation}
Thus, $1 - \Tr_{SB}\left[\rho_U H^{(i)}_{\rm L}\right] \leq 1 - \Tr_{SB}\left[\rho_U H^{(i)}_{\rm G}\right] \leq n \left( 1 - \Tr_{SB}\left[\rho_U H^{(i)}_{\rm L}\right] \right)$ as required.
\end{proof}

\begin{propos}\label{thm:purefaithful}
$C_{\rm L}^{\rm CCPS}$ is faithful for pure states, and ``close to faithful'' for states with low impurity. Specifically we have that
\begin{equation}\label{eq:impuritybound}
   n C_{\rm L}^{\rm CCPS} \geq  C_{\rm G} - \left(n - 1\right) ( \operatorname{Impurity}(\rho) + \operatorname{Impurity}(\sigma)) \, ,
\end{equation}
where $\operatorname{Impurity}(X) \coloneqq 1 - \Tr[X^2]$ for $X = \rho$ and $X = \sigma$ and we write $C_G = \left\| \rho - \sigma \right\|_2^2$ to emphasize the globality of the standard Hilbert--Schmidt distance cost.
\end{propos}
\begin{rmk}
It follows from Proposition~\ref{thm:purefaithful}, that if $C_{L}^D = 0$ then 
\begin{equation}
  \, (n- 1) \left(\operatorname{Impurity}(\rho) - \operatorname{Impurity}(\sigma) \right) \geq \left\| \rho - \sigma \right\|_2^2 \, .
\end{equation}
That is, if the target and trained state are pure, i.e., $ \operatorname{Impurity}(\sigma) =  \operatorname{Impurity}(\rho) = 0$, we have that $C_{L} = 0$ entails that $C_{G} = 0$. More generally, if the impurities of the target and trained states are low, $C_{L} = 0$ entails that $C_{G}$ is small. 
\end{rmk}

\begin{proof}
From \eqref{QAQCrel}, we have 
\begin{equation}
\begin{aligned}
    & 1 - \Tr\left[\rho_U H^{(i)}_{\rm G}\right] \leq n \left(1 - \Tr\left[\rho_U H^{(i)}_{\rm L}\right] \right) \\
    &\implies -\Tr\left[ \rho_U H^{(i)}_{\rm G}\right] \leq (n - 1) - n \Tr\left[\rho_U H^{(i)}_{\rm L}\right] \\
    &\implies -2 \sum_i p_{\phi}(i) \Tr\left[ \rho_U H^{(i)}_{\rm G}\right] \leq -2n \sum_i p_{\phi}(i) \Tr\left[\rho_U H^{(i)}_{\rm L}\right] + 2(n-1)\sum_i p_{\phi}(i), \\
\end{aligned}
\end{equation}
where $2(n-1) \sum_i p_{\phi}(i) = 2(n-1)$. Adding the purity terms to both sides gives
\begin{equation}
\begin{aligned}
    & C_G \leq \Tr\left[\rho^2\right] + \Tr\left[\sigma^2\right] - 2 n \sum_i p_{\phi}(i) \Tr\left[\rho_U H^{(i)}_{\rm L}\right] + 2(n-1) \\
    &\implies C_G \leq n C_{\rm L}^{\rm CCPS} + (n - 1)\left(2 - \Tr\left[\rho^2\right] - \Tr\left[\sigma^2\right] \right),
\end{aligned}
\end{equation}
which can be rewritten as
\begin{equation}
    nC_{\rm L}^{\rm CCPS} \geq C_G - (n - 1)(\operatorname{Impurity}(\rho) + \operatorname{Impurity}(\sigma).
\end{equation}

Accordingly, if $C_{\rm L}^{\rm CCPS} = 0$, then
\begin{equation}
    C_G \leq (n - 1)(\operatorname{Impurity}(\rho) + \operatorname{Impurity}(\sigma))
\end{equation}
Therefore if the purity of the target and trained states are zero, the cost is faithful; i.e., $C_{\rm L}^{\rm CCPS} = 0$ implies $C_{\rm G} = 0$. Similarly, for high purity states the cost is approximately faithful, i.e., $C_{\rm L}^{\rm CCPS} = 0$ implies $C_{\rm G}$ is small. 
\end{proof}
\medskip

Similar to the generalized marginal cost defined earlier, we can construct an extension of the low impurity local cost function, such that it is local on $k$ qubits, for $k \leq n$.
One simple approach is to perform $n/k$ measurements, on $k$ qubits at a time. This naturally restricts us to only choosing $k$ such that $k \mid n$. Defining
\begin{equation}
    H_{\textrm{L}}^{k (i)} \coloneqq \frac{1}{(n/k)} \sum_{m=1}^{n/k} H_{\textrm{L}_m}^{k(i)}, 
\end{equation}
with
\begin{equation}
    H_{\textrm{L}_m}^{k (i)} \coloneqq \ket{i_{\mathcal{P}_m}}\bra{i_{\mathcal{P}_m}}_{\mathcal{P}_m} \otimes \id_{\overline{\mathcal{P}_m}} ,
\end{equation}
where $\mathcal{P}_m$ contains the $k$ indices of the $k$ qubits being measured over by the $m$-th operator, such that $\mathcal{P}_1 \cup \cdots \cup \mathcal{P}_{n/k}$ spans $\{1,\ldots,n\}$; we can define a $k$-local cost function as follows:
\begin{defn}
For $k \mid n$, we have the \emph{$k$-local cost function}
\begin{equation}
    C^k_{\rm L} = \Tr\left[\rho^2\right] + \sum_{i} p_{\phi}(i)^2 - 2 \sum_{i} p_{\phi}(i) \Tr_{S} \left[ U^{\dagger}_{\vec{\theta}} \rho U^{\dagger}_{\vec{\theta}} H_{\textrm{L}}^{k (i)}\right]
\end{equation}
\end{defn}

\begin{propos}
Let $\rho$ be an arbitrary quantum state, and let $U_{\vec{\theta}}$ be a parametrized unitary matrix. Denoting $\rho_U = U^{\dagger}_{\vec{\theta}} \rho U_{\vec{\theta}}$, we have that
\begin{equation}\label{eqn:klocalrelation}
    1 - \Tr \left[ \rho_U H^{k (i)}_{\textrm{L}} \right] \leq 1 - \Tr \left[ \rho_U H_{\textrm{G}}^{(i)} \right] \leq \frac{n}{k} \left( 1 - \Tr \left[ \rho_U H^{k (i)}_{\textrm{L}} \right] \right)
\end{equation}
\end{propos}

\begin{proof}
Similarly to Proposition~\ref{QAQCThm}, we can associate events $E_i$ with the projections $H^{k (i)}_{\textrm{L}_m}$ such that $\Pr\!\left[E_i\right] = \Tr \left[ \rho_U H^{k (i)}_{\textrm{L}_m} \right] $. Then, $\Tr \left[ \rho_U \prod_{m = 1}^{n/k} H^{k (i)}_{\textrm{L}_m}\right] = \Pr\!\left[\cup_{m=1}^{n/k} E_m \right]$. We have that
\begin{equation}
\begin{aligned}
    \Pr\!\left[ \bigcup_{m=1}^{n/k} \overline{E_m} \right] &\geq \frac{1}{(n/k)} \sum_{m=1}^{n/k} \Pr \left[ \, \overline{E_m}\, \right] \\
    \implies 1 - \Pr\!\left[ \bigcap_{m=1}^{n/k} E_m \right] &\geq \frac{1}{(n/k)} \sum_{m=1}^{n/k} \left(1 - \Pr \left[ E_m \right] \right) \\
    \implies 1 - \Tr \left[\rho_U H^{(i)}_{\textrm{G}} \right] &\geq 1 - \frac{1}{(n/k)} \sum_{m = 1}^{n/k} \Tr\!\left[ \rho_U H^{k (i)}_{\textrm{L}_m} \right],
\end{aligned}
\end{equation}
forming one side of our inequality. We also have
\begin{equation}
\begin{aligned}
    \Pr\!\left[ \bigcup_{m=1}^{n/k} \overline{E_m} \right] &\leq \sum_{m=1}^{n/k} \Pr\!\left[\, \overline{E_m}\, \right] \\
    \implies 1 - \Pr\!\left[ \bigcap_{m=1}^{n/k} E_m \right] &\leq \sum_{m=1}^{n/k} \left(1-\Pr\!\left[ E_m \right]\right) \\
    \implies 1 - \Tr\!\left[\rho_U H^{(i)}_{\textrm{G}}\right] &\leq \frac{n}{k} \left(1 - \Tr\!\left[\rho_U H^{k (i)}_{\textrm{L}}\right]\right).
\end{aligned}
\end{equation}
Thus,
\begin{equation}
    1 - \Tr \left[ \rho_U H^{k (i)}_{\textrm{L}} \right] \leq 1 - \Tr \left[ \rho_U H^{(i)}_{\textrm{G}} \right] \leq \frac{n}{k} \left( 1 - \Tr \left[ \rho_U H^{k (i)}_{\textrm{L}} \right] \right).
\end{equation}
\end{proof}
\begin{propos}\label{thm:klocthm}
The $k$-local cost function $C^k_{\rm L}$ is faithful for pure states, and ``close to faithful'' for states with low purity, with this closeness increasing with $k$. Specifically we have that
\begin{equation}
    \frac{n}{k} C^k_{\rm L} \geq C_G - \left(\frac{n}{k} - 1\right) \left(\operatorname{Impurity}(\rho) + \operatorname{Impurity}(\sigma)\right) .
\end{equation}
\end{propos}
\begin{rmk}
It follows from Proposition~\ref{thm:klocthm}, that if $C^k_L = 0$, then
\begin{equation}
   \left(\frac{n}{k} - 1\right)\left( \operatorname{Impurity}(\rho) + \operatorname{Impurity}(\sigma)\right) \geq \left\| \rho - \sigma \right\|^2_2 .
\end{equation}
\end{rmk}
\begin{proof}
The proof is entirely analogous to that for Proposition~\ref{thm:purefaithful} but with $n \rightarrow n/k$.
\end{proof}

Comparing this to the inequality found for the $1$-local cost \eqref{eq:impuritybound}, we find that the $k$-local cost is `closer to faithful' at low impurities than the $1$-local cost. It becomes increasingly faithful as $k$ tends to $n$ and, trivially, perfectly faithful for $k = n$. Thus similarly to the marginal local cost, we could start training on the $1$-local cost and as we approach the solution, add in $k$-local cost terms to drive the ansatz towards the global minimum. 

\subsubsection{Local costs for SP ansatz}
Consider our global Hilbert--Schmidt cost function, in the SP ansatz framework:
\begin{equation}
\begin{aligned}
C_{\mbox{\tiny{SP}}}(\rho, \sigma(\vec{\theta},n_A)) 
&= \Tr\!\left[\rho^2\right] + \Tr\!\left[\sigma\left(\vec{\theta},n_A\right)^2\right] - 2 \Tr\!\left[\rho \sigma\left(\vec{\theta},n_A\right)\right] .
\end{aligned}
\end{equation}
Without loss of generality, we can also express $\rho$ via its purification for analysis, i.e.,
\begin{equation}
    \rho = \Tr_{A}\left[ V_{\rho} (\ket{0}\!\bra{0})^{\otimes (n+n_A)} V_{\rho} ^{\dagger} \right] ,
\end{equation}
where $V_{\rho}$ is the purifying unitary associated with the target state $\rho$.
Recall
\begin{equation}
    \Tr[\rho \sigma] = d_A \Tr_{SA}\left[U_{\vec{\theta}}^\dagger \left(\rho \otimes \frac{\id_A}{d_A} \right) U_{\vec{\theta}} (\ket{0}\!\bra{0})^{\otimes (n +n_A)} \right]
\end{equation}
where $d_A$ is the dimension of the environment system. 

% From this perspective, we are introducing the state induced by our target state $\sigma_S$ and the maximally mixed state on register $B$, evolving it (via the purifying unitary associated with the Stinespring representation of $\rho_S$) and performing a global measurement.

In order to construct a local cost, we can replace the global $(\ket{0}\!\bra{0})^{\otimes (n+n_A)}$ projector in each term with a local measurement. The structure of this local measurement may, again, be chosen freely. In the case of learning $\rho$ completely, (i.e., $R = \rank(\rho)$), one simple approach is to replace the global measurement with the average of all measurements local to one system qubit and one environment qubit, i.e.,
\begin{equation}
    H_{\rm L}^{\rm D} = \frac{1}{n} \sum^n_{j=1} \ket{0}\!\bra{0}_{S_j} \otimes \id_{S_{\overline{j}}}  \otimes \ket{0}\!\bra{0}_{A_j} \otimes \id_{A_{\overline{j}}} \, ,
\end{equation}
%\eb{return to figure out best way to write this notation wise}
with $\overline{j}$ denoting all qubits other than qubit $j$. We dub this the \emph{doubly-local SP Hamiltonian}.

When seeking to learn a low-rank approximation (i.e., compression) of $\rho$, we need not make the measurement local on the ancilla. In this case we can use the following \emph{singly-local SP Hamiltonian}, 
\begin{equation}
    H_{\rm L}^{\rm S} = \frac{1}{n} \sum^n_{j=1} \ket{0}\!\bra{0}_{S_j} \otimes \id_{S_{\overline{j}}}  \otimes \ket{0}\!\bra{0}_{A} \, .
\end{equation}

\begin{defn}
For the state purification ansatz, we can define the \emph{Doubly-} and \emph{Singly-local costs},
\begin{equation}
    C_{\textrm{L}}^{\rm X} = c_{\rm L}^{\rm X}(V_{\rho}, \rho) + c_{\rm L}^{\rm X}(U_{\vec{\theta}}, \sigma) - 2 c_{\rm L}^{\rm X}(U_{\vec{\theta}}, \rho)  \, ,
\end{equation}
where
\begin{equation}\label{eq:localoverlap}
\begin{aligned}
    c_{\rm L}^{\rm X}(U_{\vec{\theta}}, \rho) \coloneqq d_A \, \Tr_{SA}\left[U_{\vec{\theta}}^\dagger \left( \rho_S \otimes \id/d_A \right) U_{\vec{\theta}}  H_{\rm L}^{\rm X} \right] \, ,
\end{aligned}
\end{equation}
with the freedom to choose arbitrary $H^{\rm{X}}_{\rm L},$ e.g. $H^{\rm{D}}_{\rm L}$ or $H^{\rm{S}}_{\rm L}$ as defined above. 
\end{defn}

\medskip
The terms $c_{\rm L}^{\rm X}(U_{\vec{\theta}}, \sigma)$ and $c_{\rm L}^{\rm X}(U_{\vec{\theta}}, \rho)$ can be measured using Loschmidt-echo type circuits (as discussed in Section~\ref{sec:algs}) but with the global measurements replaced with local ones. However, as in general one will not have access to $V_{\rho}$, it is generally not possible to measure $c_{\rm L}^{\rm X}(V_{\rho},\rho)$. Nonetheless, as this term remains constant throughout and does not contribute to the gradient (discussed further in App. \ref{app:grads}) it can be neglected without effecting the optimization procedure. 

The methods used to prove faithfulness of $C^{\rm CCPS}_{\rm L}$ for pure states do not carry over for the singly- and doubly-local SP costs, due to the factor of $d_A$. Thus, faithfulness for pure states for $C_{\rm L}^{\rm D}$ and $C_{\rm L}^{\rm S}$ remains an open question. However, we can prove faithfulness for tensor-product states in the SP picture.

% For a successful local cost function, we desire that it remains efficiently computable and that it remains faithful to its global counterpart (i.e., the local cost function vanishes if and only if the global cost function vanishes), whilst avoiding barren plateaus that plague its global counterpart. 

% A general proof for complete faithfulness cannot be found for either case. However, we find that each proposed function lends itself to use cases for different types of quantum states.
%\paragraph*{Analysis.}
%Here we prove that these local costs are faithful, or approximately faithful, in the following special cases. In the SP ansatz picture, we find that we can construct a local cost function that is provably faithful for tensor-product states, whilst for the CCPS ansatz, we can prove faithfulness for pure states.
\begin{propos}
$C^{\rm{D}}_{\rm L}$ is faithful for tensor-product states.
\end{propos}
\begin{proof}
In the case of tensor-product mixed states, we can write $\rho =  \bigotimes_{j=1}^{n} \rho_{S_j}$, $U_{\vec{\theta}} = \bigotimes_{j=1}^{n} U_{S_jA_j}^{\vec{\theta}}$. Thus the overlap term can be written as 
\begin{equation}
\begin{aligned}
    &c_{\rm L}^{\rm D}(U_{\vec{\theta}}, \rho) \\ &= d_A \, \Tr_{SA}\left[ \left(\bigotimes_{k=1}^{n} U_{S_kA_k}^{\vec{\theta} \dag} \left( \bigotimes_{k=1}^{n} \rho_{S_k} \otimes \id_A/d_A \right) \bigotimes_{k=1}^{n} U_{S_kA_k}^{\vec{\theta}} \right)  H_{\rm L}^{\rm D} \right] \\
    &= \frac{d_A}{n} \,  \sum_{j=1}^n \Tr_{SA}\left[ \left(\bigotimes_{k=1}^{n} U_{S_kA_k}^{\vec{\theta}\dag} \left( \bigotimes_{k=1}^{n} \rho_{S_k} \otimes \id_A/d_A \right) \bigotimes_{k=1}^{n} U_{S_kA_k}^{\vec{\theta}} \right) \left( \ket{0}\!\bra{0}_{S_j} \otimes \id_{S_{\overline{j}}}  \otimes \ket{0}\!\bra{0}_{A_j} \otimes \id_{A_{\overline{j}}} \right)\right] \\
    &=  \, \frac{1}{n} \sum_{j=1}^n \Tr_{S_jA_j}\left[ \left(U_{S_jA_j}^{\vec{\theta} \dag} \left(  \rho_{S_j} \otimes \id_{A_j} \right)  U_{S_jA_j}^{\vec{\theta}} \right) \left( \ket{0}\!\bra{0}_{S_jA_j}  \right)\right] \Tr_{S_{\overline{j}}A_{\overline{j}}} \left[ \left(  \bigotimes_{k \neq j} \rho_{S_{k}} \otimes \id_{A_{\overline{j}}} \right) \left( \id_{S_{\overline{j}} A_{\overline{j}}} \right) \right]  \\
    &= \frac{d_A}{2n}  \sum_{j=1}^n \Tr_{S_jA_j}\left[U_{S_jA_j}^{\vec{\theta} \dagger}  \left( \rho_{S_j}  \otimes \id_{A_{j}} \right) U_{S_jA_j}^{\vec{\theta}}  \ket{0}\!\bra{0}_{S_j A_j} \right] \\
    &= \frac{d_A}{2n} \sum_{j=1}^n \Tr[\rho_{S_j} \sigma_{S_j} ] \, . \\ 
\end{aligned}
\end{equation}
Thus, in this case, we have that
\begin{equation}
\begin{aligned}
        C_{L}^{\rm D} &= \frac{d_A}{2n} \left( \sum_{j=1}^n \Tr[\rho_j \rho_j ]  + \sum_{j=1}^n \Tr[\sigma_j \sigma_j ] -2 \sum_{j=1}^n \Tr[\rho_j \sigma_j ]  \right) \\
        &\propto \sum_{j=1}^n \left\| \rho_j - \sigma_j \right\|_2^2  ,
\end{aligned}
\end{equation}
which vanishes if and only if $\rho_j = \sigma_j$ for all $j$, i.e., assuming $\rho$ and $\sigma$ are tensor-product states, iff $\rho = \sigma$.
\end{proof}
% We note that in the case of tensor-product states it does not make sense to use the singly local cost since the rank of an $n$-qubit tensor-product state is $2^n$, i.e., you need $k = n$ ancillas. Thus the singly local cost would exhibit exponential convergence in this case. 

\medskip

In this subsection, we have shown how by introducing local measurements directly into our cost function, we are able to construct local cost functions. For the case of the CCPS ansatz this construction is provably faithful for pure states and approximately faithful for high purity states. For the case of the SP ansatz the construction is faithful for product states. 
% We also provide bounds that guarantee both the $1$- and $k$-local constructions for the CCPS ansatz remain close to faithful in the scenario that the state being learned is close to pure.
Thus we expect these costs to prove useful for learning mixed states with relatively low impurities and low entanglement respectively. 
However, for target states that are highly entangled and/or mixed, we are unable to provide guarantees on the behaviour of the cost function. In these cases, the function no longer resembles a distance measure, as positivity cannot be guaranteed. Thus the construction of a truly faithful, yet entirely local equivalent to the Hilbert--Schmidt distance remains an open question.

\medskip

However, in practise one may create a cost that is both faithful and exhibits non vanishing gradients by taking a linear combination of the absolute value of the local cost and the global cost, i.e., by training on a cost of the form $\alpha  |C_{L}| + (1-\alpha)  C_{\rm Global} $ for some choice of the local cost $C_{L}$ and $0 \leq \alpha \leq 1$. By tuning $\alpha$ such that it is (close to) one at the start of the optimization and (close to) zero at later stages of the optimization, it should be possible to steer towards the global minimum.

\section{\label{app:ansatze} Circuit Ans{\"a}tze Particulars}

\subsection{Summarizing the circuits in our algorithm\label{app:circuit-ansatz-for-cost}}
We summarize the circuits used to evaluate our cost function to clarify in detail how one can implement our algorithm. In particular, we provide three circuits which sample the three terms in our cost function,
\begin{equation}
    C(\vec{\alpha}, R) \equiv \Tr[\rho^2] + \Tr[\sigma(\vec{\alpha}, R)^2] - 2 \Tr[\rho \sigma(\vec{\alpha}, R)]
\end{equation}
for both the state purification (SP) ansatz and the convex combination of pure states (CCPS) ansatz. 

\subsubsection{SP Ansatz Circuits}
\begin{figure}[ht]
    \centering
    \includegraphics{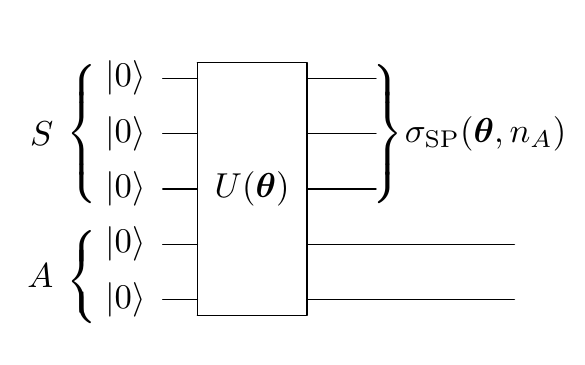}
    \includegraphics{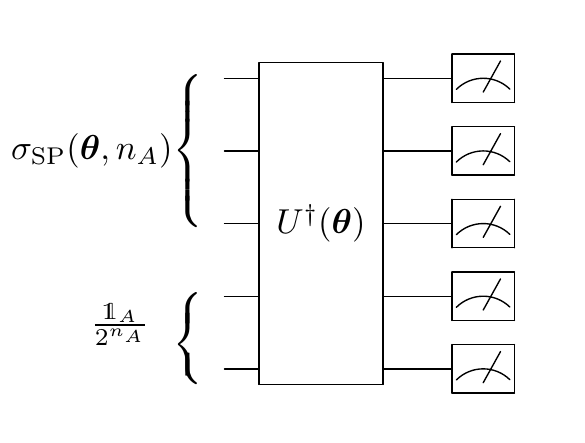}
    \includegraphics{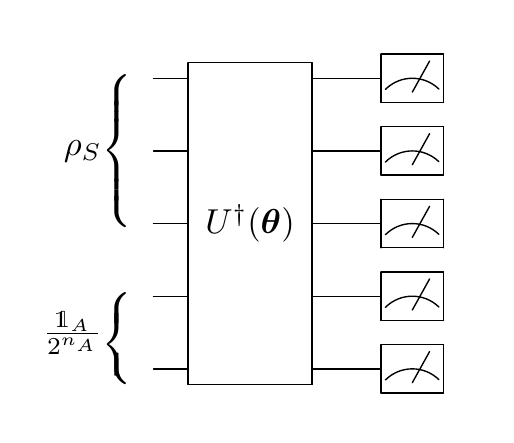}
    \caption{We summarize the circuits used in evaluating the cost function for the SP ansatz. (Left) An example circuit preparing a 3 qubit mixed state with 2 ancilla. (Middle) A Loschmidt-echo style circuit to measure the purity of $\sigma_{\text{SP}}(\vec{\theta}, n_A)$, i.e. $\Tr[\sigma_{\text{SP}}(\vec{\theta}, n_A)^2]$. (Right) The same Loschmidt-echo circuit is also capable of measuring the cross term, $\Tr[\rho_S \cdot \sigma_{\text{SP}}(\vec{\theta}, n_A)]$, when the input state is $\rho_S$.}
    \label{fig:sp-ansatz-circuit}
\end{figure}

The SP ansatz generates an $n$ qubit mixed state by applying a unitary on $n + n_A$ qubits and tracing out the $n_A$ ancilla,
\begin{equation}
    \sigma_{\mbox{\tiny SP}}(\vec{\theta},n_A) \coloneqq \Tr_A[ U_{\vec{\theta}} (\ketbra{0}{0})^{\otimes(n +n_A)} U_{\vec{\theta}}^\dagger ].
\end{equation}
The translation of this procedure into a quantum circuit is straightforward and is shown for an $n = 3$, $n_A = 2$ example in the left-most circuit of Fig.~\ref{fig:sp-ansatz-circuit}. Given $U(\vec{\theta})$, we can measure an estimate of our cost function using the remaining two Loschmidt echo style circuits shown in Fig.~\ref{fig:sp-ansatz-circuit}. The middle circuit evaluate the purity term $\Tr[\sigma_{\text{SP}}(\vec{\theta}, n_A)^2]$ and the right-most circuit evaluates the cross term $\Tr[\rho_S \cdot \sigma_{\text{SP}}(\vec{\theta}, n_A)]$ which we showed in the main text Eq.~\eqref{eq:GlobalCostTerm}. 
Note that we omitted providing a circuit to measure the purity of $\rho_S$ since this is a static term during the optimization anyway. If desired, one could simply use middle circuit and replace $\sigma_{\text{SP}}(\vec{\theta}, n_A)$ with $\rho_S$. 

In the middle purity evaluation circuit, we avoid writing out the circuit which prepares $\sigma_{\text{SP}}(\vec{\theta}, n_A)$ a second time. However, it is worth mentioning that generating $\sigma_{\text{SP}}(\vec{\theta}, n_A)$ itself takes $n_A$ ancilla, so the middle circuit takes a minimum of $n + 2 n_A$ qubits without resetting the ancilla $A$ to use twice. As for preparing the totally mixed state on the ancilla system, this can be done in two ways. In the first, we prepare one of $2^{n_A}$ basis states with a uniformly random probability for every shot that $C$ is evaluated. Alternatively, one may choose to actually prepare the uniformly mixed state which can be done by preparing any completely entangled bi-partite state (i.e. a GHZ state) on $2 n_A$ qubits. In the latter case, the middle circuit therefore uses $n + 3 n_A$ qubits. 

We also draw attention to the fact that we use $U(\vec{\theta})$ to prepare $\sigma_{\text{SP}}(\vec{\theta}, n_A)$ in the first circuit but $U^{\dagger}(\vec{\theta})$ in the two cost function evaluation circuits. The use of $U^{\dagger}(\vec{\theta})$ is fleshed out mathematically in Eq.~\eqref{eq:GlobalCostTerm} of the main text. Intuitively, it's as if $U^{\dagger}$ is undoing the preparation circuit $U$--hence the name Loschmidt-echo like circuit. In fact, the final cost term is ultimately evaluated by counting the number of $0$'s obtained at the final registers which corroborates this intuition.

\begin{figure}
    \centering
    \includegraphics{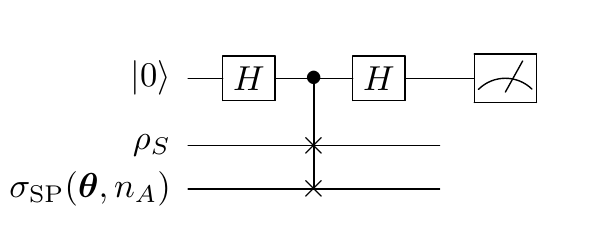}
    \hspace{0.5cm}
    \includegraphics{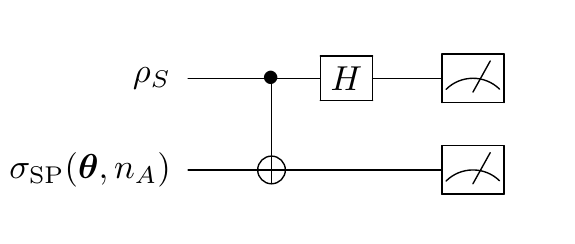}
    \caption{Two circuits to evaluate $\Tr[\rho_S \cdot \sigma_{\text{SP}}(\vec{\theta}, n_A)]$ or $\Tr[ \sigma_{\text{SP}}(\vec{\theta}, n_A)^2]$ by choosing $\rho_S = \sigma_{\text{SP}}(\vec{\theta}, n_A)$. (Left) A version of the SWAP test as first discussed in Ref.~\cite{barenco1997stabilization}. For completeness, we demonstrate that this circuit works as intended from Eq.~\eqref{eq:swap-identity} to Eq.~\eqref{eq:final-swap-line} since Ref.~\cite{barenco1997stabilization} is not explicit for our context. This requires an ancilla and a controlled swap (aka Fredkin) gate. As the proof will show, this circuit can be straightforwardly generalized to $n$ qubit states by performing an $2n$ qubit SWAP rather than a two qubit SWAP. (Right) A destructive variant of the SWAP test discovered in the context of optical systems and the Hong-Ou-Mandel effect in Ref.~\cite{garcia2013swap}. This removes the need for ancilla and a complicated Fredkin gate but destroys the prepared states in the process in what is known as a \emph{Bell measurement.} Much of of Ref.~\cite{garcia2013swap} is dedicated to showing the right circuit is equivalent to the left circuit, so we do not re-derive the entire result here. Instead, we comment on how to connect the measurements of the circuit to the desired quantity in Eq.~\eqref{eq:des-swap-11}. We then discuss how to generalize this procedure to $n$ qubit mixed states in Eq.~\eqref{eq:des-swap-gen}.}
    \label{fig:swap-and-des-swap}
\end{figure}

As discussed in the main text, the Loschmidt echo circuits we cooked up in Fig.~\ref{fig:sp-ansatz-circuit} are NISQ friendly but incur a poor shot scaling as $n_A$ grows. If in practice, $n_A$ is on the order of $n$ and both are large, then it's best to instead use SWAP test circuits~\cite{barenco1997stabilization} or their destructive variant~\cite{garcia2013swap}. The SWAP and destructive SWAP circuits to evaluate $\Tr[\rho_S \cdot \sigma_{\text{SP}}(\vec{\theta}, n_A)]$ are shown in Fig.~\ref{fig:swap-and-des-swap}. Note that these also can be used for $\Tr[ \sigma_{\text{SP}}(\vec{\theta}, n_A)^2]$ by replacing $\rho_S$ with $\sigma_{\text{SP}}(\vec{\theta}, n_A)$.  While the references ultimately contain sufficient information to deduce that these circuits work as claimed, it is not obvious at a glance. For posterity and completeness, we provide a tailored derivation of the claim in the present context. 

We begin by showing that the left SWAP test circuit is sufficient to measure $\Tr[\rho \sigma]$ (we've dropped subscripts for simplicity). The first thing we need to know for this derivation is that
\begin{equation}
    \label{eq:swap-identity}
    \Tr[(\rho \otimes \sigma) \text{SWAP}] = \Tr[\rho \sigma].
\end{equation}
As we'll see, our goal will be to find an observable on the ancilla qubit alone whose expectation value gives us the left-hand side of Eq.~\eqref{eq:swap-identity}. From this identity, the desired outcome follows, and this is why we call it a SWAP test method.  

Consider the state right after the controlled SWAP (aka Fredkin) gate which we shall call $\omega$ for concreteness. By simple Dirac notation manipulations, we arrive at, 
\begin{align}
    \omega &\equiv \left( \ketbra{0}{0} \otimes \mathds{1} + \ketbra{1}{1} \otimes \text{SWAP} \right) \left( \ketbra{+}{+} \otimes \rho \otimes \sigma \right)  \left( \ketbra{0}{0} \otimes \mathds{1} + \ketbra{1}{1} \otimes \text{SWAP} \right) \\
    %------------------------
    &= \frac{1}{2} \left\{ \ketbra{0}{0} \otimes \rho \otimes \sigma + \ketbra{0}{1} \otimes (\rho \otimes \sigma \ \text{SWAP}) + \ketbra{1}{0} \otimes (\text{SWAP} \ \rho \otimes \sigma) + \ketbra{1}{1} \otimes \sigma \otimes \rho \right\}.
\end{align}

The second line can be thought of as block matrix,
\begin{equation}
    \omega = 
       \frac{1}{2}  \begin{pmatrix}
         \rho \otimes \sigma & (\rho \otimes \sigma) \text{SWAP} \\
         %--------
        \text{SWAP}(\rho \otimes \sigma) & \sigma \otimes \rho 
        \end{pmatrix}.
\end{equation}

Right before measurement we are then left with the state $H \omega H$. Since we then only make a measurement on the ancilla qubit, we will only need the diagonal entries of this block matrix to choose the right observable to measure. In particular, we find
\begin{equation}
    \label{eq:state-before-measurement}
    H\omega H = 
       \frac{1}{4}  \begin{pmatrix}
         \rho \otimes \sigma + (\rho \otimes \sigma) \text{SWAP} + \text{SWAP}(\rho \otimes \sigma) + \sigma \otimes \rho  & \cdots \\
         %--------
       \cdots & \rho \otimes \sigma - (\rho \otimes \sigma) \text{SWAP} - \text{SWAP}(\rho \otimes \sigma) + \sigma \otimes \rho
        \end{pmatrix}.
\end{equation}

Staring at Eq.~\eqref{eq:state-before-measurement} in light of Eq.~\eqref{eq:swap-identity}, we see that the right observable is $\sigma_z$. Indeed,
\begin{equation}
    \label{eq:final-swap-line}
    \Tr[\sigma_z H \omega H] = \Tr[\rho \otimes \sigma \text{SWAP}] = \Tr[\rho \sigma].
\end{equation}
Note that the above derivation really only relied on the fact that $\text{SWAP}(\rho \otimes \sigma)\text{SWAP} = \sigma \otimes \rho$. Hence, this circuit works to evaluate $\Tr[\rho \sigma]$ when $\rho$ and $\sigma$ are composed of an arbitrary number of qubits despite the way our diagram implies they are single qubit states. We simply chose to write it this way since it's easier to interpret and reason in this way afterwards then try to worry about multiple qubits form the start. 

Next, we discuss the right destructive SWAP variation. Let's again begin by considering that $\rho$ and $\sigma$ are single qubit states. It turns out, the relevant measurement is the projection into the $\ketbra{11}{11}$ subspace. In particular, let
\begin{equation}
    P_{11} = \Tr[\ketbra{11}{11} \cdot  \text{H} \cdot \text{CNOT} \cdot \rho \otimes \sigma \cdot \text{CNOT} \cdot \text{H}]
\end{equation}
be the probability that the final result from the destructive SWAP circuit is $\ket{11}$. Then,
\begin{equation}
    \label{eq:des-swap-11}
    \Tr[\rho \sigma] = 2(1 - P_{11}) - 1
\end{equation}
gives us the desired quantity. 

An $n$ qubit generalization is not quite as straightforward as the SWAP test generalization. In particular, we replace the single CNOT and Hadamard with a transversal application of CNOTs and Hadamards--i.e. Fig.11 in Ref.~\cite{garcia2013swap}. Further, we don't just project onto $\ketbra{1\ldots 1}{1\ldots 1}$. In fact, the augmented procedure is actually easier to state at the level of individual measurements rather than projectors. Supposing $\rho$ and $\sigma$ are $n$ qubit states, then we can label the measurement outcomes for the first $n$ $\rho$ registers with a bitstring $\vec{a}$ and of the $\sigma$ registers $\vec{b}$. We say the test \enquote{fails} when the bit-wise and of the two bit-strings has odd parity, i.e. $|\vec{a} \wedge \vec{b}|$ is odd. Identifying $P_f$ as this failure probability obtained by repeating this procedure ad infinitum, we again find
\begin{equation}
    \label{eq:des-swap-gen}
    \Tr[\rho \sigma] = 2 (1 - P_f) - 1.
\end{equation}
As a sanity check, we can confirm that $P_f$ corresponds to $P_{11}$ for the single qubit case. Here, the bit-strings each have one element, so the condition reduces to $|a \wedge b| = 1$ which occurs if and only if $a = b = 1$. An
alternative derivation for the n-qubit destructive SWAP test viewed as a Bell basis measurement can be found
in \cite{agarwal2021estimating} (page 8) whose intuition can be understood from \cite{brun2004measuring} (page 6). 

\begin{comment}
In this case, the circuit is precise, and we can label the measurement outcomes $A_i$ and $B_i$ for shot $i$. Much of Ref.~\cite{garcia2013swap} is concerned with building this circuit as an equivalent circuit to the SWAP test better suited for optical systems, so we'll immediately write down the right way to interpret $A_i$ and $B_i$ to compute $\Tr[\rho \sigma]$. The correct choice turns out to be to take the and of the arguments
\begin{equation}
    M_i = \frac{A_i \wedge B_i},
\end{equation}
and hence that
\begin{equation}
    \overline{M} \equiv \frac{1}{2 N} \sum_{i=1}^N M_i \approx \Tr[\rho \sigma].
\end{equation}
In particular, $\overline{M}$ is an unbiased estimator of $\Tr[\rho \sigma]$, and the convergence is the usual statistical $\frac{1}{\sqrt{N}}$ one. 
\end{comment}

\subsubsection{CCPS Ansatz Circuits}
\begin{figure}[ht]
    \centering
    \includegraphics{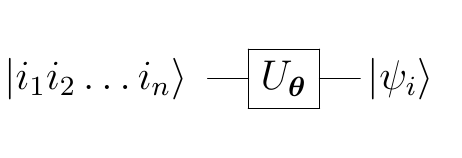}
    \hfil
    \includegraphics{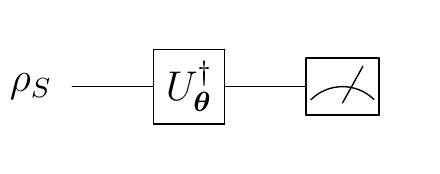}
    \caption{We summarize the circuits used to evaluate the cost function for the CCPS ansatz. (Left) A generic way to prepare a randomly sampled eigenvector of the CCPS state. Here, the state $\ket{i_1 i_2\ldots i_n}$ is a short-hand for an $n$ qubit computational basis state, so each $i_j$ is either $0$ or $1$. Given the randomly sampled bitstring, we can prepare the desired state with single qubit $X$ gates, and then we apply the same unitary $U_{\vec{\theta}}$ regardless of the randomly sampled input. (Right) Here, we provide a circuit to sample the distribution $q_{\vec{\theta}}(i)$ defined in \eqref{eq:app-compute-cross-term-ccps}. We simply prepare $\rho_S$, apply $U^{\dagger}_{\vec{\theta}}$, and measure each qubit separately in the computational basis. By \eqref{eq:app-compute-cross-term-ccps} and the fact that $p_{\vec{\phi}}(i)$ is stored classically, we can therefore compute the desired cross term using a \emph{classical SWAP test} as described in the main text around \eqref{eq:classical-swap-q-dist}.}
    \label{fig:ccps-ansatz-circuits}
\end{figure}

A mixed state is often thought as a probabilistic mixture of pure states. The CCPS ansatz is a direct implementation of this idea, 
\begin{equation}
    \sigma_{\mbox{\tiny CCPS}}( \vec{\alpha} , R) \coloneqq \sum_{i=0}^{R-1} p_{\vec{\phi}}(i) U_{\vec{\theta}} \ketbra{i}{i} U_{\vec{\theta}}^\dagger.
\end{equation}
Namely, experimental observables are obtained by averaging over many experiments in which the input state is $U_{\vec{\theta}}\ket{i}$ chosen with probability $p_{\vec{\phi}}(i)$~\footnote{When $R$ is small, we can think of $p_{\vec{\phi}}(i)$ as a probability vector with entries $p_i$ stored in classical memory. The funny $\vec{\phi}$ notation conveys the fact that $R$ can be exponentially large in general, and more involved means of storing and sampling this distribution must be used in this case.}. The ensemble of circuits can thus be represented by $U_{\vec{\theta}}$ acting on an input basis vector $\ket{i}$ as in Fig.~\ref{fig:ccps-ansatz-circuits}. One can choose any set of basis vector $\{\ket{i}\}_{i = 0}^{R-1}$, but for our work, we choose the set of computational basis states~\footnote{This is not just a matter of simplicity. The value in this choice is that we can reach any computational basis state with a depth 1 circuit--namely one applying an $X$ gate on those qubits initialized in $\ket{1}$ and identity otherwise. Thus, to reach a \enquote{complex state,} we must use a deep circuit ansatz $U_{\vec{\theta}}$. A different choice for $\ket{i}$ could lead to hiding the necessary complexity of the circuit ansatz in practice where $\ket{0}^{\otimes n}$ is the initial state.}. Given a the preparation unitary $U_{\vec{\theta}}$, we can use the right circuit to sample the distribution
\begin{equation}
    \label{eq:app-q-dist}
q_{\vec{\theta}}(i)\coloneqq\langle i|U^\dagger_{\vec{\theta}}\rho U_{\vec{\theta}}|i\rangle.
\end{equation}

As discussed in the text surrounding Eq.~\eqref{eq:classical-swap-q-dist}, this distribution--alongside the classically stored $p_{\vec{\phi}}(i)$--allow us to compute the cross term
\begin{equation}
    \label{eq:app-compute-cross-term-ccps}
\operatorname{Tr}[\rho\sigma]=\sum_{i}p_{\vec{\phi}}(i)q_{\vec{\theta}}(i)
\end{equation}
by using the \emph{classical SWAP test}. 

As before, we omit a procedure to estimate $\Tr[\rho_S^2]$ which doesn't affect the optimization. This time, however, we've also omitted a circuit to estimate $\Tr[\sigma_{\text{CCPS}}^2]$ since this is done entirely classically again using the classical SWAP test. Other than that, the only subtlety in our circuits is simply that we've not explicitly written out different lines for different qubits. The reason is that for this ansatz, there are no ancilla qubits necessary, so it's understood that each line is for $n$ qubits. As mentioned in the caption, the measurement is a single qubit measurement on each qubit.  

\subsection{The parameterized circuits used for $U(\vec{\theta})$}

So far, we have described all the circuits where $U(\vec{\theta})$ was understood to be some parameterized quantum circuit (PQC). Here, we define and justify the choices we make for the different classes of states we consider. We begin with an abstract description of a hardware efficient tiling. We then discuss how this tiling is applied for the parameterized circuits in the SP ansatz and the CCPS ansatz. Then we move into specifics for Bures random states, XY thermal states, and hardware noise induced states. As discussed in the main text, the CCPS ansatz automatically provides a description of the principal components of the target state $\rho$ in the computational basis. This is not generally true for the SP ansatz, and we conclude with a discussion of how to generate an SP ansatz that does allow for extraction of the principal components but note that it is generally not practical.

\subsubsection{Hardware Efficient Tiling and its Use in the CCPS and SP Ans{\"a}tze}

\begin{figure}
    \centering
    \includegraphics[width=0.4\textwidth]{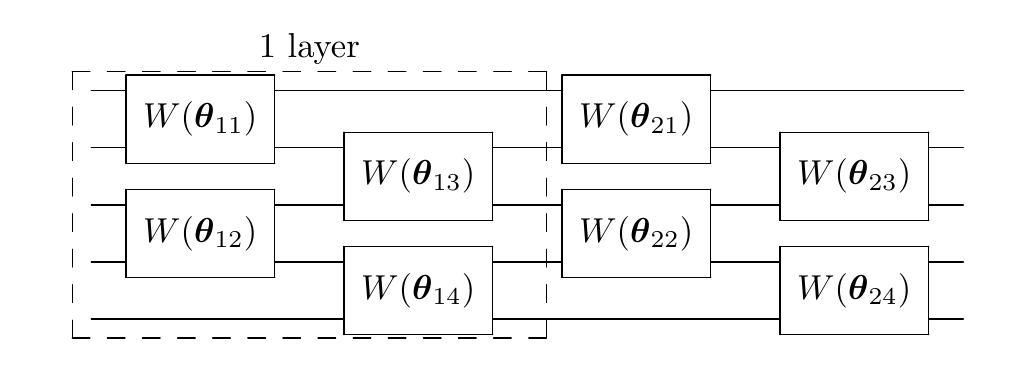}
    \caption{A hardware efficient tiling of a two-qubit gate $W(\boldsymbol{\theta})$ across a linearly connected five-qubit circuit. The notation $\vec{\theta}_{lk}$ is meant to convey the parameters of the $k^{\text{th}}$ gate in the $l^{\text{th}}$ layer. Each layer consists of $n - 1$ applications of $W$ for $n$ qubits.}
    \label{fig:hef-tiling}
\end{figure}

Let $W(\boldsymbol{\theta})$ be an unspecified two-qubit gate parameterized by a vector of angles, $\vec{\theta}$. A hardware efficient tiling of $W$ is given in Fig.~\ref{fig:hef-tiling}. A single layer consists of what is shown in the \enquote{dotted rectangle} on the left. The name hardware-efficient comes from the fact that for a linearly connected device, only neighboring qubits need to be coupled, and so no swaps are needed. Furthermore, each layer consists of a depth-$2D_W$ circuit only, where $D_W$ is the depth of $W$ itself.

For the CCPS ansatz, we simply apply a hardware efficient tiling on the $n$ qubits. That is, 
\begin{equation}
    U^{\text{CCPS}}_{\vec{\theta}} = \prod_{l=1}^{L} \prod_{k=1}^{n-1} W(\vec{\theta}_{lk}),
\end{equation}
where the product over $k$ represents the application of each gate in layer $l$ and the product over $l$ is for $L$ layers, which is chosen depending on the class of state. For the SP ansatz, we simply apply the same tiling but to the $n +n_A$ system plus ancilla qubits,
\begin{equation}
    U^{\text{SP}}_{\vec{\theta}} = \prod_{l=1}^{L} \prod_{k=1}^{n +n_A - 1} W(\vec{\theta}_{lk}).
\end{equation}

\subsubsection{Bures Random State Ansatz}

There are no non-trivial symmetry operators $S$ for which $[\rho, S] = 0$  for all $\rho$ drawn from the Bures measure. Without any inherent symmetry structure, we choose the most generic $W$: an arbitrary two-qubit gate; i.e., $W$ can express any rotation in \textbf{U}(4). By the KAK decomposition~\cite{tucci2005introduction, shende2004minimal, vatan2004optimal}, we can decompose $W$ using three CNOTs and 15 elementary single qubit rotations (see Fig.~7 in \cite{vatan2004optimal} for example). Next, note that one way to generate $\rho$ from the Bures distribution is through applying a Haar random unitary on $2n$ qubits, and then putting this state in coherent superposition with the same state changed by a local transformation on the $n$ system qubits and then tracing out the ancilla~\cite{zyczkowski2011generating} (see Eq.~\eqref{eq:bures-intuitive-generation}). This suggests that no low-depth circuit exists to faithfully generate $\rho$, so we choose $L = n$. This gives a total of $15 n (n-1)$ trainable parameters. As discussed in Cerezo \textit{et al}.~\cite{cerezo2020cost}, we expect a linear depth alternating ansatz with this many parameters to exhibit a barren plateau and thus to not be scalable. However, in fact, we know that trying to learn a Haar random unitary induces a barren plateau regardless of the choice of ansatz when no other information is known  \cite{holmes2021barren}. Hence, this class of states is likely not scalable beyond the small sizes testable in the NISQ era anyway. To that end, it serves as a proof of principle that even the most difficult states (for tractable sizes) can be learned by our method. 

\subsubsection{XY Model Ansatz}

The XY model does exhibit symmetries. In particular, the spin model is particle conserving, and as a chain in 2D, it is invariant under any global rotation. As discussed in Ref.~\cite{verstraete2009quantum}, the structure of the XY model can be used to design a generically good ansatz using $\mathcal{O}(n^2)$ gates with circuit depth $\mathcal{O}(n \log n)$. However, as discussed in Appendix~3c of~\cite{gibbs2022dynamical}, one can instead use $n$~alternating layers of Givens rotations which still obeys the symmetries but results in a lower $2n$ depth with a gate count of $n^2 - n$. The Givens rotation is a single parameter gate,
\begin{equation}
    G(\theta) = 
    \begin{pmatrix}
    1 & 0 & 0 & 0 \\
    %------
    0 & \cos(\theta / 2) & -\sin(\theta / 2) & 0 \\
    %-----
    0 & \sin(\theta / 2) & \cos(\theta / 2) & 0 \\
    %------
    0 & 0 & 0 & 1
    \end{pmatrix},
\end{equation}
which rotates in the subspace where $\ket{00}$ and $\ket{11}$ are fixed.  

In our work, however, we want to associate the state $\ket{0}^{\otimes n}$ in the CCPS ansatz to the first principal component of $\rho$. Hence, $\ket{0}^{\otimes n}$ should not in general be preserved. As a fix, we can consider other Givens rotations that differ from $G(\theta)$ by an arbitrary permutation of basis elements. For example,
\begin{equation}
    G'(\theta) = 
    \begin{pmatrix}
    0 & 1 & 0 & 0 \\
    %------
    \cos(\theta / 2) & 0 & 0 & \sin(\theta / 2) \\
    %-----
    -\sin(\theta / 2) & 0 & 0 & \cos(\theta / 2) \\
    %------
    0 & 0 & 1 & 0
    \end{pmatrix},
\end{equation}
is also a valid Givens rotation where the set of fixed states is permuted. To avoid attaching  to a fixed choice, we define a Givens gate as
\begin{equation}
    W(\theta_1, \theta_2, \theta_3) = G(\theta_1) G'(\theta_2) G(\theta_3),
\end{equation}
which is analogous to an Euler decomposition of a rotation in 3D as $R_X(\theta_1) R_Y(\theta_2) R_X(\theta_3)$. Using a tiling of this $W$, we find empirically that a depth $L = \log{n}$ is sufficient, and so use a total of $3 (n - 1) \log(n)$ single-parameter gates with depth $3 \log(n)$, which represents yet another improvement over  \cite{gibbs2021long}.

\subsubsection{Ansatz for Hardware Implementation}

To learn the purification (i.e., the SP ansatz) of the single qubit state $\rho_+$, we used an arbitrary two-qubit gate built up with the previously mentioned KAK decomposition. For the pure state approximation, we simply forgo using an ancilla. In other words, for the $R = 1$ approximation of $\rho_+$, we used a single qubit circuit ansatz,
\begin{equation}
    W^{R=1}_{\rho_+}(\theta_1, \theta_2, \theta_3) = R_Z(\theta_1) R_Y(\theta_2) R_Z(\theta_3),
\end{equation}
on the system qubit.

To learn the principal components and values of $\rho_{\Phi^+}$ with the CCPS ansatz, we use an arbitrary two-qubit circuit $U_2$. The principal components are then obtained as $\{U_2 \ket{00}, U_2\ket{01}, U_2 \ket{10}, U_2\ket{11}\}$ whereas the principal values are stored as a vector we train over, $\{p_{00}, p_{01}, p_{10}, 1 - (p_{00} + p_{01}, + p_{10})\}$.  

\subsection{SP PCA Ansatz}

\begin{figure}
    \centering
    \includegraphics[width=0.33\textwidth]{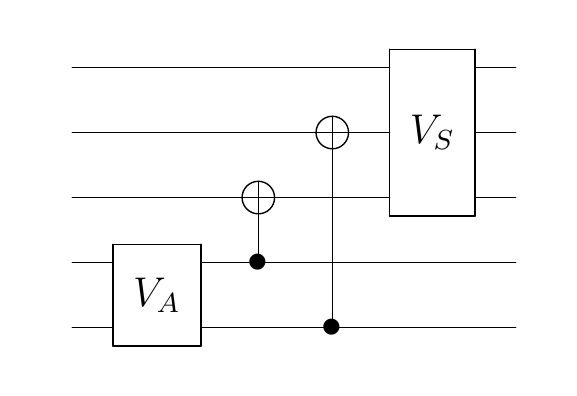}
    \caption{\textbf{SP PCA Ansatz.} An example SP PCA ansatz acting on $3 + 2$ qubits. We first apply a gate $V_A$ on the $n_A = 2$ ancilla, connect the two sets of qubits with a CNOT cascade, and then apply a gate $V_S$ on the system qubits.}
    \label{fig:SP_PCA_ansatz}
\end{figure}

We note that it is possible to construct an SP ansatz that allows for the principal components of $\rho$ to be extracted using measurements in the computational basis. For clear reasons, we call this the PCA ansatz, and we show a generic example of such an ansatz in Fig.~\ref{fig:SP_PCA_ansatz}. The idea bears some similarities to the ansatz  recently presented in \cite{liu2022mitigating}. 

To see why this is a PCA ansatz, we first consider the action of a generic version of the circuit in Fig.~\ref{fig:SP_PCA_ansatz}, proceeding step by step. The state before the CNOT cascade is as follows:
\begin{subequations}
    \begin{align}
        \ket{\phi_1} &\equiv \left(\id_S \otimes V_A \right) \ket{0}_S^{\otimes n_S} \ket{0}_A^{\otimes n_A} \\
        %-------------
        &= \ket{0}_S^{\otimes n_S} \sum_{i_1, i_2, \ldots, i_{n_A} = 0}^{1} c_{i_1, i_2, \ldots, i_{n_A}} \ket{i_1, i_2, \ldots, i_{n_A}}_A.
    \end{align}
\end{subequations}
After applying the CNOTs, we get
\begin{subequations}
    \begin{align}
        \ket{\phi_2} &\equiv \prod_{i=0}^{n_A-1}\text{CNOT}_{n_S + i, n_S - (i + 1)} \ket{\phi_1} \\
        %--------
        &= \ket{0}_S^{\otimes n_S - n_A} \sum_{i_1, i_2, \ldots, i_{n_A} = 0}^{1} c_{i_1, i_2, \ldots, i_{n_A}}\ket{i_1, i_2, \ldots, i_{n_A}}_S \ket{i_1, i_2, \ldots, i_{n_A}}_A \\
        %---------
        &= \sum_{i_1, i_2, \ldots, i_{n_A} = 0}^{1} c_{i_1, i_2, \ldots, i_{n_A}}\ket{0, \ldots, 0, i_1, i_2, \ldots, i_{n_A}}_S \ket{i_1, i_2, \ldots, i_{n_A}}_A \\
        %------------
        &= \sum_{k=0}^{2^{n_A} - 1} c_k \ket{\Tilde{k}}_S \ket{k}_A,
    \end{align}
\end{subequations}
where we introduced an arbitrary simpler indexing at the end. The choice of \emph{tilde} on the $S$ basis ket is to emphasize that there is a leading $n_S - n_A$ qubits in the all-zeros state. 

By applying a system local unitary $V_S$ and then tracing out the ancilla, we get
\begin{subequations}
    \begin{align}
        \sigma_S &\equiv \Tr_A \left[\sum_{k, j = 0}^{2^{n_A} - 1} c_k c^*_j (V_S \otimes \id_A) \ket{\Tilde{k}}_S\ket{k}_A \bra{\Tilde{j}}_S \bra{j}_A (V_S^{\dagger} \otimes \id_A) \right] \\
        %---------
        &= \sum_{k=0}^{2^{n_A} - 1} p_k V_S \ketbra{\Tilde{k}}{\Tilde{k}} V^{\dagger}_S,
    \end{align}
\end{subequations}
where in the last step we simply identified $|c_k|^2 = p_k$ as probabilities. For arbitrary $V_A$ and $V_S$, the state $\sigma_S$ is an arbitrary rank-$2^{n_A}$ density matrix, and hence, this is clearly a legitimate ansatz for a density matrix on $S$.

At the same time, this ansatz prepares the state
\begin{equation}
    \sigma_A \equiv \Tr_S[\ketbra{\phi_2}{\phi_2}] = \sum_{k=1}^{2^{n_A}} p_k \ketbra{k}{k}_A
\end{equation}
on the ancilla system. By inspection of $\sigma_A$ and $\ket{\phi_2}$ we see that a measurement of the ancilla in the computational basis yields the state
\begin{equation}
    \ket{u_k} \equiv V_S \ket{\Tilde{k}}_S \ket{k}_A
\end{equation}
with probability $p_k$. Hence, a measurement of the ancilla prepares the eigenstates of $\sigma_S$, $V_S \ket{\Tilde{k}}$ with corresponding probability $p_k$. Provided $\sigma_S \approx \rho$ as a result of a successful training, this gives us a way to probabilistically prepare the first $2^{n_A}$ principal components of $\rho$. In particular, by repeatedly preparing $\sigma_S$ and measuring the ancilla, we prepare the $k^{\text{th}}$ principal component with probability $p_k$.

\section{Gradient Analysis for State Purification Ansatz}\label{app:grads}

% \zo{Computing the gradient for the CCPS ansatz can trivially be done using the standard parameter shift rule/basic differentiation, but we should probably state as such for completeness.}

% \paragraph*{State Purification Ansatz.}

The Hilbert--Schmidt distance cost for the state purification ansatz takes the form $C_{\mbox{\tiny SP}(\vec{\theta},n_A)} = \Tr[\rho^2] + \Tr[\sigma(\vec{\theta},n_A)^2] - 2 \Tr[\rho \sigma(\vec{\theta},n_A)]$, where the trial state $\sigma(\vec{\theta},n_A)$ is found via its purification, i.e., $\sigma_{\mbox{\tiny SP}}(\vec{\theta},n_A) \coloneqq \Tr_A[ U_{\vec{\theta}} (\ket{0}\!\bra{0})^{\otimes(n +n_A)} U_{\vec{\theta}}^\dagger ]$, for our vector of training parameters $\vec{\theta}$.
The gradient with respect to $\vec{\theta}$ is given by 
\begin{equation}
    \nabla C_{\mbox{\tiny SP}\left(\vec{\theta},n_A\right)} = \nabla \Tr\!\left[\rho ^2\right] + \nabla \Tr\!\left[\sigma\left(\vec{\theta},n_A\right)^2\right] - 2 \nabla \Tr\!\left[\rho \sigma\left(\vec{\theta},n_A\right)\right] \, .
\end{equation}
The $\Tr[\rho ^2]$ term vanishes trivially due to a lack of dependence on $\vec{\theta}$, whilst the overlap term $\Tr[\rho \sigma(\vec{\theta},n_A)]$ obeys the standard parameter shift rule~\cite{mitarai2018quantum, schuld2019evaluating}. In our work, we specifically use the \enquote{Pauli parameter shift rule} in which the circuit ans{\"a}tze are described using single qubit gates of the form $P_k = e^{-i \theta \sigma_k / 2}$ for Pauli $\sigma_k$ and parameter free CNOTs. In this case, the shift refers to running the circuit with $\theta \rightarrow \theta \pm \pi / 2$ (this will become clear in the next equation). This choice is motivated by our use of IBM devices where this is the appropriate gate set. 

We can still use the Pauli parameter shift rule for the $\Tr[\sigma(\vec{\theta},n_A) ^2]$ term after applying a matrix differentiation rule~\cite[Eq.~(11.175)]{wilde2011classical}. Namely, given some function $f(x)$ we have that $\frac{\partial}{\partial \theta} \Tr[f(A(\theta))] = \Tr \left[  g(A(\theta) ) \frac{\partial A}{\partial \theta} \right]$ where $g(x) \coloneqq \frac{\partial f}{\partial x}$. Thus we have that 
\begin{equation}
\begin{aligned}
      \frac{\partial  \Tr\left[\sigma(\vec{\theta},n_A) ^2\right]}{\partial \theta_k }  &=  2 \Tr \left[ \sigma(\vec{\theta},n_A)   \frac{\partial  \sigma(\vec{\theta},n_A) }{\partial \theta_k }  \right] \\ &= \Tr \left[ \sigma(\vec{\theta},n_A) \sigma\left(\vec{\theta} + \frac{\pi}{2} \cdot \vec{e}_k,n_A \right)\right] - \Tr\!\left[\sigma(\vec{\theta},n_A) \sigma\left(\vec{\theta} - \frac{\pi}{2} \cdot \vec{e}_k,n_A \right)  \right] \, ,
\end{aligned}
\end{equation}
where in the second line we use the Pauli parameter shift rule (hence the $\pm \pi / 2$) as applied directly to an operator rather than an expectation value.
Thus we have the complete analytic expression to compute $ \nabla C_{\mbox{\tiny SP}\left(\vec{\theta},n_A\right)}$. We note that the gradient analysis for the local costs is entirely analogous.

\section{Description of State Ensembles}

\subsection{\label{app:bures} Bures Random States}

Before diving into what a Bures random state is specifically, it is helpful to discuss a few general considerations regarding generating random quantum states, as is done in the brief introduction of \cite{hall1998random}. Generally, random quantum states are selected from a distribution that is invariant under global unitary transformations. For pure states, this single property uniquely defines a probability measure known as the \emph{Haar measure}. But for mixed states, this property does not uniquely specify the measure. Assuming the distribution of eigenvectors and eigenvalues is independent, then we can write the probability measure over mixed states as~\cite{zyczkowski2011generating}
\begin{equation}
    d \mu = d \nu(\lambda_1, \lambda_2, \ldots, \lambda_N) \times d \mu_V,
\end{equation}
where $d \mu_V$ is the Haar measure and $d \nu$ is the measure over the normalized eigenvalues. So the problem is that we must also specify $d \nu$, which, while trivial for pure states, is unclear for mixed states.  

There are many protocols one can follow to determine $d\nu$, but the most mathematically straightforward is to define a measure from the normalized volume elements of a metric~\cite{hall1998random, zyczkowski2011generating}. In this formalism, the key is then to choose an appropriate metric. The Bures distance, given by 
\begin{equation}
    \label{eq:bures}
    D_B(\rho, \sigma) \coloneqq \sqrt{2 - 2 \Tr \left[ \sqrt{ \sqrt{\rho } \sigma \sqrt{\rho }} \right]},
\end{equation}
has many attractive properties as an unbiased choice. Many of these properties were first pointed out by Bures himself~\cite{bures1969extension}, but simpler explanations are provided in Hall~\cite{hall1998random} and Zyczkowski \textit{et al}.~\cite{zyczkowski2011generating}. The summary in Zyczkowski \textit{et al}.~\cite{zyczkowski2011generating} is especially concise: the Bures metric (i) has an interpretation as a distinguishability measure~\cite{bengtsson2006An, hayashi2016quantum}, (ii) is the minimal monotone metric under quantum channels~\cite{petz1996geometries}, and (iii) gives the statistical distance when applied to two diagonal operators~\cite{zyczkowski2011generating}. The form of $d \nu$ is given in Eq.~(13) of Zyczkowski \textit{et al}.~\cite{zyczkowski2011generating}, and for brevity, we call states drawn from this measure \emph{Bures random} states. 

For numerically tractable system sizes, generating a Bures random state is straightforward~\cite{zyczkowski2011generating}. First, we generate a $2^n \times 2^n$ Ginibre random matrix $G$ \cite{ginibre1965statistical} with complex entries\footnote{The exact description is just a matrix whose entries are $G_{ij} = x + i y$ where $x, y \sim \mathcal{N}(0, 1)$.}. Then, we generate a Haar random unitary matrix $U$ with the same dimensions. With these two matrices, the random state is given by
\begin{equation}
    \label{eq:bures-random}
    \rho_B = \frac{(\id + U)G G^{\dagger}(\id + U^{\dagger})}{\Tr[(\id + U)G G^{\dagger}(\id + U^{\dagger})]}.
\end{equation}
An alternative, more physically motivated way to generate the states is to construct a superposition of a random bipartite state $\ket{\psi_1} = U_{AB} \ket{0,0}$ with a local transformation of the same state, $\ket{\psi_2} = (V_A \otimes \id) \ket{\psi_1}$ and then trace out the $B$ degrees of freedom (see Fig.~5 in Ref.~\cite{zyczkowski2011generating}), 
\begin{equation}
    \label{eq:bures-intuitive-generation}
    \rho_A = \frac{\Tr_B \ketbra{\phi}{\phi}}{\braket{\phi}{\phi}} \ \ \ \ket{\phi} \equiv [(\id + V_A) \otimes \id] \ket{\psi_1}. 
\end{equation}

\subsection{\label{app:xy-thermal} XY Thermal States}

\begin{figure}
    \centering
    \includegraphics[width=0.45\textwidth]{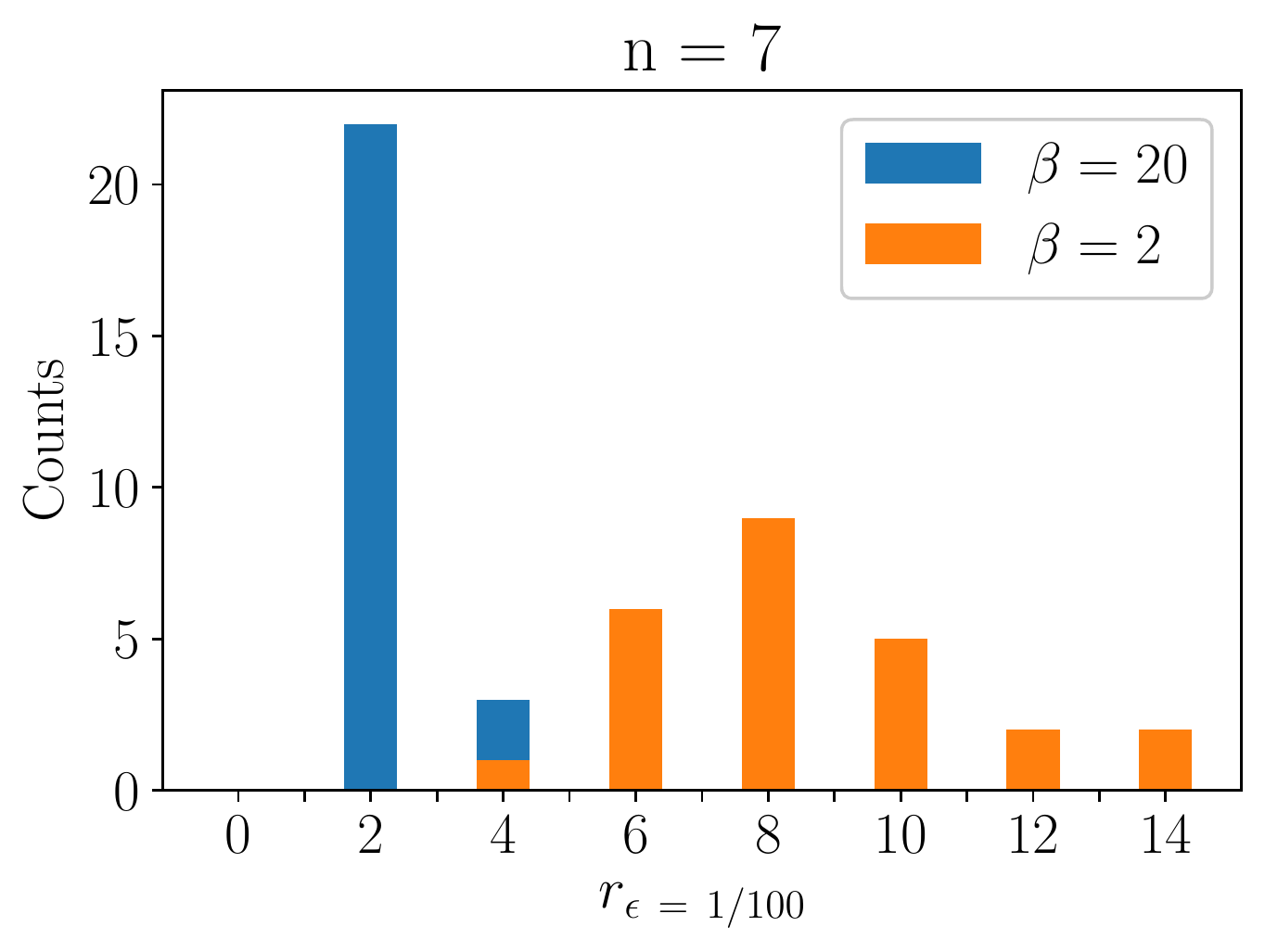}
    \includegraphics[width=0.45\textwidth]{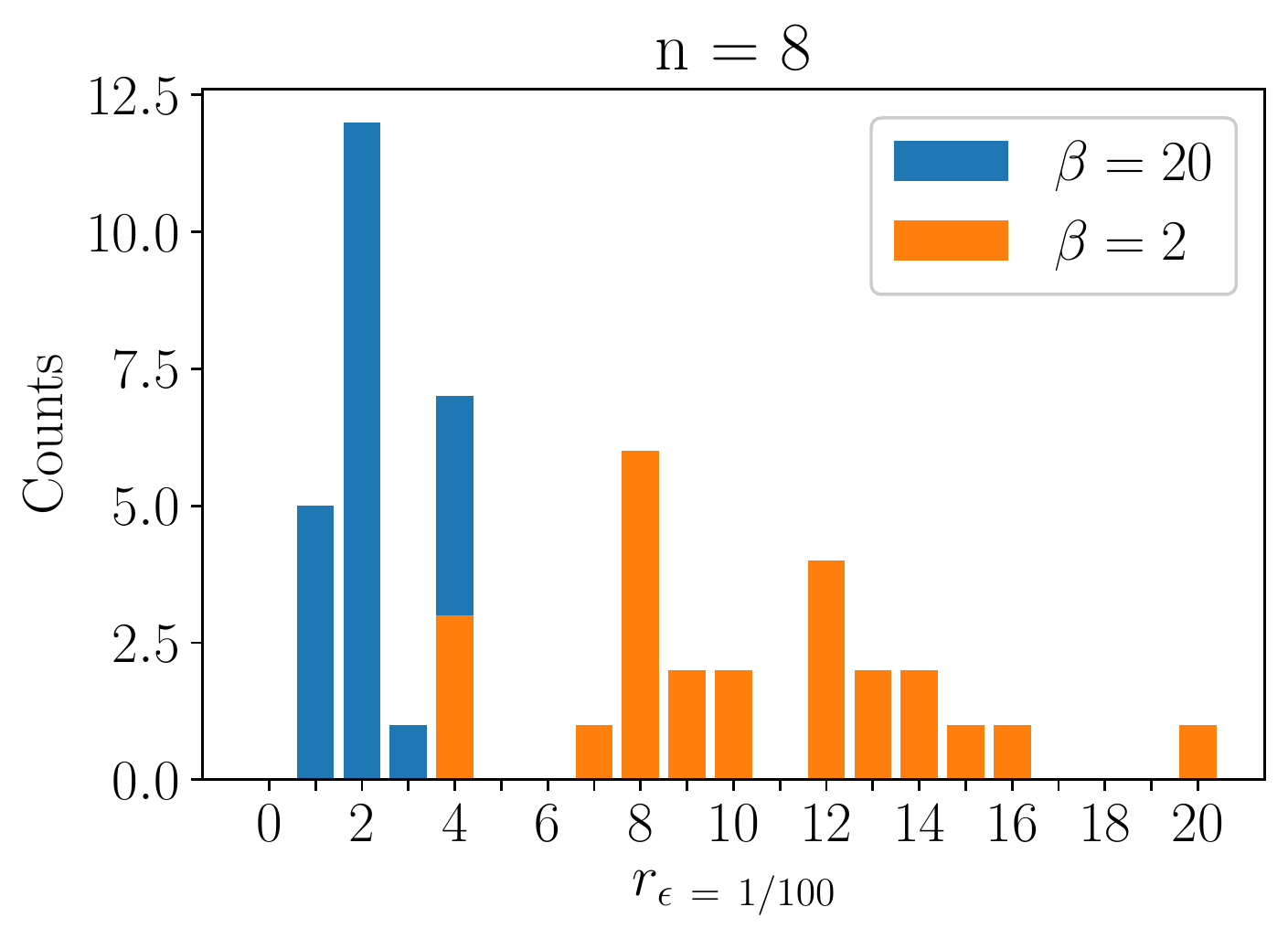}
    \caption{\textbf{XY thermal state $\boldsymbol{\epsilon}$-rank depends on $\boldsymbol{\beta}$.} We show the distribution of $\epsilon$-ranks $(r_\epsilon)$ for the $n = 7$ and $n = 8$ XY thermal state instances (25 per $(n, \beta)$). At $\beta = 20$ (low temperature), the resulting rank never goes beyond $r_\epsilon = 4$. For $n = 7$, the ground state is doubly degenerate, so $r_\epsilon \geq 2$ as well. For $n = 8$, the ground state is unique, so $r_\epsilon = 1$ is reached for some of the states. For $\beta = 2$, the ranks go from $r_\epsilon = 4$ all the way up to $r_\epsilon = 14, 20$ for $n = 7, 8$, respectively.}
    \label{fig:xy-eps-rank}
\end{figure}

By XY thermal states, we mean states of the form
\begin{subequations}
    \begin{align}
        \rho^{(XY)}_n &= \frac{e^{- \beta H_{\text{XY}}}}{\Tr[e^{-\beta H_{\text{XY}}}]},\\
        H_{\text{XY}} &= \sum_{i=1}^{n-1} J_i X_i X_{i+1} + K_i Y_i Y_{i+1} ,
        %------------------
    \end{align}
\end{subequations}
where $J_i, K_i \sim \mathcal{N}(0, 1)$ are i.i.d.~normal random variables. By controlling $\beta$, we control the effective rank $r_\epsilon$ of the random states generated in this way, which we demonstrate in Fig.~\ref{fig:xy-eps-rank} with $\epsilon = 1 / 100$. The reason is simple: At $\beta = \infty$ (i.e., $T = 0$), we expect a system with a non-degenerate ground state to be in its pure ground state which has $r_\epsilon = 1$. For a degenerate ground state, we get a totally mixed state in the ground-state subspace, so if the degeneracy is $g_0$, we find $r_\epsilon = g_0$. For small but non-zero temperatures ($\beta = 20$ here), the state is a convex combination of low-lying energy states, so $r_\epsilon > g_0$ for most choices of $\epsilon$. As $\beta$ decreases, the state becomes closer to the totally mixed state until it actually reaches it at $\beta = 0$. By choosing a larger intermediate temperature ($\beta = 2$ here), we interpolate between the two extremes and simply get a Gibbs state with large $\epsilon$-rank but without being completely mixed.  

Despite the random states having different coefficients (and subsequent rank), the model still respects important symmetries. Most importantly, our XY model is a chain in 2D. Hence, it is invariant under any global rotation of all spins, and furthermore, the number of spins is constant. These symmetries allow us to greatly simplify the choice of ansatz to one that respects these symmetries, as discussed in Appendix~\ref{app:ansatze}.

\section{\label{app:computing-fig-of-merit} Computing Figure of Merit}

Our figure of merit comes from the solution to the quantum low-rank approximation problem (QLRAP)~\cite{ezzell2022quantum}. Namely, the QLRAP is to find the state $\sigma_{\text{opt}}(R)$ that satisfies
\begin{equation}
    \label{eq:quantum-low-rank-approx-prob}
    \sigma_{\text{opt}}(R) = \argmin_{\sigma \geq 0,\rank(\sigma) \leq R, \Tr(\sigma)=1} D(\rho, \sigma)^2.
\end{equation}
As shown in Ref.~\cite{ezzell2022quantum}, the unique optimal solution for the Hilbert--Schmidt distance is given by
\begin{subequations}
    \begin{align}
    \sigma_{\text{opt}}(R) &= \tau_R + N_R, \\ 
    %---------------
    \tau_R &\equiv \Pi_R \rho \Pi_R, \\
    %---------------
    N_R &\equiv \frac{1 - \Tr[\tau_R]}{R} \Pi_R,
    \end{align}
\end{subequations}
with corresponding minimal Hilbert--Schmidt cost 
\begin{equation}
    \label{eq:exact-opt-cost}
    C(\vec{\alpha}_{\text{opt}}, R) = \Tr\left[\left(I - \Pi_R\right) \rho^2\right] + \Tr\left[N_R^2\right]. 
\end{equation}
With a little algebra, it is easy to see that the first term corresponds to the sum of squares of the $r - R$ eigenvalues of $\rho$ not approximated (assuming $\rank(\rho) = r)$, and the second term accounts for the constant offset between the first $R$ eigenvalues of $\rho$ and the $R$ re-normalized eigenvalues of $\sigma_{\text{opt}}$.

\begin{figure}
    \centering
    \includegraphics[width=0.45\textwidth]{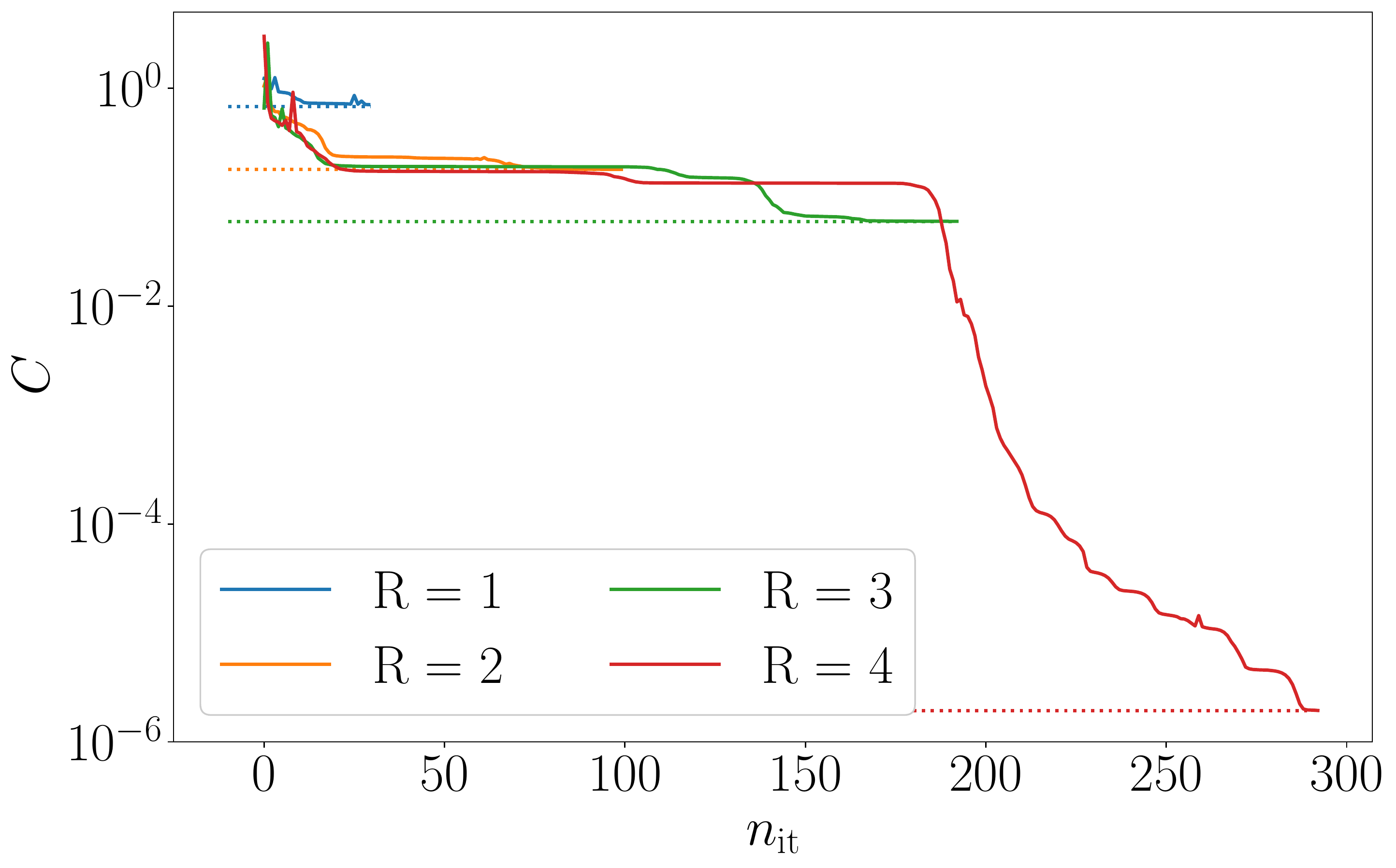}
    \includegraphics[width=0.45\textwidth]{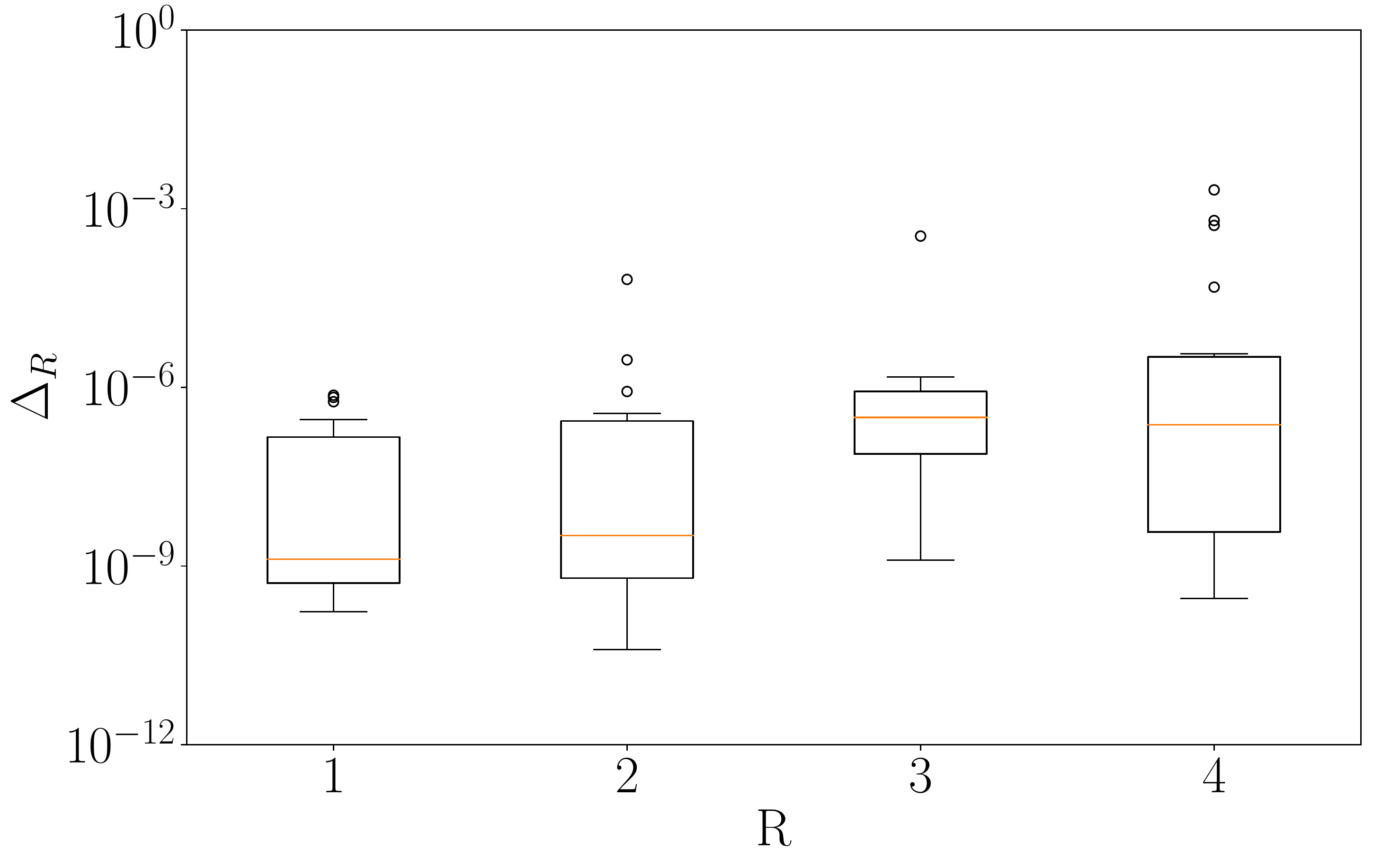}
    \caption{\textbf{Demonstrating meaning of $\boldsymbol{\Delta_R}$ performance metric.} On the left, we show an example optimization of a seven-qubit XY model thermal state which illustrates what $\Delta_R$ means pictorially. Namely, for the same fixed state, we use the CCPS ansatz to find a rank $R=1, 2, 3, 4$ approximation. The resulting optimizations terminate when the learned cost is approximately equal to the optimal costs represented by horizontal lines. The difference between the final cost and the optimal cost is what is plotted on the right box-plots. The variation in performance for each $R$ comes from sampling over 25 random states (see Appendix~\ref{app:xy-thermal}).} 
    \label{fig:figure-of-merit}
\end{figure}

Any compilation of $\rho$ as described throughout our paper will find an empirical cost $C(\vec{\alpha}^*, R) \geq C(\vec{\alpha}_{\text{opt}}, R)$, so a simple figure of merit is just their difference:
\begin{equation}
    \Delta_R \equiv C(\vec{\alpha}^*, R) - C(\vec{\alpha}_{\text{opt}}, R). 
\end{equation}
In Fig.~\ref{fig:figure-of-merit}, we help clarify how we compute this difference for an example optimization of a seven-qubit XY thermal state using the CCPS ansatz with $R \in \{ 1, 2, 3, 4\}$. Pictorially, $\Delta_R$ is the vertical distance between the final cost (solid line) and the optimal cost (horizontal dashed line). 

\section{Additional hardware results \label{app:additional-hardware-results}}

\begin{figure}
    %---------------------------------
    \subfigure[Direct compilation of $\Tilde{\rho}_{+}$ using the SP ansatz]{
    \centering
    \label{fig:addres-sp}
    \includegraphics[width=0.48\textwidth]{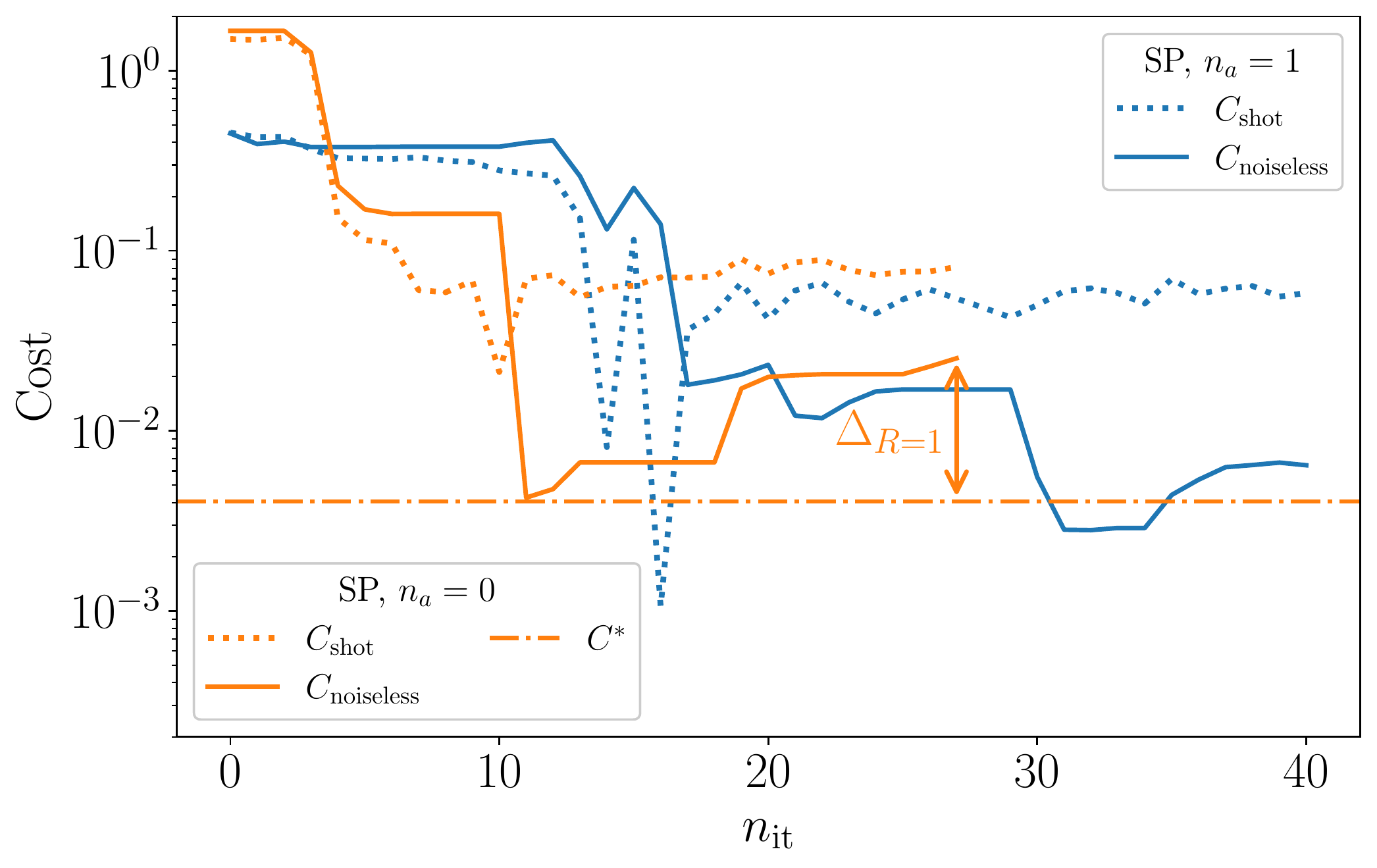}
     }
    %-----------------
    \subfigure[Direct compilation of $\rho_{\text{HS}}$ using the CCPS ansatz]{
    \centering
    \label{fig:addres-ccps}
    \includegraphics[width=0.48\textwidth]{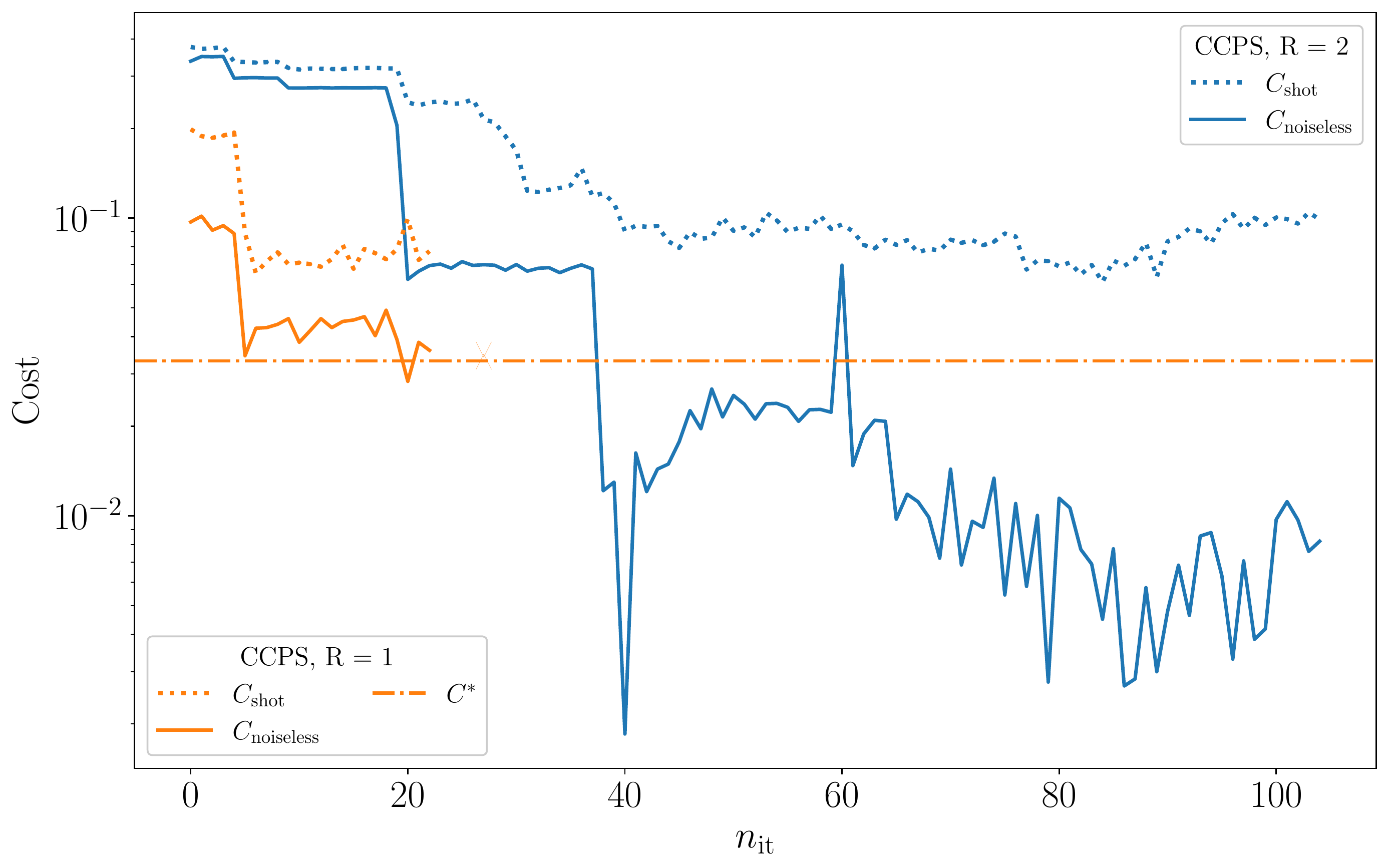}
     }
      %---------------------------------
    \caption{\textbf{Direct low rank compilations of single qubit hardware states.} We perform a direct full-rank and low-rank compilation of $\rho_{\text{HS}}$ with the SP ansatz on the (left) and of $\Tilde{\rho}_+$ with the CCPS ansatz on the (right). By direct we mean we performed a separate and independent compilation of the same state with an ansatz supporting a rank two or rank one approximation. Since these are one qubit states, this encompasses all interesting possibilities. For the rank one approximation, we also include a horizontal line, $C^*$, which denotes the lowest possible value of the noiseless cost. The difference between the final value of $C_{\text{noiseless}}$ and this line denotes $\Delta_{R}$ as shown explicitly on the (left) plot.}
    \label{fig:addres-direct}
\end{figure}

\begin{figure}[ht]
    %-----------------
    \subfigure[A direct full rank ($R = 4$) compilation of $\Tilde{\rho}_{\Phi^+}$ with the CCPS ansatz and the corresponding indirect compilation cost of the $R = 2$ approximation by truncation]{
    \centering
    \label{fig:jakarta-a}
    \includegraphics[width=0.5\textwidth]{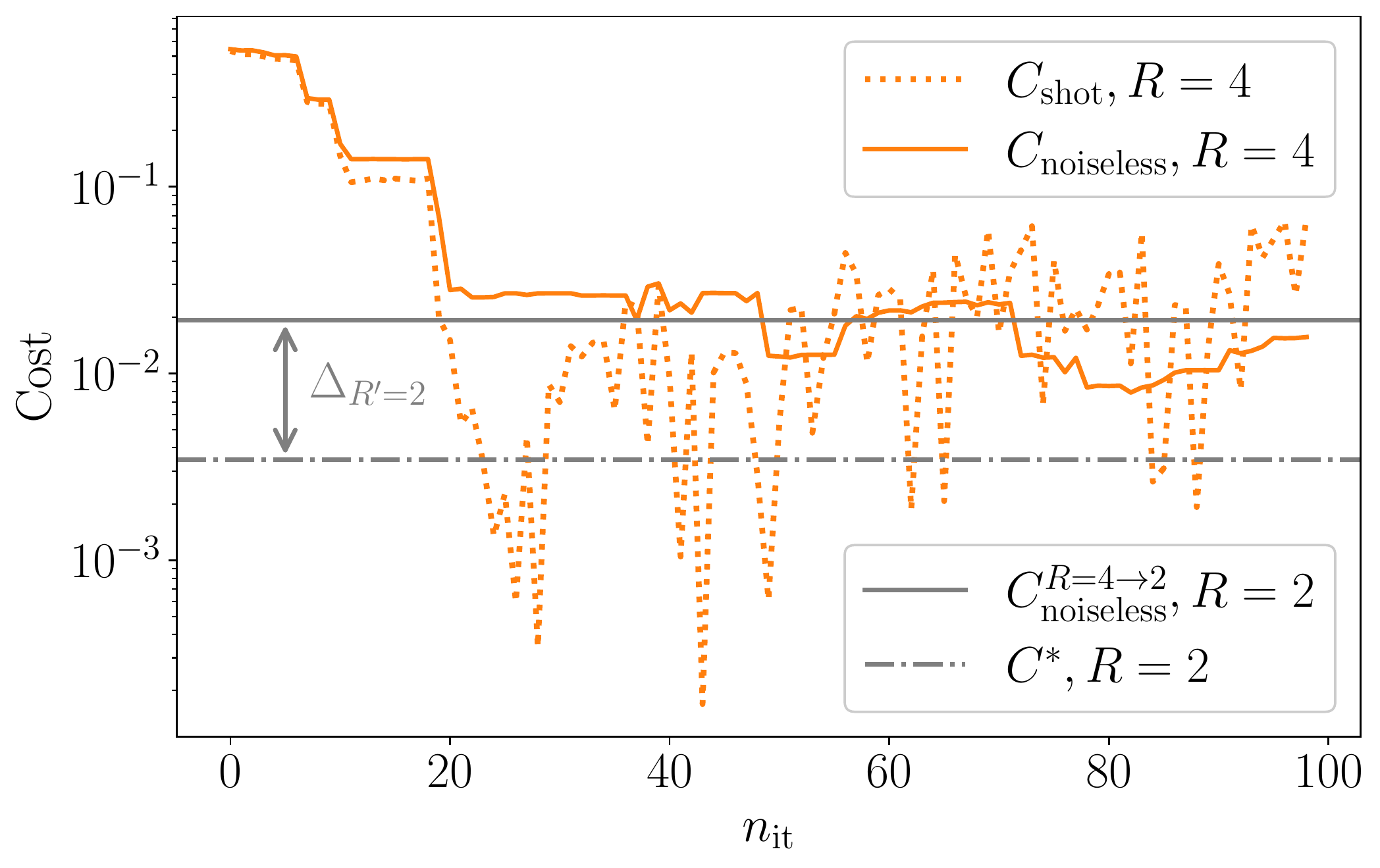}
     }
    %---------------------------------
    \subfigure[Effect of truncation from $R = 4 \rightarrow R'$ on the numerically optimal cost $C(\vec{\alpha}^*, R \rightarrow R')$ and the optimal possible cost $C(\vec{\alpha}_{\text{opt}}, R \rightarrow R')$]{
    \centering
    \label{fig:jakarta-b}
    \includegraphics[width=0.48\textwidth]{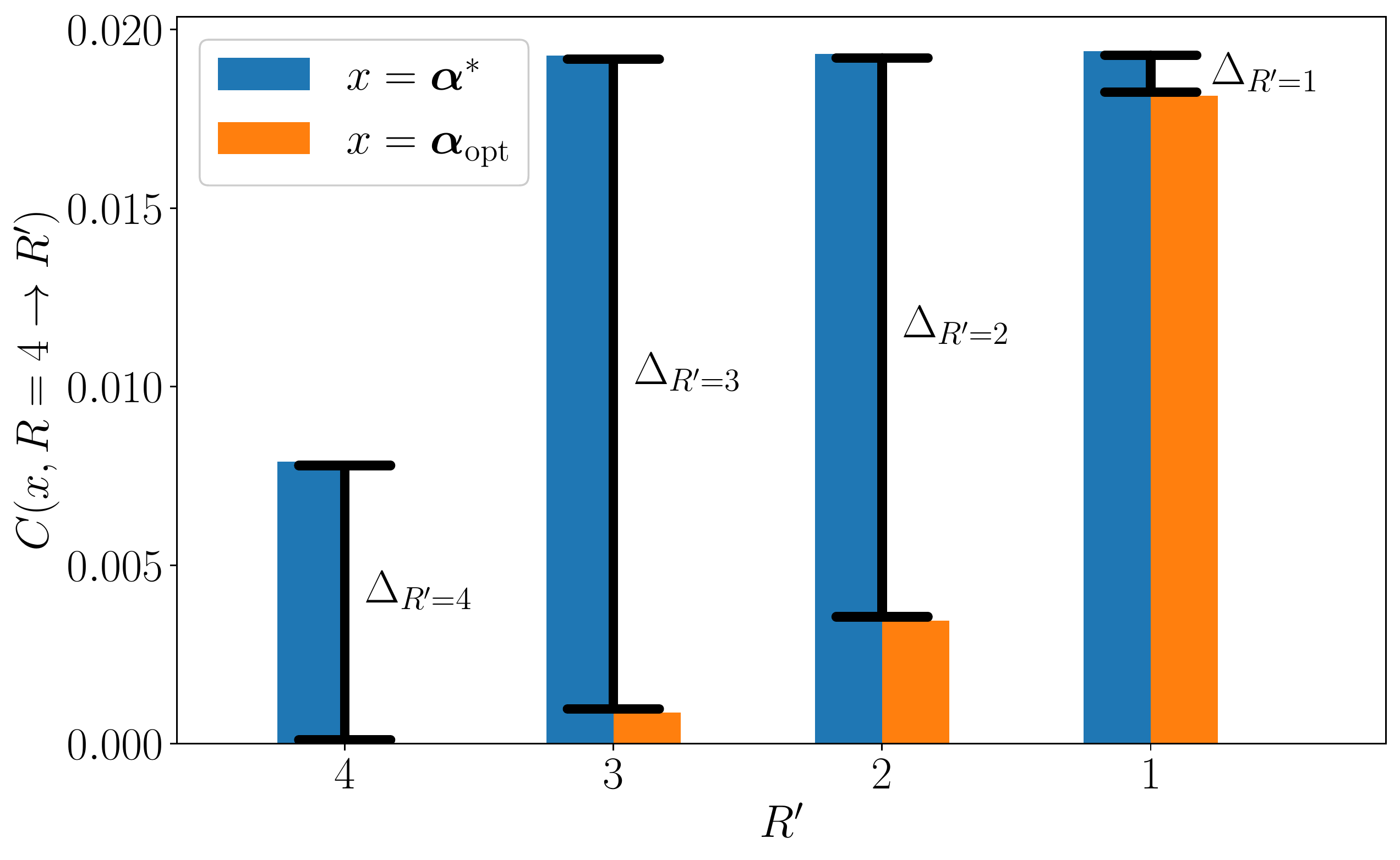}
     }
    %---------------------------------
    \subfigure[Principal values (ordered spectrum) of target state $\Tilde{\rho}_{\Phi^+}$ and the learned CCPS state $\sigma_{\text{CCPS}}(\vec{\alpha}^*, R = 4)$]{
    \centering
    \label{fig:jakarta-c}
    \includegraphics[width=0.48\textwidth]{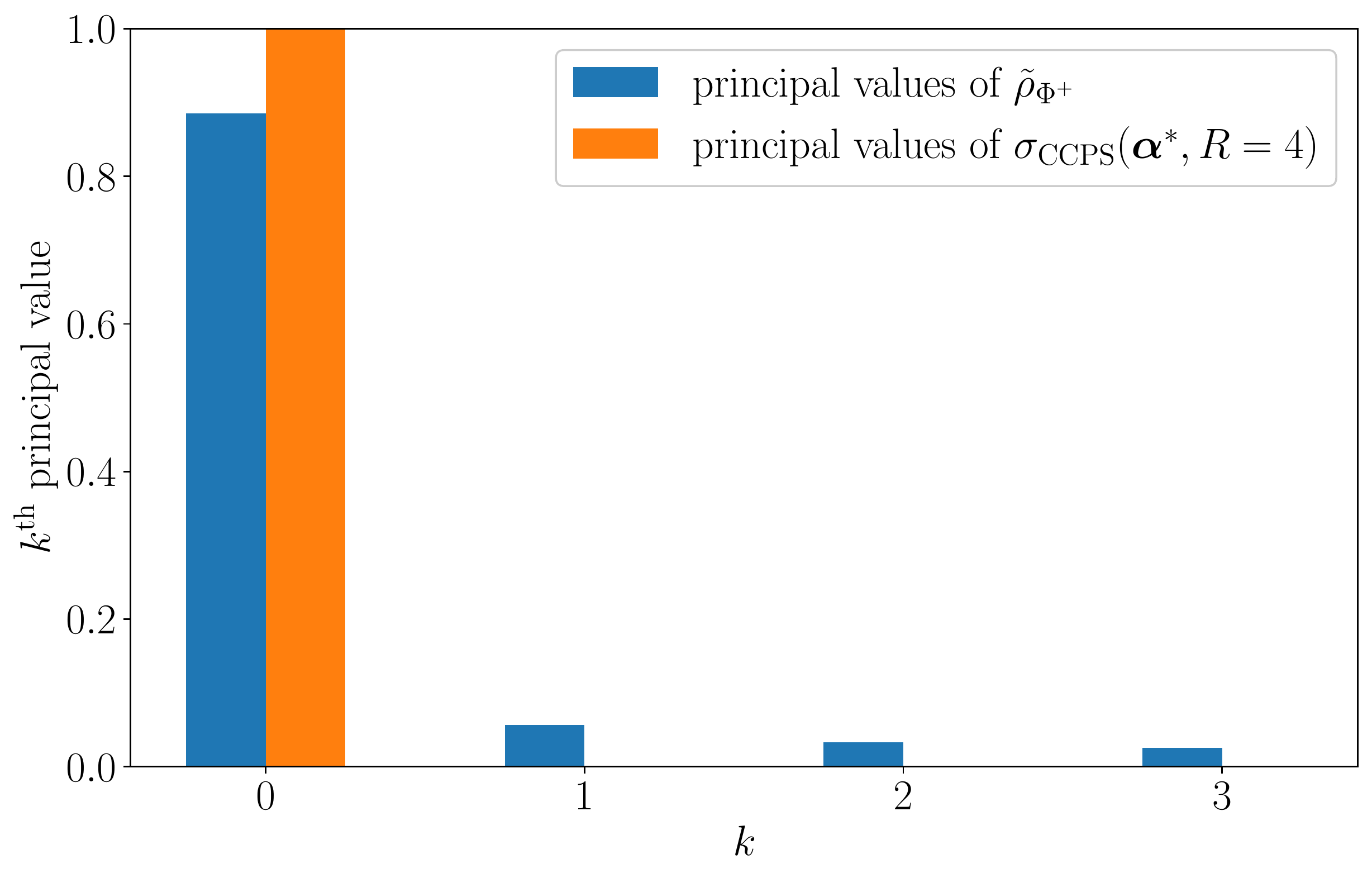}
    }
    %---------------------------------
    \caption{\textbf{PCA analysis of $\boldsymbol{\Tilde{\rho_{\Phi^+}}}$ when compiled on hardware using CCPS ansatz.} In (a), we show the optimization result when finding a direct full-rank ($R = 4)$ compilation of $\Tilde{\rho}_{\Phi^+}.$ Given this CCPS ansatz, we can obtain any $R' < 4$ approximation for free by truncation. As an example, we plot the found cost when truncating to a rank two approximation, $C_{\text{noiseless}}^{R = 4 \rightarrow 2}$, as a solid horizontal line. The optimal possible value for this cost, $C^*$, is shown in a dot-dashed line. The vertical distance between these lines denotes their difference, $\Delta_{R' = 2}$.\\
    %------------------------------------------
    In (b), we show the quality of our truncation state for different truncation levels from $R' = 4$ (no truncation) to $R' = 1$ (maximum truncation to a pure state). For each $R'$, we plot the value of the found truncated cost $C(\vec{\alpha}^*, R \rightarrow R')$ (left, blue) as well as the optimal possible cost $C(\vec{\alpha}_{\text{opt}}, R \rightarrow R')$ (right, orange). The difference between these costs is $\Delta_{R'}$ and is pictorially the height difference in the bars.\\
    %------------------------------------------
    In (c), we show the principal values (ordered eigenvalues) of the target state $\Tilde{\rho}_{\Phi^+}$ and the learned CCPS state $\sigma_{\text{CCPS}}(\vec{\alpha}^*, R = 4)$.
    }
    \label{fig:addres-indirect}
\end{figure}

In Sec.~\ref{subsec:results-qh}, we successfully compiled full rank approximations of the one qubit states $\rho_{\text{HS}}$ and $\Tilde{\rho}_{+}$ and the two qubit state $\Tilde{\rho}_{\Phi^+}$ (see the referenced section for state definitions). In particular, within an allotted budget of 100 Powell iterations, we optimized to a cost value on the order of $10^{-3} \leq C_{\text{noiseless}} \leq  10^{-2}$ which is close to the precision floor allowed by our use of $10^{5}$ shots to evaluate the cost function. Here, we present additional hardware results where we compile lower rank approximations of these states either directly as in Fig.~\ref{fig:addres-direct} or indirectly through truncation of the learned full rank CCPS state as in Fig.~\ref{fig:addres-indirect}. For convenience, we have included the full-rank optimizations shown in Sec.~\ref{subsec:results-qh} alongside the lower rank optimizations.  

In Fig.~\ref{fig:addres-sp}, we compile a full rank (i.e., rank two using $n_A = 1$ ancilla) and a pure state ($n_A = 0$) approximation of the single qubit state $\Tilde{\rho}_+$ using the SP ansatz. In the full rank case, we find $C_{\text{noiseless}}(n_A = 1) = \Delta_{2}  \approx 6.44 \times 10^{-3}$. In the pure state case, we find $C_{\text{noiseless}}(n_A = 0) = 2.5 \times 10^{-2}$, but the optimal possible cost is $C^*(n_A = 0) = 4.1 \times 10^{-3}$. Thus, $\Delta_1  = 2.1 \times 10^{-2}$, which is an acceptable final value on the order of $10^{-2}$. It is also interesting to note that around $n_{\text{it}} = 10$, the difference is minimal, reaching $C_{\text{noiseless}}(n_A = 0) = 4.3 \times 10^{-3} \implies \Delta_1 = 2.0 \times 10^{-3}$. However, it is not forthright to report this as the final found value since it relies on knowledge of $C_{\text{noiseless}}$ to pick the right $n_{\text{it}}$ whereas our optimization stopping condition does not (i.e., it relies on $C_{\text{shot}}$ only). 

In Fig.~\ref{fig:addres-ccps}, we compile a full rank ($R = 2$) and a pure state ($R = 1$) approximation of the single qubit state $\rho_{\text{HS}}$ using the CCPS ansatz. In the full rank case, we find $C_{\text{noiseless}}(R = 2) = \Delta_{2} = 8.2 \times 10^{-3}$. In the pure state case, we find $C_{\text{noiseless}}(R = 1) = 4.3 \times 10^{-2}$, but the optimal possible cost is $C^*(R = 1) = 3.3 \times 10^{-2}$, so $\Delta_1  = 9.9 \times 10^{-3}$ which is an acceptable final value on the order of at least $10^{-2}$. 

Overall, we find that direct pure state compilations of $\rho$ for both ans\"{a}tze are learned to an acceptable value of $\Delta_1$. In addition, we find that finding this value takes fewer iterations for both ans\"{a}tze suggesting that finding a lower-rank approximation is easier. This observations was also found for the Bures random and XY random thermal states in idealized noiseless, infinite shot classical simulations. Given that the lower-rank optimization requires fewer learnable parameters alongside these empirical results, we suspect this to be a general property of our algorithm. Namely, learning a lower rank approximation is easier.

In Fig.~\ref{fig:addres-indirect} we explore the quality of indirect compilations of $\Tilde{\rho}_{\Phi^+}$ by truncating a full-rank compilation. The results here mirror those in Sec.~\ref{subsec:pca-example} but for the hardware optimization of $\Tilde{\rho}_{\Phi^+}$. The story here is a bit more interesting on account of finite shot noise, however. The punchline is that due to having only a precision of roughly $10^{-2}$, our $R = 4$ optimization found a local minimum, $C_{\text{noiseless}}(R = 4) = 6.8 \times 10^{-2}$ (see Fig.~\ref{fig:jakarta-a}), that is effectively a rank one approximation. In other words, a pure state approximation of $\Tilde{\rho}_{\Phi^+}$ is sufficient to reach the cost noise floor, and this pure state solution was found in our hardware optimization.

With the punchline stated, the empirical quality of the truncated states is summarized in Fig.~\ref{fig:jakarta-b}. Evidently, the pure state approximation ($R' = 1)$ is closest to its optimal value since $\Delta_{R' = 1} = 1.2 \times 10^{-3}$. On the other hand, $\Delta_{R' > 1}$ is an order of magnitude higher in all cases, i.e., $\Delta{R' = 2} = 1.6 \times 10^{-2}$, $\Delta_{R' = 3} = 1.8 \times 10^{-2}$, and $\Delta_{R' = 4} = 6.8 \times 10^{-2}$. In fact, the quality monotonically decreases with $R'$ increasing. This suggests that the performance of our $R = 4$ compilation is dominated by the quality of the estimate of the first principal component. In Fig.~\ref{fig:jakarta-c}, we verify this intuition by seeing that the numerically optimal CCPS state, $\sigma_{\text{CCPS}}(\vec{\alpha}^*, R = 4)$, is essentially a pure state. Namely, it only has non-trivial support on the largest principal value. This interesting observation aside, we ultimately find that truncated costs have a comparable $\Delta_{R'}$ to the original $\Delta_R$, so the indirect compilation of the full-rank CCPS state into lower rank approximations works well.

\end{document}